\documentclass[12pt]{article}

\usepackage{graphics,amssymb,float}
\usepackage{graphicx}
\usepackage{rotating}
\usepackage{dcolumn}
\usepackage{bm}
\usepackage{cite}
\usepackage{amsmath}
\usepackage{indentfirst} 
\usepackage{epstopdf}

\def\br{{\mathcal B}}
\def\brinv{\br_{\rm inv}}
\def\CU{C_U}
\def\CD{C_D}

\def\CV{C_V}
\def\CG{C_g}
\def\CP{C_\gamma}
\def\cu{\CU}
\def\cd{\CD}
\def\cv{\CV}
\def\cg{\CG}
\def\cp{\CP}
\def\sina{\sin\alpha}
\def\sa{s_\alpha}
\def\ca{c_\alpha}
\def\cosa{\cos\alpha}
\def\tanb{\tan\beta}
\def\sinb{\sin\beta}
\def\cosb{\cos\beta}
\def\cotb{\cot\beta}

\def\cf{C_F}
\def\chimin{\chi^2_{\rm min}}

\def\gam{\gamma}
\def\ie{{\it i.e.}}
\def\eg{{\it e.g.}}

\def\dcg{\Delta \CG}
\def\dcp{\Delta \CP}

\def\gev{~{\rm GeV}}
\def\chisq{\chi^2}
\def\anti{\overline}
\def\fbi{~{\rm fb}^{-1}}

\def\bit{\begin{itemize}}
\def\eit{\end{itemize}}
\def\ben{\begin{enumerate}}
\def\een{\end{enumerate}}
\def\beq{\begin{equation}}
\def\eeq{\end{equation}}

\def\Eq#1{Eq.~(\ref{#1})}
\def\dchisq{\Delta\chisq}

\def\honepr{H_1^{0\,\prime}}
\def\hone{H_1^0}

\def\hpr{H'}
\def\mhpr{m_{\hpr}}

\def\hthreepm{H_3^\pm}
\def\wmp{W^\mp}

\def\hthree{H_3^0}

\def\mhfive{m_{H_5}}
\def\mhthree{m_{H_3}}
\def\ch{c_H}
\def\sh{s_H}
\def\vev#1{\langle #1 \rangle}
\def\lam{\lambda}
\def\rta{\rightarrow}
\def\bea{\begin{eqnarray}}
\def\eea{\end{eqnarray}}
\def\epem{e^+e^-}

\def\lsim{\mathrel{\raise.3ex\hbox{$<$\kern-.75em\lower1ex\hbox{$\sim$}}}}
\def\gsim{\mathrel{\raise.3ex\hbox{$>$\kern-.75em\lower1ex\hbox{$\sim$}}}}
\def\ifmath#1{\relax\ifmmode #1\else $#1$\fi}

\begin{document}
\begin{titlepage}
\begin{center}

\vspace*{-1cm}
\begin{flushright}
LAPTH-032/13\\
LPSC1319\\
LPT Orsay 13-40\\
UCD-2013-3\\
\end{flushright}

\vspace*{1.6cm}
{\Large\bf Global fit to Higgs signal strengths and couplings\\[3mm] and implications for extended Higgs sectors} 

\vspace*{1cm}\renewcommand{\thefootnote}{\fnsymbol{footnote}}

{\large 
G.~B\'elanger$^{1}$\footnote[1]{Email: belanger@lapth.cnrs..fr},
B.~Dumont$^{2}$\footnote[2]{Email: dumont@lpsc.in2p3.fr},
U.~Ellwanger$^{3}$\footnote[3]{Email: Ulrich.Ellwanger@th.u-psud.fr},
J.~F.~Gunion$^{4}$\footnote[4]{Email: jfgunion@ucdavis.edu}, 
S.~Kraml$^{2}$\footnote[5]{Email: sabine.kraml@lpsc.in2p3.fr}
} 

\renewcommand{\thefootnote}{\arabic{footnote}}

\vspace*{1cm} 
{\normalsize \it 
$^1\,$LAPTH, Universit\'e de Savoie, CNRS, B.P.110, F-74941 Annecy-le-Vieux Cedex, France\\[2mm]
$^2\,$Laboratoire de Physique Subatomique et de Cosmologie, UJF Grenoble 1,
CNRS/IN2P3, INPG, 53 Avenue des Martyrs, F-38026 Grenoble, France\\[2mm]
$^3\,$Laboratoire de Physique Th\'eorique, UMR 8627, CNRS and
Universit\'e de Paris--Sud, F-91405 Orsay, France\\[2mm]
$^4\,$Department of Physics, University of California, Davis, CA 95616, USA}

\vspace{1cm}

\begin{abstract}
The most recent LHC data have provided a considerable improvement in the
precision with which various Higgs 
production and decay channels
have been measured. Using all available public results from ATLAS, CMS
and the Tevatron, we derive for each final state the
combined confidence level contours for the signal strengths in the
(gluon fusion + ttH associated production) versus (vector boson fusion + VH associated
production) space.
These ``combined signal strength ellipses'' can be used in a simple, generic 
way to constrain a very wide class of New Physics models in which the couplings of 
the Higgs boson deviate from the Standard Model prediction.
Here, we use them to constrain the reduced
couplings of the Higgs boson to up-quarks, down-quarks/leptons and
vector boson pairs. We also consider New Physics contributions to the
loop-induced gluon-gluon and photon-photon couplings of the Higgs, as
well as invisible/unseen decays.
Finally, we apply our fits to some simple models with an extended Higgs
sector, in particular to Two-Higgs-Doublet models of Type~I and Type~II,
the Inert Doublet model, and the Georgi--Machacek triplet Higgs model.  
\end{abstract}

\end{center}

\end{titlepage}

\section{Introduction} \label{sintro}

That the mass of the Higgs boson is about 125--126~GeV is a very fortunate circumstance in that we can  
detect it in many different production and decay channels~\cite{atlas:2012gk,cms:2012gu}.
Indeed, many distinct signal strengths, defined as production$\times$decay rates relative to Standard Model (SM) expectations, $\mu_i\equiv (\sigma\times {\rm BR})_i/(\sigma\times {\rm BR})_i^{\rm SM}$, 
have been measured with unforeseeable 
precision already with the 7--8~TeV LHC run~\cite{ATLAS-CONF-2013-034,CMS-PAS-HIG-13-005}. 
From these signal strengths one can obtain information about the couplings of 
the Higgs boson to electroweak gauge bosons, fermions (of the third generation) and 
loop-induced couplings to photons and gluons.

According to the latest measurements presented at 
the 2013 Moriond~\cite{ATLAS-CONF-2013-011,ATLAS-CONF-2013-012,ATLAS-CONF-2013-013,ATLAS-CONF-2013-014,
ATLAS-CONF-2013-030,ATLAS-CONF-2013-034,
CMS-PAS-HIG-13-001,CMS-PAS-HIG-13-002,CMS-PAS-HIG-13-003,
CMS-PAS-HIG-13-004,CMS-PAS-HIG-13-005,CMS-PAS-HIG-13-009,CMS-PAS-HIG-12-053,
Chatrchyan:2013yea,Chatrchyan:2013lba} and 
LHCP~\cite{Aaltonen:2013kxa,CMS-PAS-HIG-13-012,CMS-PAS-HIG-13-015} 
conferences, these couplings seem to coincide well with those expected in the SM. 
This poses constraints on various beyond the Standard Model (BSM) theories, in which
these couplings can differ substantially from those of the SM. The Higgs
couplings can be parametrized in terms of effective Lagrangians
\cite{Carmi:2012yp,Azatov:2012bz,Espinosa:2012ir,Klute:2012pu,Azatov:2012wq,Low:2012rj,Corbett:2012dm,Giardino:2012dp,Alanne:2013dra,Ellis:2012hz,Montull:2012ik,Espinosa:2012im,Carmi:2012in,Banerjee:2012xc,Bertolini:2012gu,Bonnet:2012nm,Plehn:2012iz,Elander:2012fk,Djouadi:2012rh,Dobrescu:2012td,Moreau:2012da,Cacciapaglia:2012wb,Corbett:2012ja,Masso:2012eq,Azatov:2012qz,Belanger:2012gc,Cheung:2013kla,Celis:2013rcs,Belanger:2013kya,Falkowski:2013dza,Cao:2013wqa,Giardino:2013bma,Ellis:2013lra,Djouadi:2013qya,Chang:2013cia,Dumont:2013wma,Bechtle:2013xfa} 
whose structure depends, however, on the class of models considered, such as
extended Higgs sectors, 
extra fermions and/or scalars contributing to loop diagrams, composite
Higgs bosons and/or fermions, nonlinear realizations of electroweak
symmetry breaking, large extra dimensions, Higgs--dilaton mixing and
more.

When such generalized couplings are used to fit the  large
number of measurements of signal strengths now available in different
channels, one faces the problem that the
experimentally defined signal categories (based on combinations of cuts) nearly always
contain  superpositions of different production modes 
and thus errors (both systematic and statistical) in different channels are correlated. 
Ideally one would like to fit not to experimentally defined categories but rather to
the different production and decay modes which lead to distinct final states and kinematic distributions.
The five usual theoretically ``pure'' production modes are gluon--gluon fusion (ggF),   
vector boson fusion (VBF), associated production with a $W$ or $Z$ boson (WH and ZH, commonly denoted as VH), 
and associated production with a top-quark pair (ttH). 
The scheme conveniently adopted by the experimental collaborations is to group these five 
modes into just two effective modes ggF + ttH and VBF + VH  and present 
contours of constant likelihood ${\cal L}$ for particular final states in the 
$\mu({\rm ggF + ttH})$ versus $\mu({\rm VBF + VH})$ plane.
This is a natural choice for the following reasons: 
\bit
\item Deviations from custodial symmetry, which implies a SM-like ratio
of the couplings to $W$ and $Z$ gauge bosons, are strongly constrained by the Peskin--Takeuchi $T$ parameter~\cite{Peskin:1990zt,Peskin:1991sw} from electroweak fits~\cite{Baak:2012kk}. Furthermore, there is no indication of such deviation from the Higgs measurements performed at the LHC~\cite{CMS-PAS-HIG-13-005,ATLAS-CONF-2013-034}. Hence, one can assume that the VBF and VH
production modes both depend on a single generalized coupling of the Higgs
boson to $V=W,Z$ and it is therefore appropriate to combine results for these two channels.
\item Grouping ggF and ttH together is more a matter of convenience in order to be able to present two-dimensional likelihood plots. Nonetheless, there are some physics motivations for considering this combination, the primary one being that, in the current data set, ggF and ttH are statistically independent since they are probed by different  final states:  ttH via $H\to b\bar b$ and ggF via a variety of other final states such as $\gam\gam$ and $ZZ^*$. While the ttH production rate depends entirely on the $Ht\bar t$ coupling, ggF production occurs at one loop and is sensitive to both the $Ht\bar t$ coupling and the $H b\bar b$ couplings as well as to BSM loop diagrams. Although in the SM limit ggF is roughly 90\% determined by the $Ht\bar t$ coupling, leading to a strong correlation with the ttH process, this need not be the case in models with suppressed $Ht\bar t$ coupling and/or enhanced $Hb\bar b$ coupling
and most especially in models with BSM loops.
\eit
The final states in which the Higgs is observed include $\gam\gam$,
$ZZ^{(*)}$, $WW^{(*)}$, $b\bar{b}$ and $\tau\tau$. However, they do not all scale independently.  
In particular, custodial symmetry implies that the branching fractions into $ZZ^{(*)}$ and
$WW^{(*)}$ are rescaled by the same factor with respect to the SM.  
We are then left  with two independent production modes (VBF+VH) and (ggF+ttH), and four 
independent final states $\gam\gam$, $VV^{(*)}$, $b\bar{b}$, $\tau\tau$.
In addition, in many models there is a common coupling to down-type fermions and hence the branching
fractions into $b\bar{b}$ and $\tau\tau$ rescale by a common factor, leading to identical $\mu$ values for the $b\bar b$ and $\tau\tau$ final states.

The first purpose of the present paper is to combine the information
provided by ATLAS, CMS and the Tevatron experiments on the 
$\gam\gam$, $ZZ^{(*)}$, $WW^{(*)}$, $b\bar{b}$ and $\tau\tau$  final states 
including the error correlations
among the (VBF+VH) and (ggF+ttH) production modes.
Using a Gaussian approximation, we derive 
for each final state a combined likelihood in the  
$\mu({\rm ggF + ttH})$ versus $\mu({\rm VBF + VH})$ plane, 
which can then simply be expressed as a $\chi^2$.
(Note that this does {\em not} rely on ggF production being dominated by the top loop.) 
We express this $\chi^2$ as
\beq\label{eq:1}
\chi_i^2 = a_i(\mu_i^{\rm{ggF}}-\hat{\mu}_i^{\rm{ggF}})^2
+2b_i(\mu_i^{\rm{ggF}}-\hat{\mu}_i^{\rm{ggF}})
(\mu_i^{\rm{VBF}}-\hat{\mu}_i^{\rm{VBF}})
+c_i(\mu_i^{\rm{VBF}}-\hat{\mu}_i^{\rm{VBF}})^2 \,,
\eeq
where the upper indices ggF and VBF stand for (ggF+ttH) and (VBF+VH), respectively, 
the lower index $i$ stands for $\gam\gam$, $VV^{(*)}$, $b\bar{b}$ and $\tau\tau$ (or $b\bar{b}=\tau\tau$),  
and $\hat{\mu}_i^{\rm{ggF}}$ and $\hat{\mu}_i^{\rm{VBF}}$ denote the
best-fit points obtained from the measurements.
We thus obtain ``combined likelihood ellipses'', which can be used in a simple, generic way to 
constrain non-standard Higgs sectors and new contributions to the loop-induced processes, provided they 
have the same Lagrangian structure as the SM.  

In particular, these likelihoods can be used to derive constraints on a
model-dependent choice of generalized Higgs couplings, the implications of which we study
subsequently for several well-motivated models. The choice of models is
far from exhaustive, but we present our results for the likelihoods as a
function of the independent signal strengths $\mu_i$ in such a manner that these
can easily be applied to other models.

We note that we will not include correlations between different final states 
but identical production modes which originate from common theoretical errors on the production cross 
sections~\cite{Giardino:2013bma,Bechtle:2013xfa} nor correlations between systematic errors due to common detector components (like EM calorimeters) sensitive to different final states (such as
$\gam\gam$ and $e^-$ from $ZZ^{(*)}$ and $WW^{(*)}$). A precise treatment of these `2nd order' corrections to our contours is only possible if performed by the experimental collaborations.
It is however possible to estimate their importance,
\eg, by reproducing the results of coupling fits performed by ATLAS
and CMS, as done for two representative cases in Appendix B. The
results we obtain are in good agreement with the ones published by the
experimental collaborations.

In the next Section, we will list the various sources of information used
for the determination of the  
coefficients $a_i$, $b_i$, $c_i$, $\hat{\mu}_i^{\rm{ggF}}$ and $\hat{\mu}_i^{\rm{VBF}}$, and
present our results for these parameters. 
In Section~3, we parametrize the signal strengths $\mu_i$ in terms of
various sets of Higgs couplings, and use our results from Section~2 to
derive $\chi^2$ contours for these couplings. 
In Section~4, we apply our fits to some concrete BSM models, which provide 
simple tree-level relations between the generalized Higgs couplings to 
fermions and gauge bosons. Our conclusions are presented in Section~5.
The Appendix contains clarifying details on Eq.~(\ref{eq:1}) as well as a comparison 
with coupling fits performed by ATLAS and CMS.

\section{Treatment of the experimental results and combined signal strength ellipses} \label{ssellipse}

The aim of the present section is to combine the most recent available information
on signal strengths from the ATLAS, CMS and Tevatron experiments for the various
Higgs decay modes. In most cases, these include error correlations in the plane of the (VBF+VH) and (ggF+ttH) production modes. For practical purposes it is very useful to represent the
likelihoods in these planes in the Gaussian approximation. 
Once the expressions for the various $\chi_i^2$ are given in the form of
Eq.~(\ref{eq:1}), it becomes straightforward to evaluate the numerical
value of $\chi^2=\sum_i \chi_i^2$ in any theoretical model with SM-like Lagrangian structure, 
in which predictions for the Higgs branching fractions and the (VBF+VH) and (ggF+ttH) production 
modes (relative to the SM) can be made.

From the corresponding information provided by the experimental collaborations one
finds that the Gaussian approximation is justified in the neighborhood (68\% confidence level (CL)
contours) of the best fit points. Hence we parametrize these 68\% CL contours, 
separately for each experiment, as in Eq.~(\ref{eq:1}).\footnote{This corresponds to fitting a bivariate normal distribution to the 68\% CL contours. We have verified that this reproduces sufficiently well the best fit points as well as the 95\% CL contours; see Section~2 of Ref.~\cite{Boudjema:2013qla} for more detail.}
Occasionally, only a single
signal rate including error bars for a specific final state is given. Using the relative contributions from the various production modes, this kind of information can still be represented in the form of Eq.~(\ref{eq:1}), leading to an ``ellipse'' which reduces to a strip in the plane of the (VBF+VH) and (ggF+ttH) production modes.

Subsequently these expressions can easily be combined and be represented again in the form of Eq.~(\ref{eq:1}). 
We expect that the result is reliable up to $\chi_i^2 \lsim 6$ (making it possible to derive
95\% CL contours), but its extrapolation to (much) larger values of
$\chi_i^2$ should be handled with care.

Starting with the $H\to \gamma\gamma$ final state, we treat in this way
the 68\%~CL contours given by ATLAS in
\cite{ATLAS-CONF-2013-012,ATLAS-CONF-2013-014, ATLAS-CONF-2013-034}, by
CMS in \cite{CMS-PAS-HIG-13-001,CMS-PAS-HIG-13-005,
CMS-PAS-HIG-13-015}\footnote{Note that we are using the MVA analysis for CMS $H \to \gamma\gamma$. The cut-based analysis (CiC) also presented by CMS [10]---that leads to higher but compatible signal strengths---is unfortunately not available in the form of contours in the plane of the (VBF+VH) and (ggF+ttH) production modes. Moreover, no information is given on the sub-channel decomposition, so in fact the CMS CiC analysis cannot be used for our purpose.}
and the Tevatron in \cite{Aaltonen:2013kxa}. 
(In the case of the Tevatron, for all final states only a strip in the plane of the
(VBF+VH) and (ggF+ttH) production modes is defined.) 
For the combination of the $ZZ$ and $WW$ final states, we use the 68\%~CL 
contours given by ATLAS for $ZZ$ in \cite{ATLAS-CONF-2013-013,ATLAS-CONF-2013-014,ATLAS-CONF-2013-034}, by CMS for $ZZ$ in \cite{CMS-PAS-HIG-13-002,CMS-PAS-HIG-13-005},
by ATLAS for $WW$ in \cite{ATLAS-CONF-2013-030,ATLAS-CONF-2013-034}, by CMS for $WW$
in \cite{CMS-PAS-HIG-13-005,CMS-PAS-HIG-13-003,CMS-PAS-HIG-13-009} and by the Tevatron for $WW$ in
\cite{Aaltonen:2013kxa}. 
For the combination of the $b\bar{b}$ and $\tau\tau$ final states, we use the
``strip'' defined by the ATLAS result for $b\bar{b}$ in associated VH
production from \cite{ATLAS-CONF-2012-161}, the 68\%~CL contour
given by CMS for $b\bar{b}$ in \cite{CMS-PAS-HIG-13-012}, the Tevatron result
for $b\bar{b}$ from \cite{Aaltonen:2013kxa} and combine
them with the ATLAS 68\%~CL contour for $\tau\tau$ from
\cite{ATLAS-CONF-2012-160,ATLAS-CONF-2013-034} and the CMS 68\%~CL contours for
$\tau\tau$ from \cite{CMS-PAS-HIG-13-005,CMS-PAS-HIG-13-004}.
We also use the ATLAS search for $ZH\to \ell^+\ell^-\!+{\rm invisible}$, extracting the likelihood from Fig.~10b of \cite{ATLAS-CONF-2013-011}. 
All the above 68\%~CL likelihood contours are parametrized by
ellipses (or strips) in
$\chi^2$ as in Eq.~(\ref{eq:1}), which can subsequently be
combined. (In Appendix A we clarify how these combinations are
performed.)

\begin{figure}[t]\centering
\includegraphics[scale=0.4]{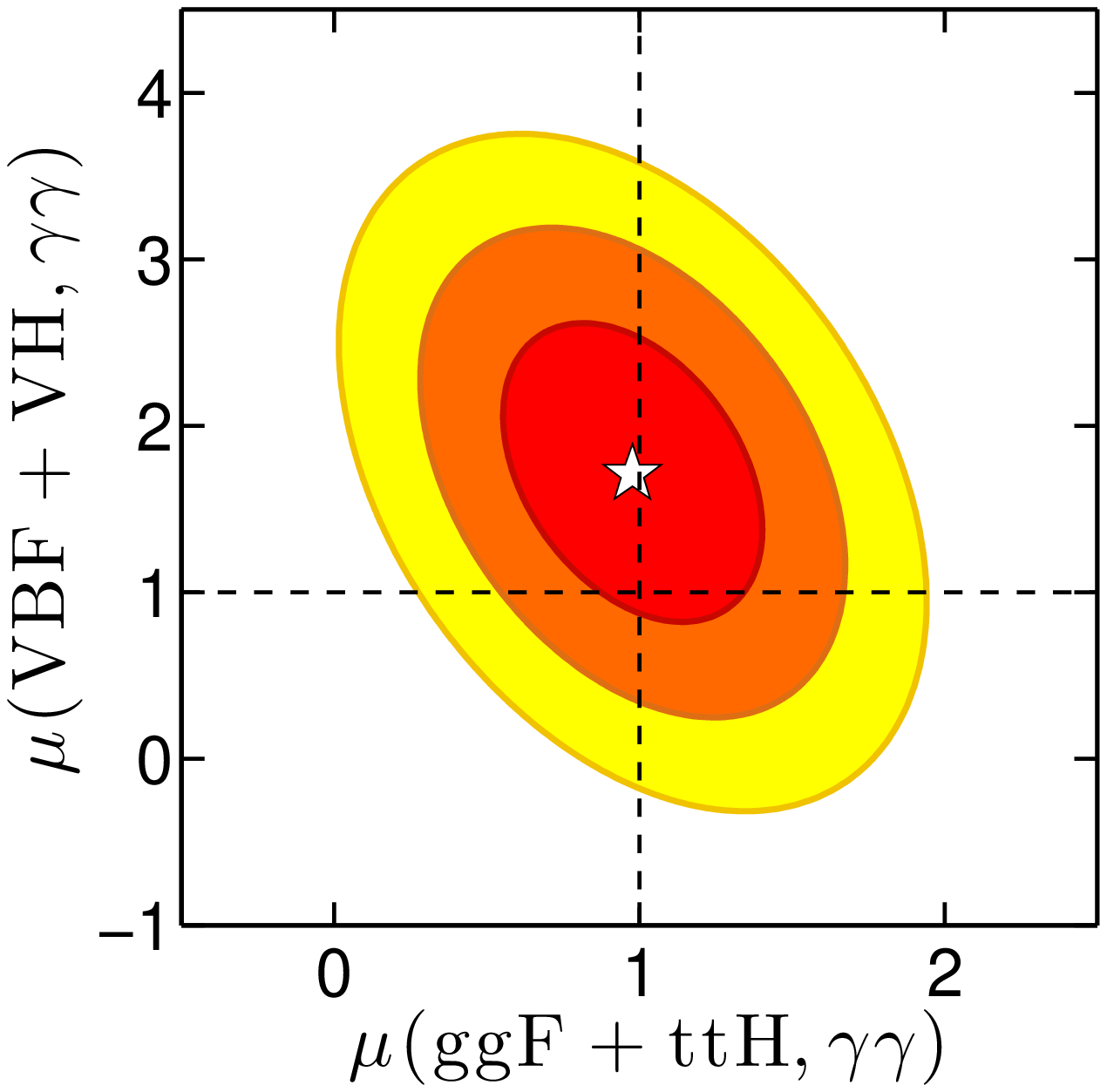}
\includegraphics[scale=0.4]{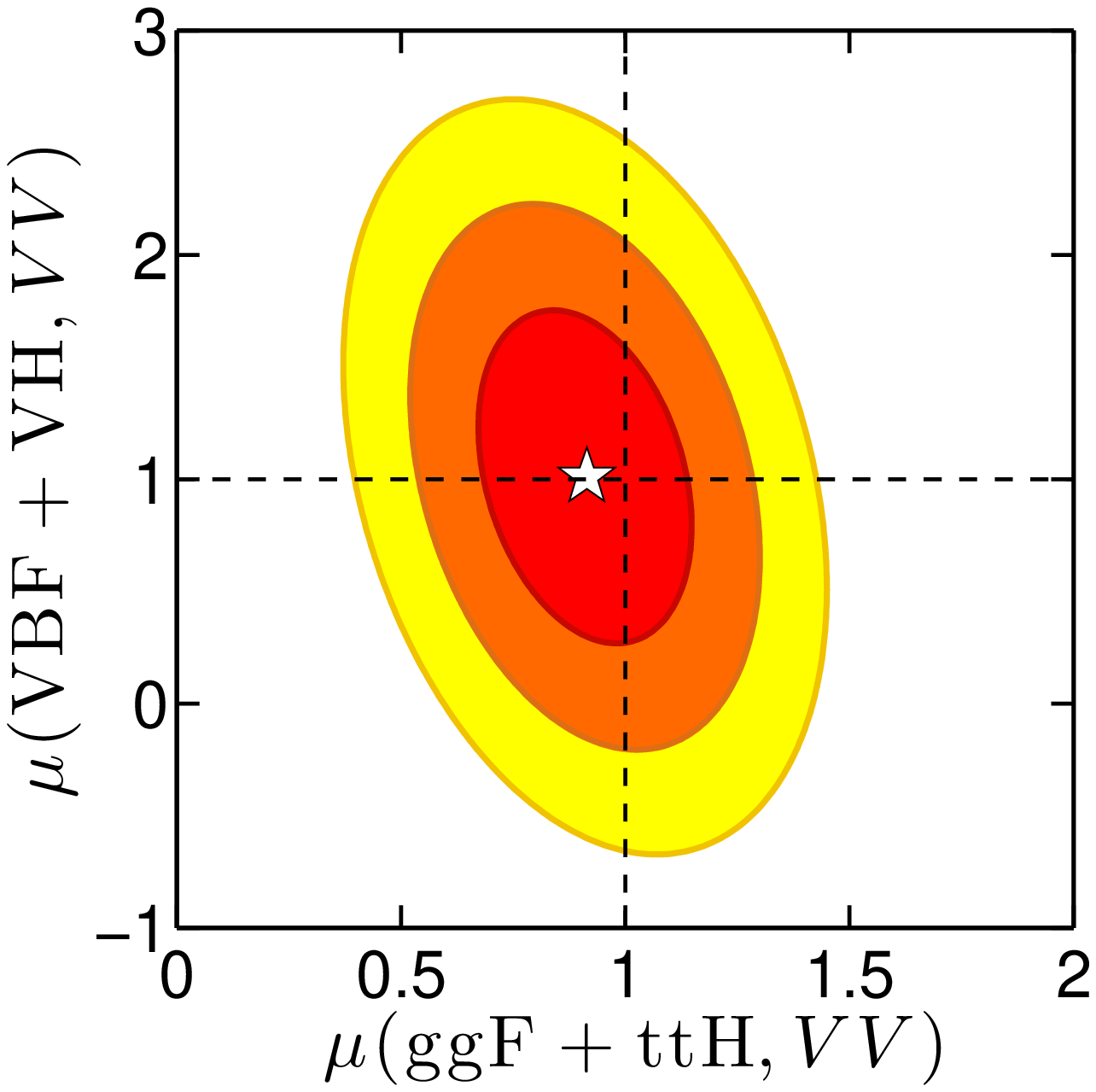}
\includegraphics[scale=0.4]{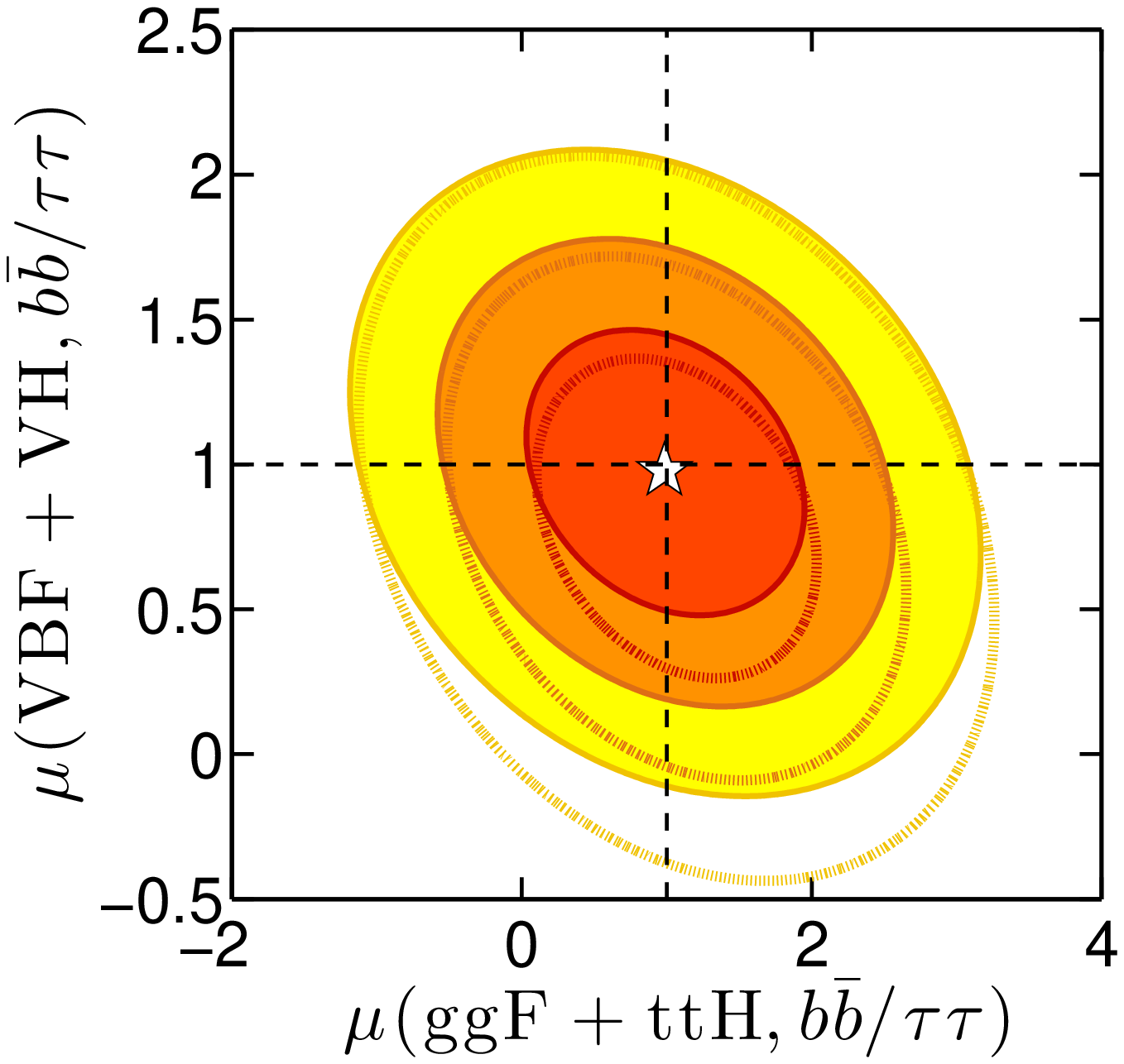}
\caption{Combined signal strength ellipses for the $\gamma\gamma$, $VV=ZZ,WW$ and $b\bar b=\tau\tau$ channels. 
The filled red, orange and yellow ellipses show the 68\%, 95\% and 99.7\%  CL regions, respectively, derived by combining the ATLAS, CMS and Tevatron results. The red, orange and yellow line contours in the right-most plot show how these ellipses change  when neglecting the Tevatron results. 
The white stars mark the best-fit points.
\label{fig:ellipses1} }
\end{figure}

The resulting parameters $\hat{\mu}^{\rm{ggF}}$, $\hat{\mu}^{\rm{VBF}}$, $a$, $b$ and
$c$ for Eq.~(\ref{eq:1}) (and, for completeness, the correlation coefficient $\rho$) for the different 
final states are listed in Table~\ref{tab:1}. The corresponding 68\%, 95\% and 99.7\%  CL ellipses are represented graphically in Fig.~\ref{fig:ellipses1}.

\begin{table}[t]
\center
\renewcommand{\arraystretch}{1.1}
\begin{tabular}{|c|c|c|c||c|c|c|}
\hline
& $\hat{\mu}^{\rm{ggF}}$ & $\hat{\mu}^{\rm{VBF}}$ & $\rho$ & $a$ & $b$ & $c$ \\
\hline 
$\gamma\gamma$ & $\phantom{-}0.98 \pm 0.28$ & $1.72 \pm 0.59$ & $-0.38$ & 14.94 & 2.69 & 3.34 \\
\hline
$VV$ & $\phantom{-}0.91 \pm 0.16$ & $1.01 \pm 0.49$ & $-0.30$ & 44.59 & 4.24 & 4.58 \\
\hline
$b\bar{b}/\tau\tau$ & $\phantom{-}0.98 \pm 0.63$ & $0.97 \pm 0.32$ & $-0.25$ & \phantom{0}2.67 & 1.31 &
10.12 \\
\hline
$b\bar{b}$ & $-0.23 \pm 2.86$ & $0.97 \pm 0.38$ & $0$ & \phantom{0}0.12 & 0 & 7.06 \\
\hline
$\tau\tau$ & $\phantom{-}1.07 \pm 0.71$ & $0.94 \pm 0.65$ & $-0.47$ & \phantom{0}2.55 & 1.31 & 3.07 \\
\hline
\end{tabular}
\caption{Combined best-fit signal strengths $\hat{\mu}^{\rm{ggF}}$, $\hat{\mu}^{\rm{VBF}}$ 
and correlation coefficient $\rho$ for various final states, as well as the coefficients 
$a$, $b$ and $c$ for the $\chi^2$ in Eq.~(\ref{eq:1}).}
\label{tab:1}
\end{table}

We see that, after combining different experiments, the best fit signal
strengths are astonishingly close to their SM values, the only 
exception being the $\gamma\gamma$ final state produced via (VBF+VH) for which
 the SM is, nonetheless, still within the 68\% CL contour. Therefore, these
results serve mainly to constrain BSM contributions to the properties of the Higgs boson.

\begin{figure}[t]\centering
\includegraphics[width=5cm]{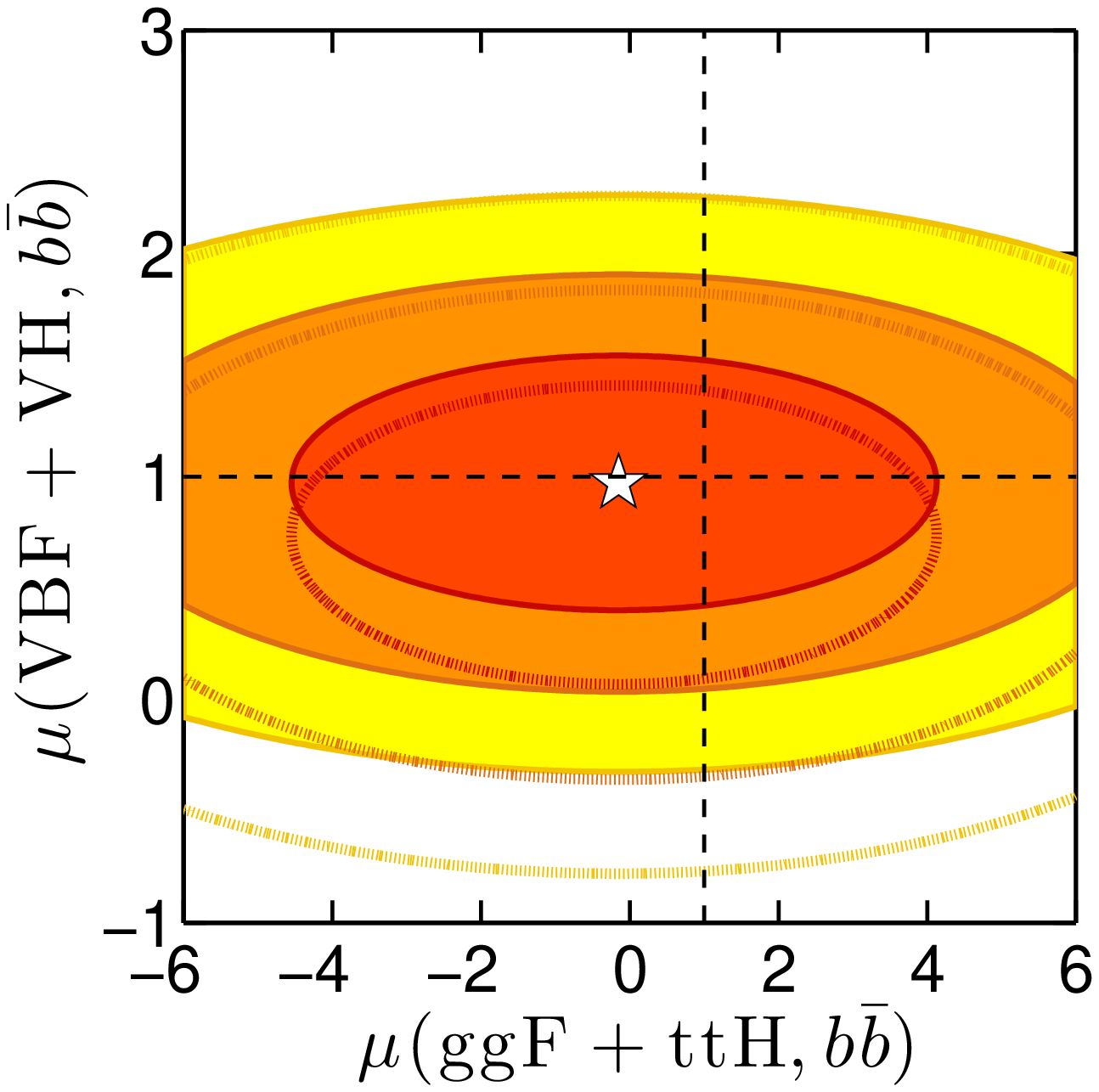}\quad
\includegraphics[width=5cm]{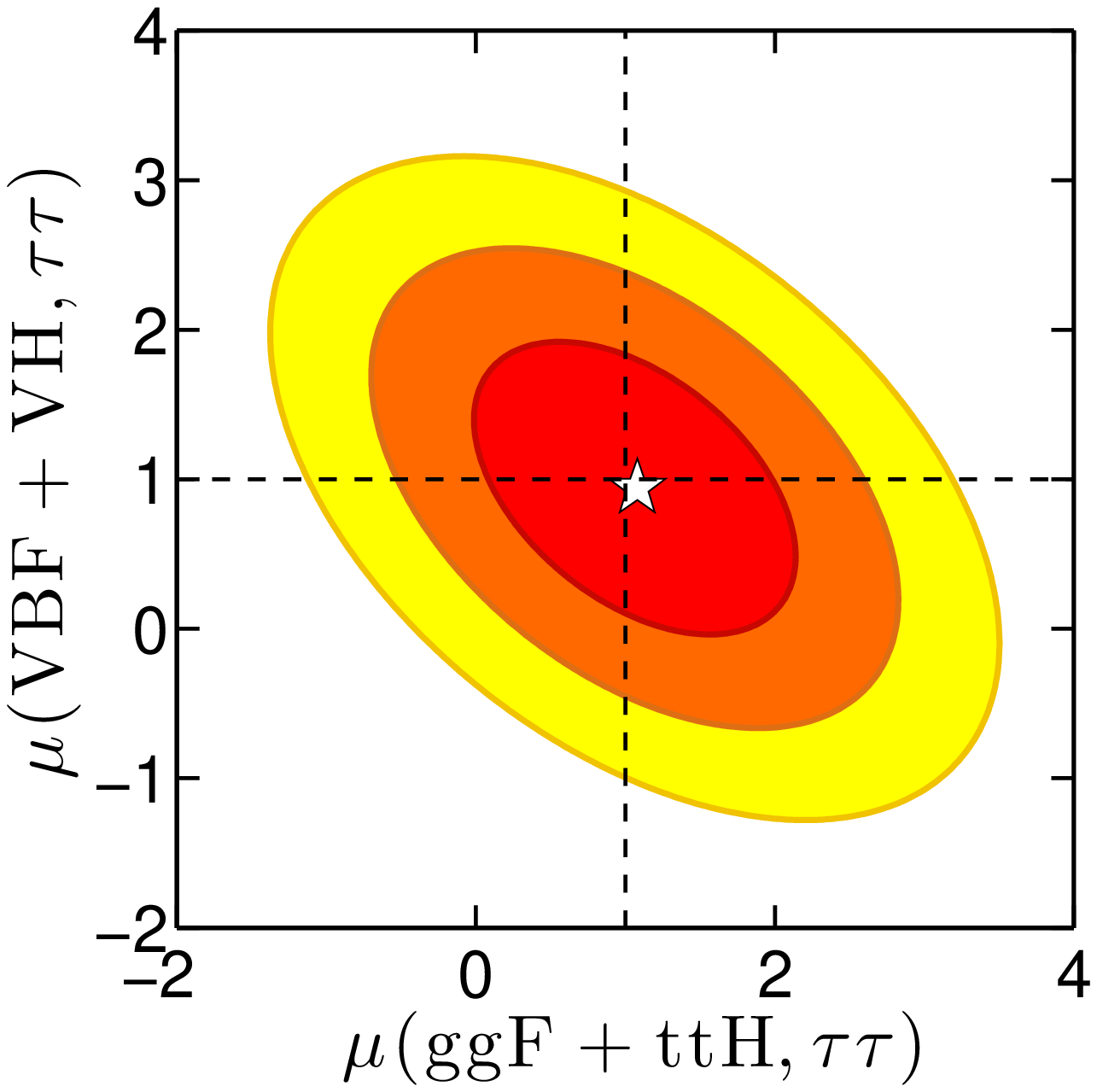}
\caption{Combined signal strength ellipses as in Fig.~\ref{fig:ellipses1} but treating the couplings 
to $b\bar b$ and $\tau\tau$ separately.
\label{fig:ellipses2} }
\end{figure}

The combination of the $b\bar{b}$ and $\tau\tau$ final states is justified, in principle,
in models where one specific Higgs doublet has the same reduced couplings (with respect
to the SM) to down-type quarks and leptons. However, even in this case QCD corrections
and so-called $\Delta_b$ corrections (from radiative corrections, notably at large $\tan\beta$, 
inducing couplings of another Higgs doublet to $b$~quarks, see {\it e.g.} \cite{Carena:1999py,Eberl:1999he}) 
can lead to deviations of the reduced $Hbb$ and $H\tau\tau$ couplings from a common value. Therefore, for completeness we show
the result for the $b\bar{b}$ final state only (combining ATLAS, CMS and
Tevatron results as given in the previous paragraph) in the fourth line of
Table~\ref{tab:1}, and the resulting 68\%, 95\% and 99.7\%  CL contours in the left plot in  
Fig.~\ref{fig:ellipses2}. The result for the $\tau\tau$ final state only (combining ATLAS and CMS results as given in the previous paragraph) is shown in the fifth line of
Table~\ref{tab:1}, and the resulting 68\%, 95\% and 99.7\%  CL contours in the right plot in Fig.~\ref{fig:ellipses2}.

Before proceeding, a comment is in order regarding the impact of the Tevatron results. 
While for the $\gamma\gamma$ and $VV$ final states, our combined likelihoods are completely 
dominated by the LHC measurements, to the extent that they are the same with or without including 
the Tevatron results, this is not the case for the $b\bar{b}$ final state. For illustration, in the plots for the 
$b\bar{b}$ final state in Figs.~\ref{fig:ellipses1} and \ref{fig:ellipses2} we also show 
what would be the result neglecting the Tevatron measurements.

\section{Fits to reduced Higgs couplings}

Using the results of the previous section, it is straightforward to determine constraints 
on the couplings of the observed Higgs boson to various particle pairs, 
assuming only a SM-like Lagrangian structure. 
As in \cite{Belanger:2012gc}, we define $\cu$, $\cd$ and $\cv$ to be
ratios of the $H$ coupling to up-type quarks, down-type quarks and
leptons, and vector boson pairs, respectively, relative  to that
predicted in the case of the SM Higgs boson (with $\CV>0$ by convention).
In addition to these tree-level couplings there are also the one-loop
induced couplings of the $H$ to $gg$ and $\gam\gam$.  Given values for
$\cu$, $\cd$ and $\cv$ the contributions of SM particles to the $gg$ and
$\gam\gam$ couplings, denoted $\anti\cg$ and $\anti \cp$ respectively,
can be computed. We take into account NLO corrections to $\anti\cg$ and $\anti \cp$ as recommended by the  LHC Higgs Cross Section Working Group~\cite{LHCHiggsCrossSectionWorkingGroup:2012nn}. 
In particular we include all the available QCD corrections for $C_g$ using \texttt{HIGLU}~\cite{Spira:1995mt,Spira:1996if} 
and for $C_\gamma$ using \texttt{HDECAY}~\cite{Spira:1996if,Djouadi:1997yw}, and we switch off the
electroweak corrections. 
In some of the fits below, we will also allow for
additional New Physics contributions to $\cg$ and $\cp$ by writing
$\cg=\anti\cg+\dcg$ and $\cp=\anti\cp+\dcp$. 

We note that in presenting one- (1D) and two-dimensional (2D) distributions of $\dchisq$, those quantities among $\cu$, $\cd$, $\cv$, $\dcg$ and $\dcp$ not plotted, but that are treated as variables, are being profiled over.
The fits presented below will be performed with and without allowing for invisible decays of the Higgs boson. In the latter case, only SM decay modes are present. In the former case, the new decay modes are assumed to produce invisible or undetected particles that would be detected as missing transverse energy at the LHC. A direct search for invisible decays of the Higgs boson have been performed by ATLAS in the $ZH \to \ell^+\ell^- + E_T^{\rm miss}$ channel~\cite{ATLAS-CONF-2013-011} and is implemented in the analysis. Thus, the total width is fully calculable from the set of $C_i$ and ${\cal B}(H \to {\rm invisible})$ in all the cases we consider. (We will come back to this at the end of this section.)

\begin{figure}[t]\centering
\includegraphics[width=4.75cm]{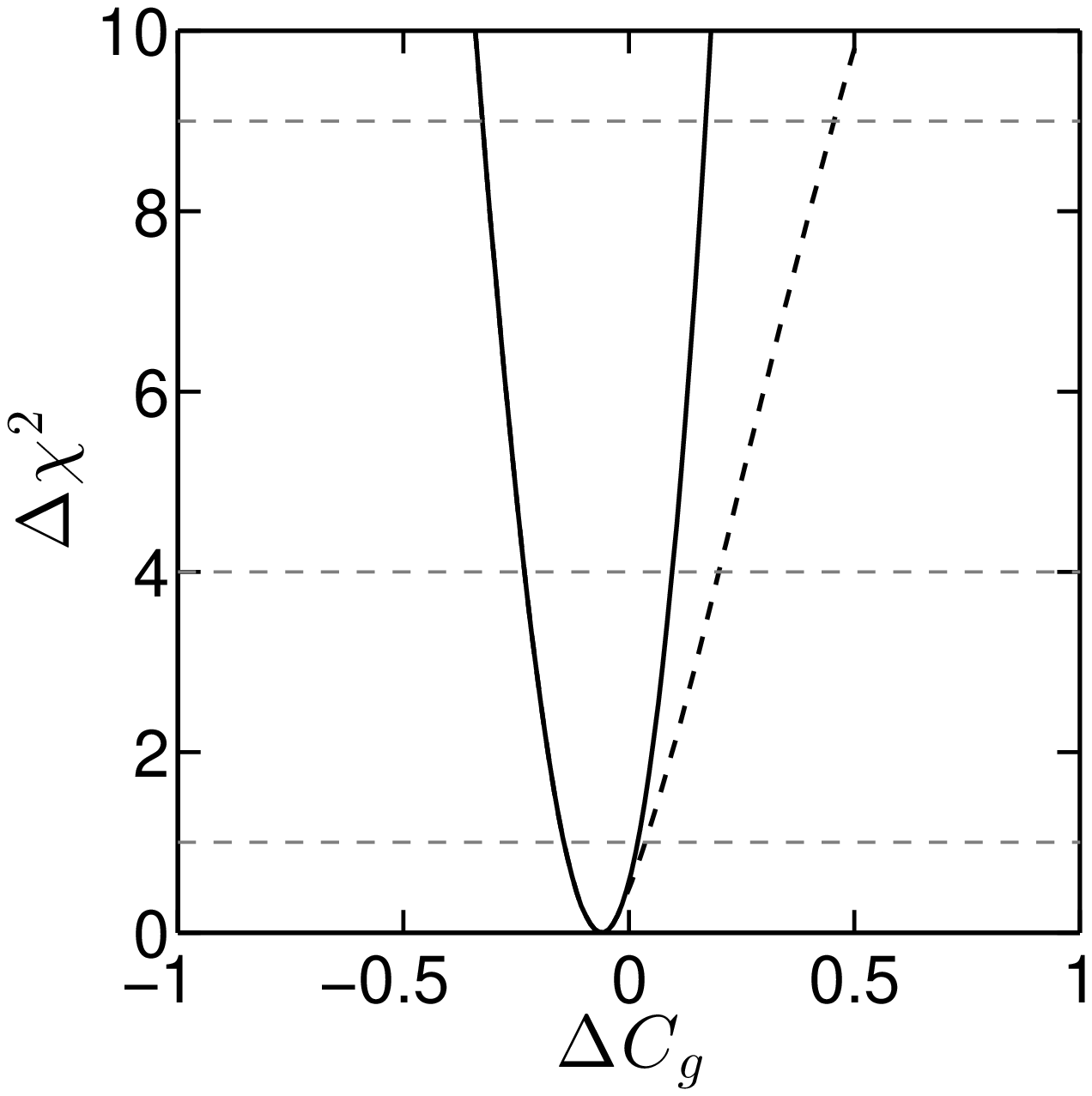}\quad
\includegraphics[width=5cm]{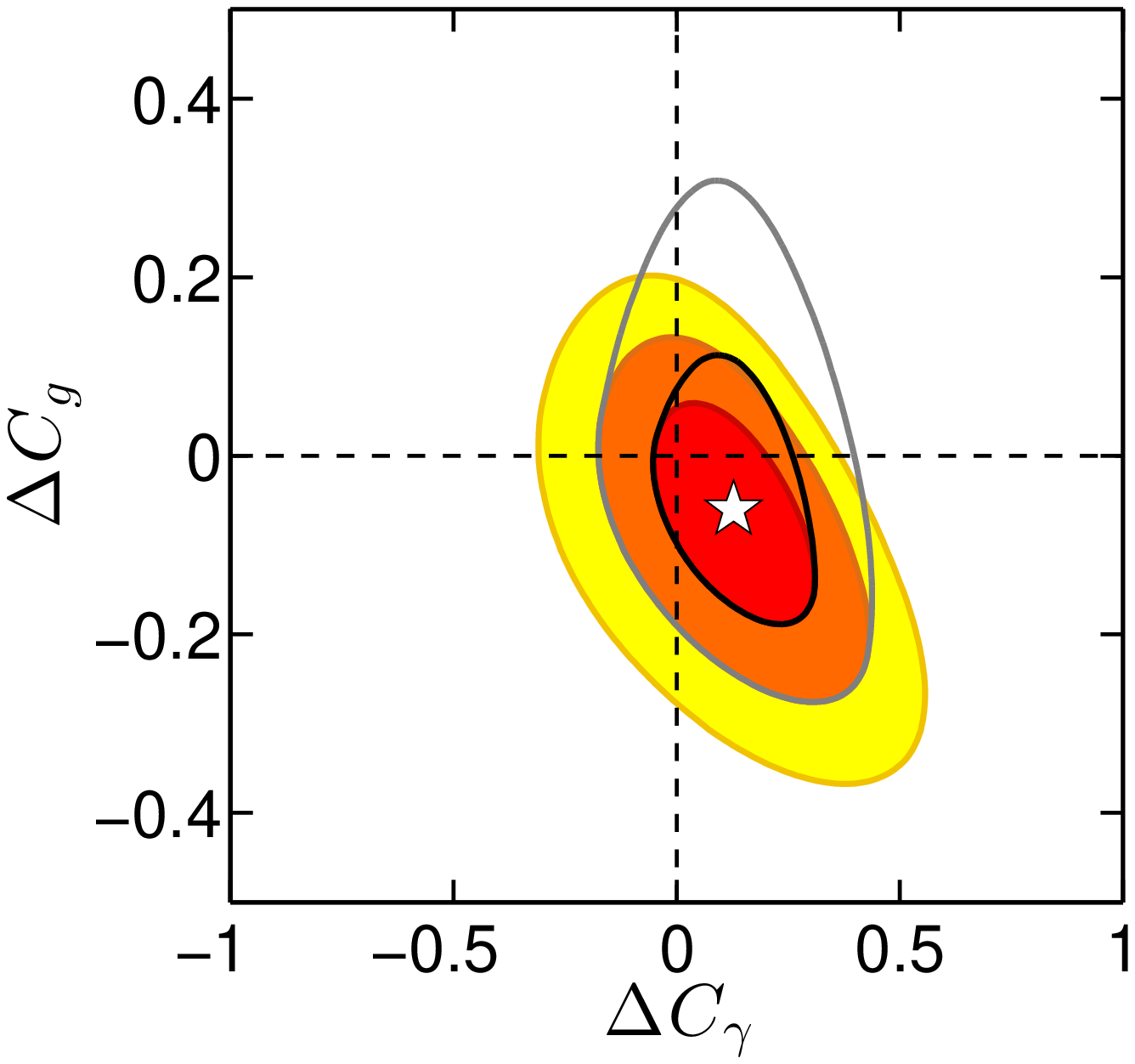}\quad
\includegraphics[width=4.75cm]{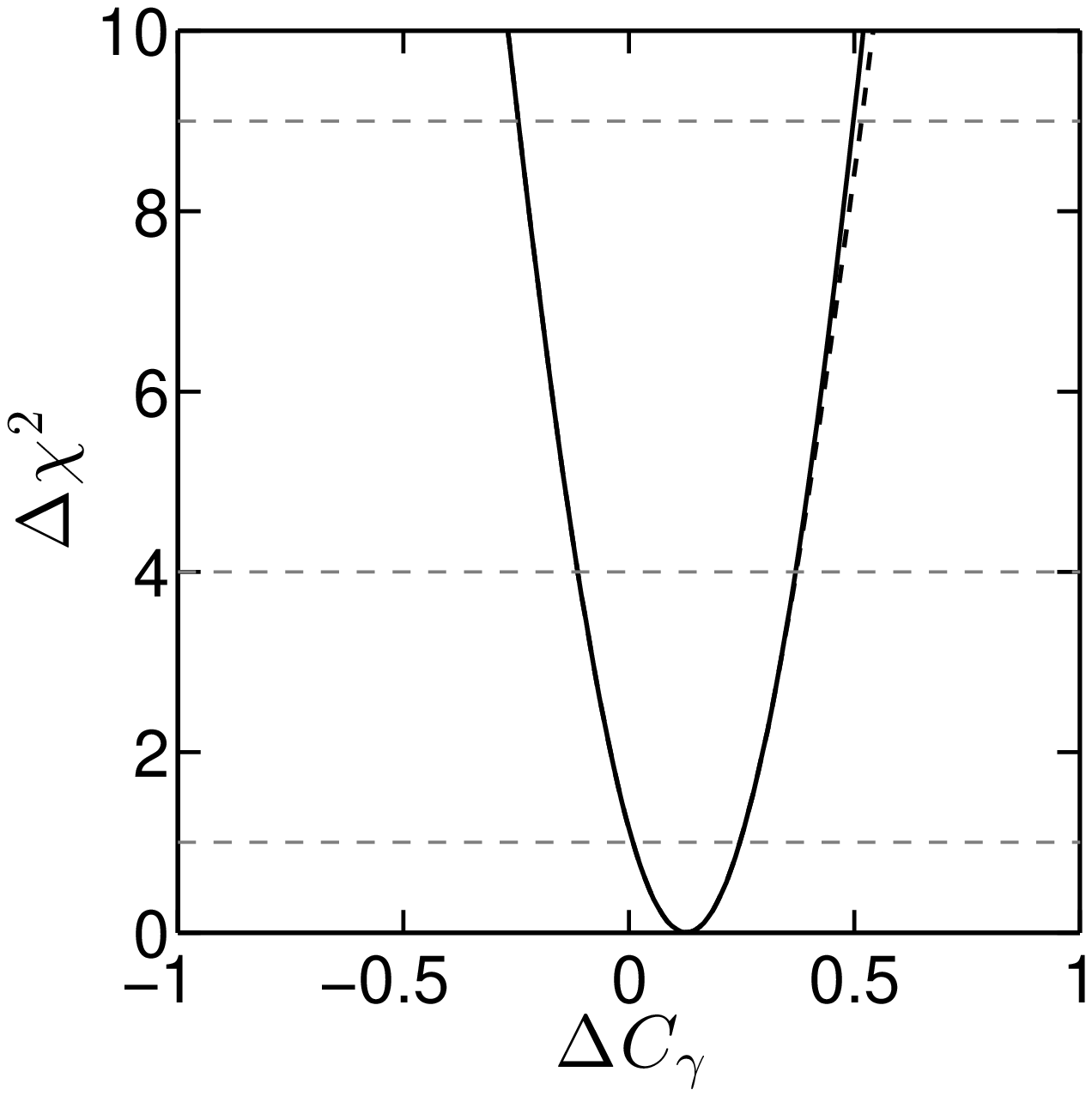}\quad
\caption{$\Delta \chi^2$ distributions in 1D and 2D for the fit of $\dcg$ and $\dcp$ for $\CU=\CD=\CV=1$. 
In the 1D plots, the solid (dashed) lines are for the case that invisible/unseen decays are absent (allowed).
In the 2D plot, the red, orange and yellow areas are the 68\%, 95\% and 99.7\% CL regions, respectively, assuming invisible decays are absent. The white star marks the best-fit point. The black and grey lines show the 68\% and 95\% CL contours when allowing for invisible decays.
\label{fig:CPadd-CGadd} }
\end{figure}

We begin by taking SM values for the tree-level couplings to fermions and vector bosons, \ie\  $\cu=\cd=\cv=1$, 
but allow for New Physics contributions to the couplings to $gg$ and $\gam\gam$.  The fit results with and without 
allowing for invisible/unseen Higgs decays are shown in  Fig.~\ref{fig:CPadd-CGadd}. We observe that the SM point of $\dcg=\dcp=0$ is well within the 68\% contour with the best fit points favoring a slightly positive (negative) value for $\dcp$ ($\dcg$). 
Allowing for invisible/unseen decays expands the 68\%, 95\% and 99.7\% CL regions by only a modest amount. 
This is in contrast to the situation at the end of 2012~\cite{Belanger:2012gc,Belanger:2013kya}, where some New Physics contribution to both $\dcg$ and $\dcp$ was preferred, and allowing for invisible decays had a large effect; 
with the higher statistics and with the reduced $\gamma\gamma$ 
signal strength from CMS~\cite{CMS-PAS-HIG-13-001}, $\dcg$ and $\dcp$ are now much more constrained. 
The best fit is obtained for $\dcg=-0.06$, $\dcp=0.13$, $\brinv\equiv \br({H\to\rm invisible})=0$ and has $\chimin=17.71$ for 21 d.o.f.\ (degrees of freedom)\footnote{There are in total 23 measurements entering our fit, and we adopt the simple definition of the number of d.o.f.\ as number of measurements minus number of parameters.}, 
as compared to $\chisq=18.95$ with 23 d.o.f.\ for the SM, so allowing for additional loop contributions  
does not 
improve the fit.

\begin{figure}[t!]\centering
\includegraphics[width=5cm]{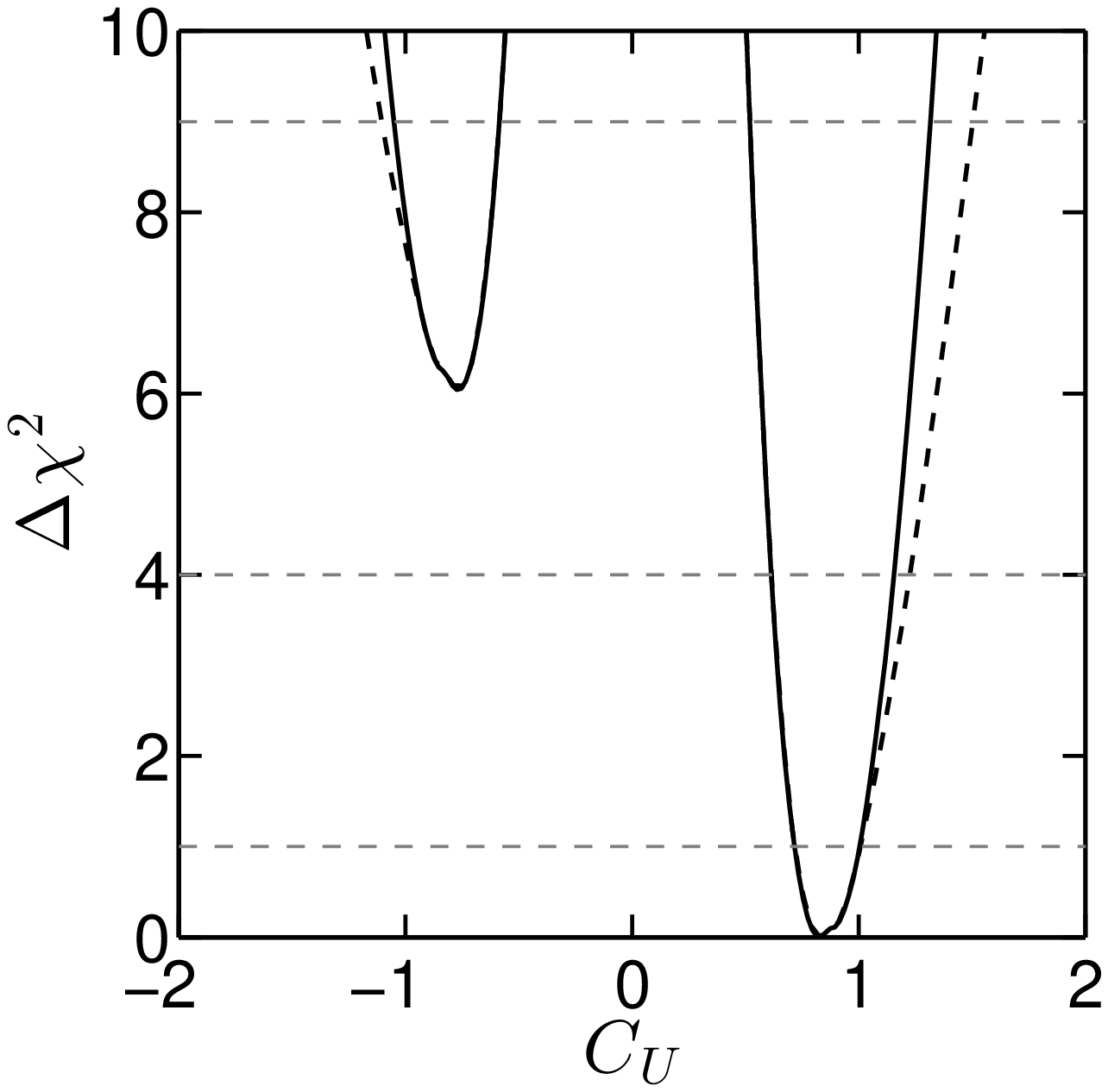}\quad
\includegraphics[width=5cm]{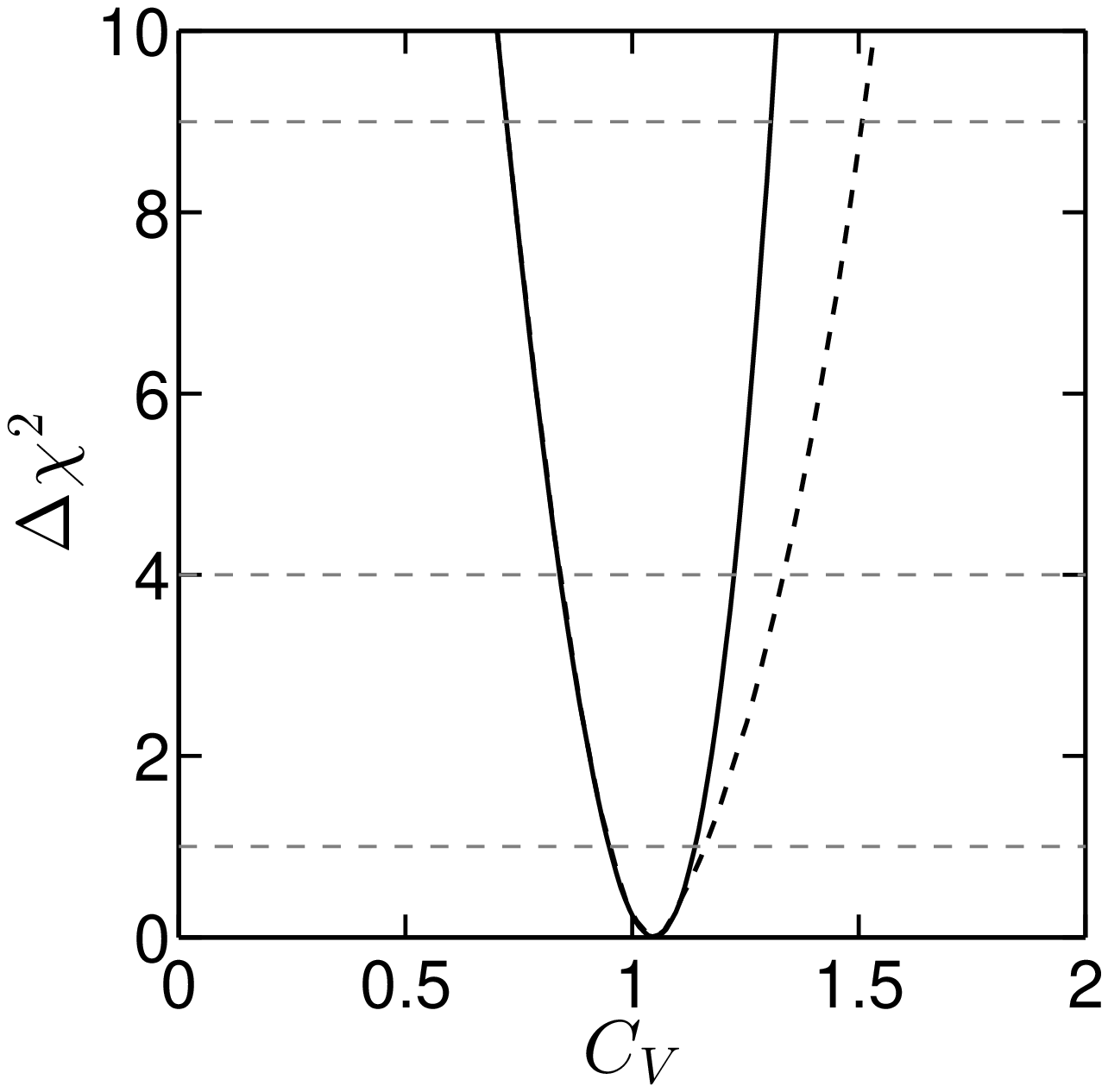}
\caption{Fit of $\CU$, $\CD$, $\CV$ for $\dcg=\dcp=0$. The plots show the 1D $\dchisq$ distribution as a function of $\cu$ (left) and $\cv$ (right). The solid (dashed) lines are for the case that invisible/unseen decays are absent (allowed).
\label{fig:CU-CD-CV} }
\end{figure}

Next, we allow $\cu$, $\cd$ and $\cv$ to vary but assume that there is no New Physics in the $gg$ and $\gam\gam$ loops, \ie\ we take $\dcg=\dcp=0$. Results for this case are shown in Fig.~\ref{fig:CU-CD-CV}.  We observe that, contrary to the situation at the end of 2012 \cite{Belanger:2012gc}, the latest data prefer a positive value of $\cu$ close to 1. 
This is good news, as a negative sign of $\cu$---in the convention where $m_t$ is positive---is quite problematic in the context of most theoretical models.\footnote{If the top quark and Higgs
bosons are considered as fundamental fields, it would require that the
top quark mass is induced dominantly by the vev of at least one
additional Higgs boson which is not the Higgs boson considered here, and 
typically leads to various consistency problems as discussed, \eg, in
\cite{Choudhury:2012tk}.}
(We do not show the distribution for $\cd$ here but just remark that
$|\CD|\simeq 1\pm0.2$ with a sign ambiguity following from
the weak dependence of the $gg$ and $\gam\gam$ loops on the bottom-quark
coupling.)  For $\CV$, we find a best-fit value slightly above 1, at
$\CV=1.07$, but  with  the SM-like value of $\cv=1$ lying well within one
standard deviation.

\begin{figure}[t!]\centering
\includegraphics[width=5cm]{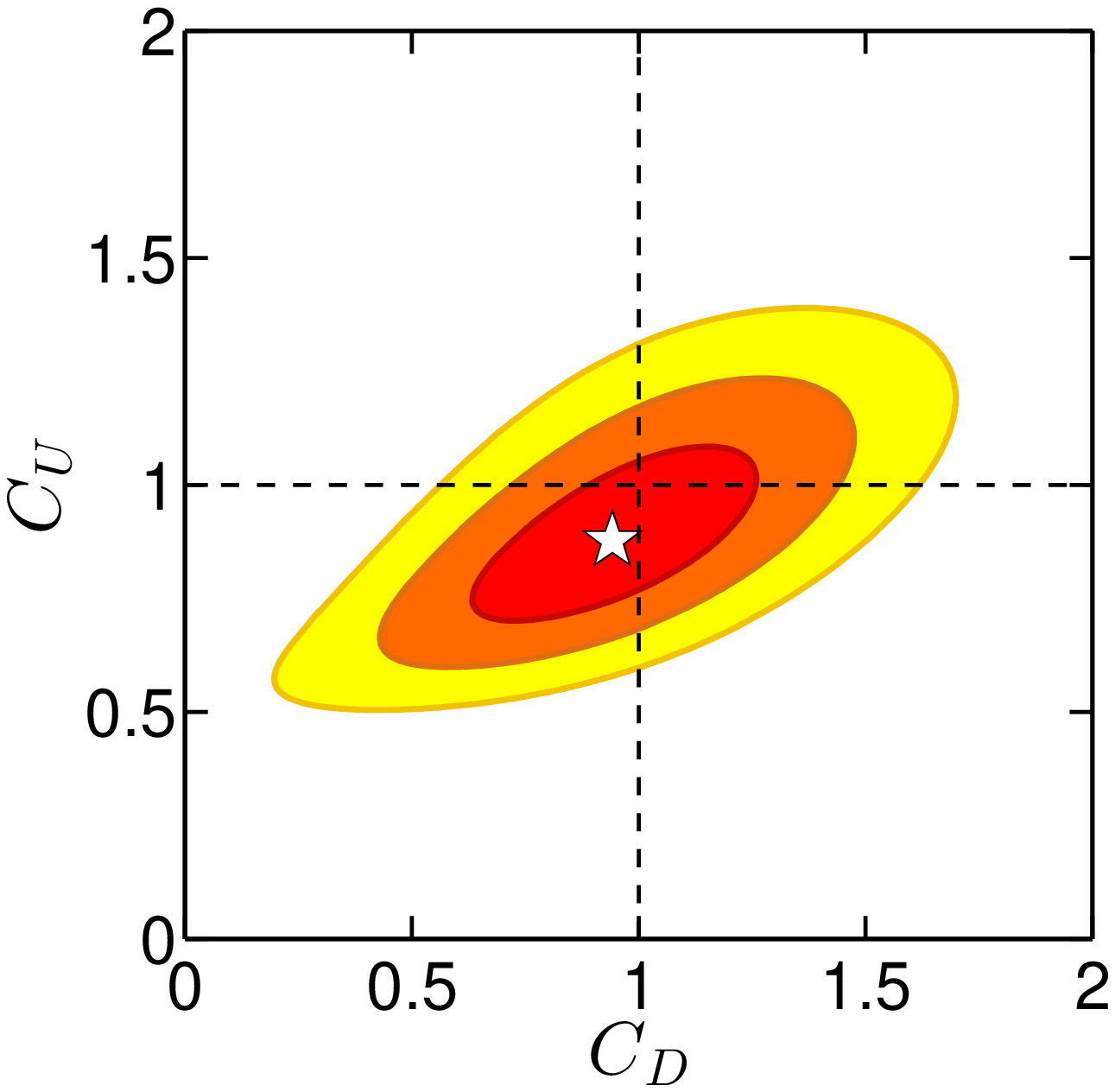}\quad
\includegraphics[width=5cm]{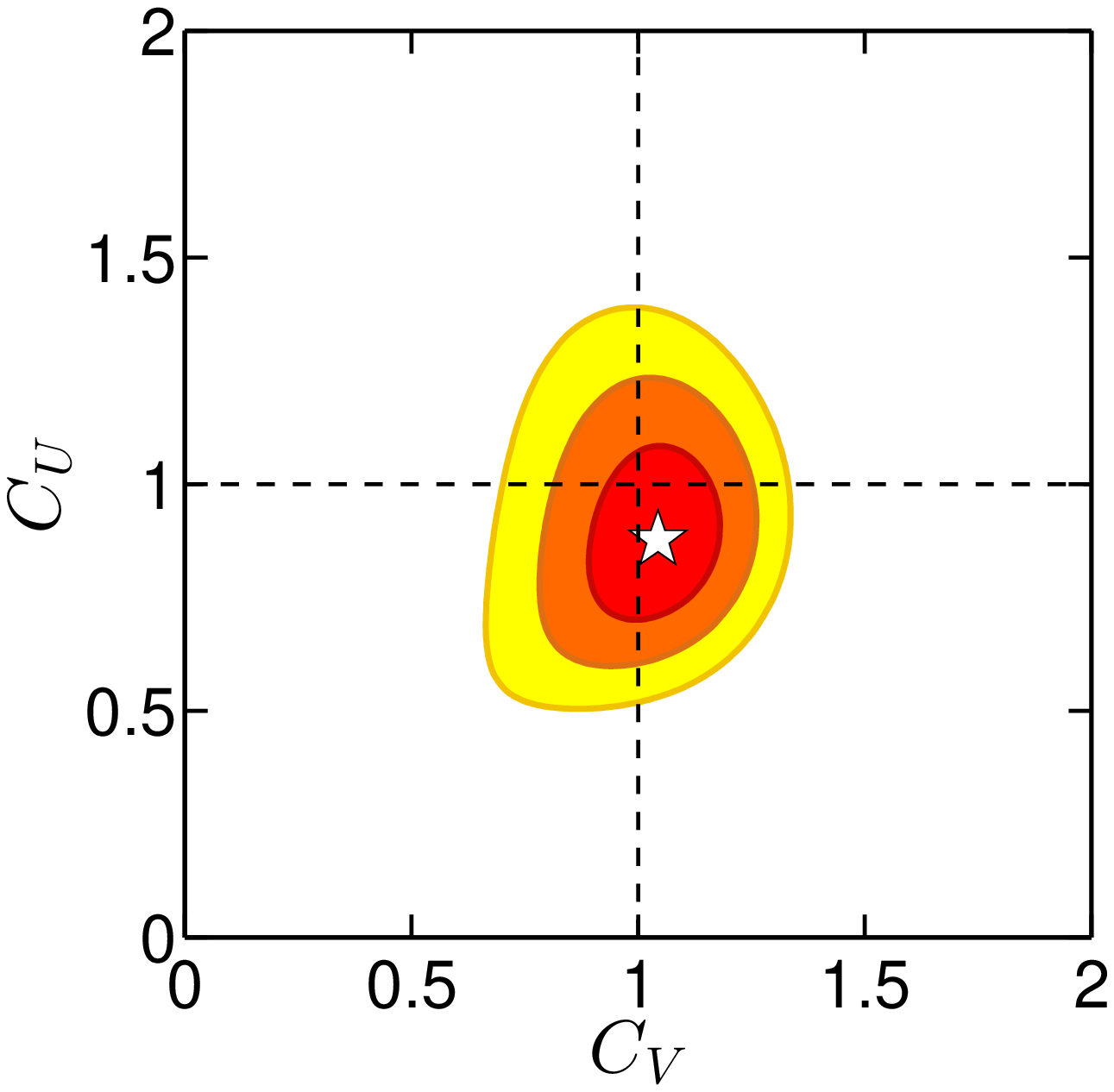}\quad
\includegraphics[width=5cm]{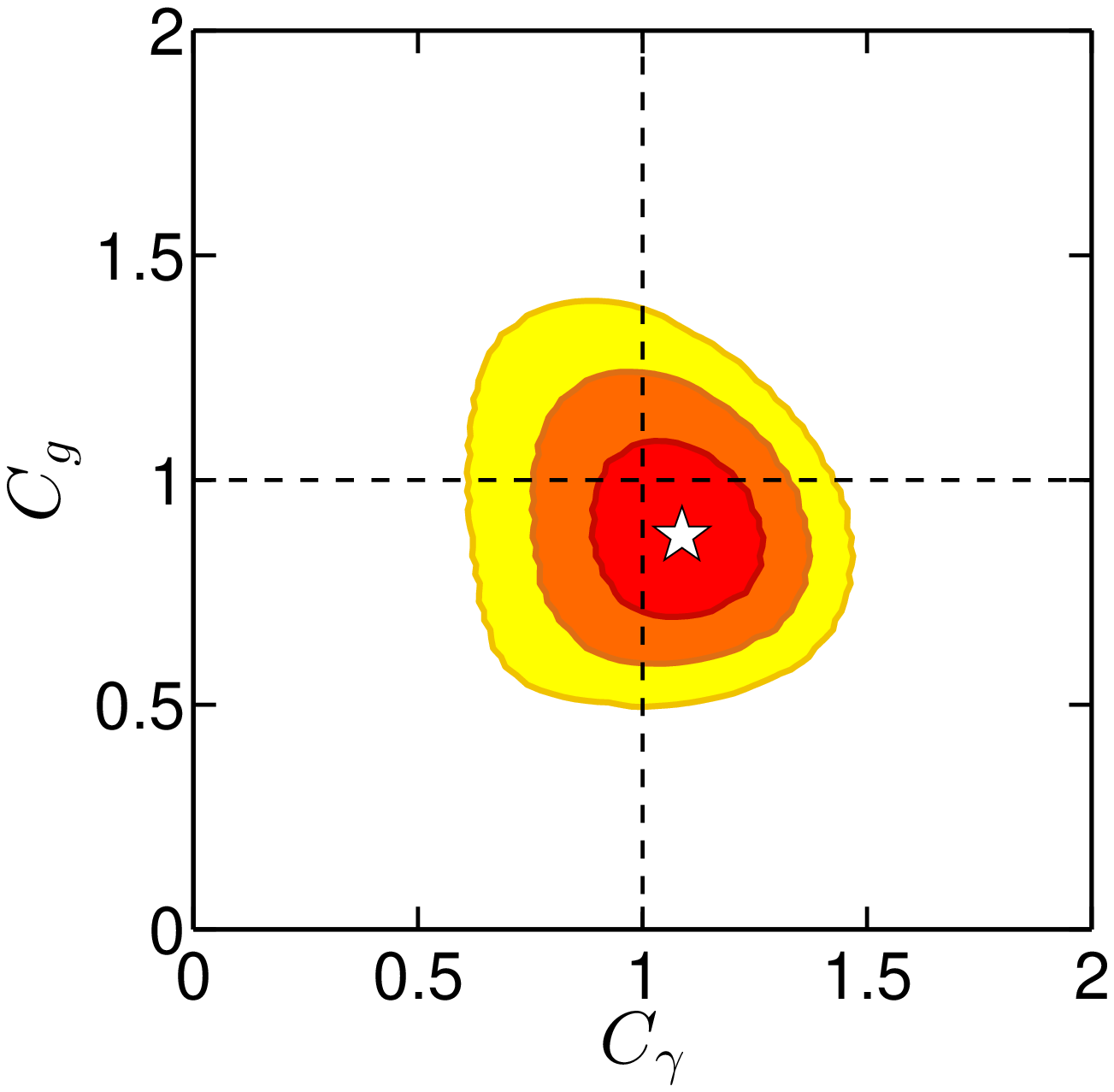}\quad
\caption{Fit of $\CU>0$, $\CD>0$ and $\CV$ for $\dcg=\dcp=0$. 
The red, orange and yellow areas are the 68\%, 95\% and 99.7\% CL regions, respectively, assuming 
invisible decays are absent. The white star marks the best-fit point. 
\label{fig:CUpos-CDpos-CV} }
\end{figure}

Since $\cu<0$ is now disfavored and the sign of $\CD$ is irrelevant, we
confine ourselves subsequently to $\cu,\cd>0$. In
Fig.~\ref{fig:CUpos-CDpos-CV}  we show $\dchisq$ distributions in 2D
planes confined to this range, still assuming $\dcg=\dcp=0$.

The mild correlation between $\CU$ and $\CD$ in the leftmost plot of
Fig.~\ref{fig:CUpos-CDpos-CV} follows from the very SM-like signal
rates in the $VV$ and $\gamma\gamma$ final states in ggF: varying $\CD$
implies a variation of the partial width $\Gamma(H\to bb)$ which
dominates the total width. Hence, the branching fractions $\br(H\to VV)$
and $\br(H\to\gamma\gamma)$ change in the opposite direction, decreasing
with increasing total width (\ie\ with increasing $\CD$) and vice versa. In
order to keep the signal rates close to~1, the ggF production cross
section, which is  roughly proportional to $ \CU^2$,  has to vary in the same direction as $\CD$.

The best fit is obtained for $\CU=0.88$, $\CD=0.94$, $\CV=1.04$, $\cp=1.09$, $\cg=0.88$
(and, in fact, $\brinv=0$).  
Note that if $\cv>1$ were confirmed, this would imply that the observed Higgs boson must have a significant  triplet (or higher representation) component~\cite{Logan:2010en,Falkowski:2012vh}.
Currently the coupling fits are, however, perfectly consistent with SM values. 
Again, with a $\chimin=17.79$ (for 20 d.o.f.) as compared to $\chisq=18.95$ for the SM, allowing for deviations from the SM does not significantly improve the fit.

\begin{figure}[t!]\centering
\includegraphics[width=5cm]{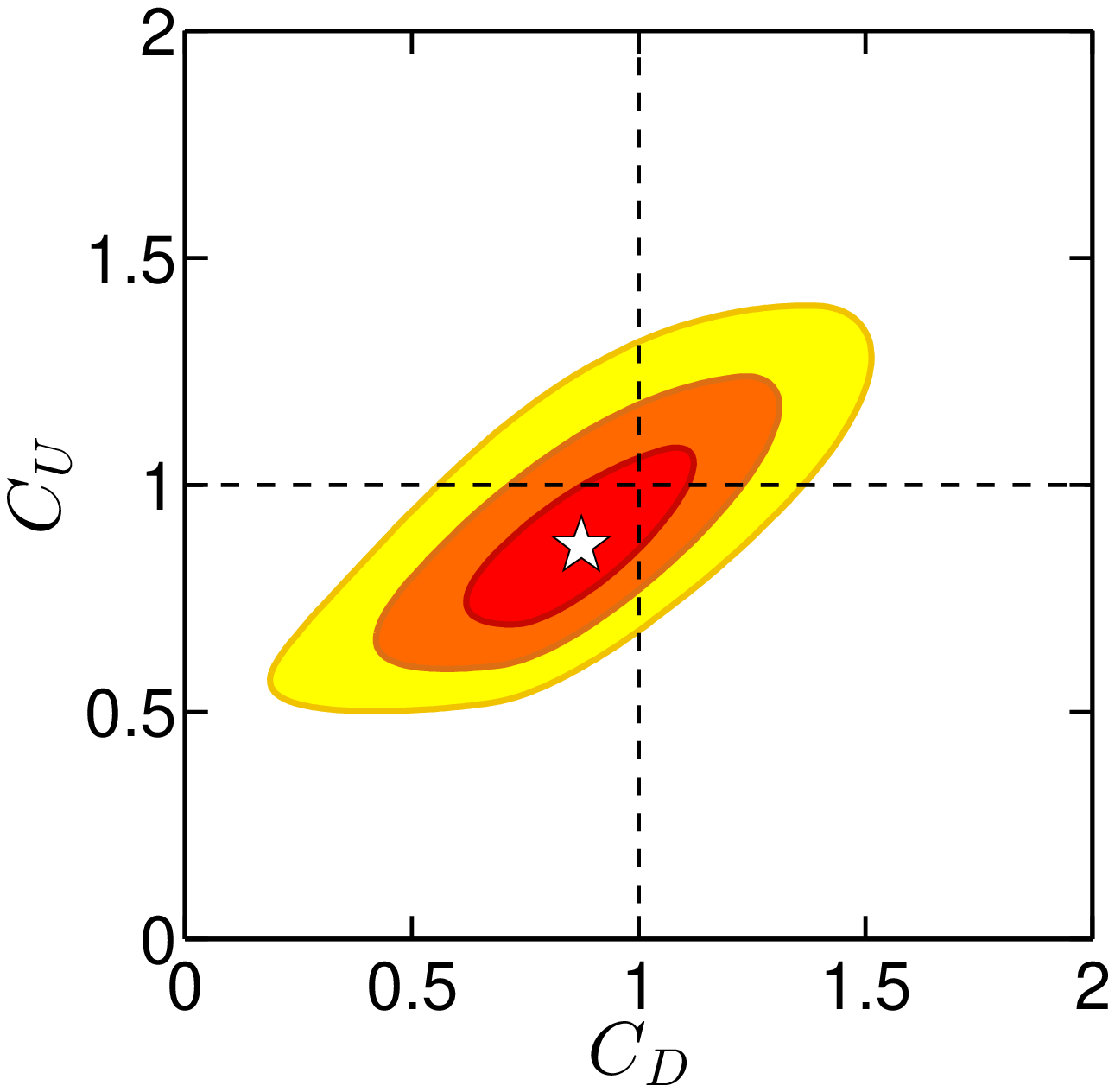}\quad
\includegraphics[width=5cm]{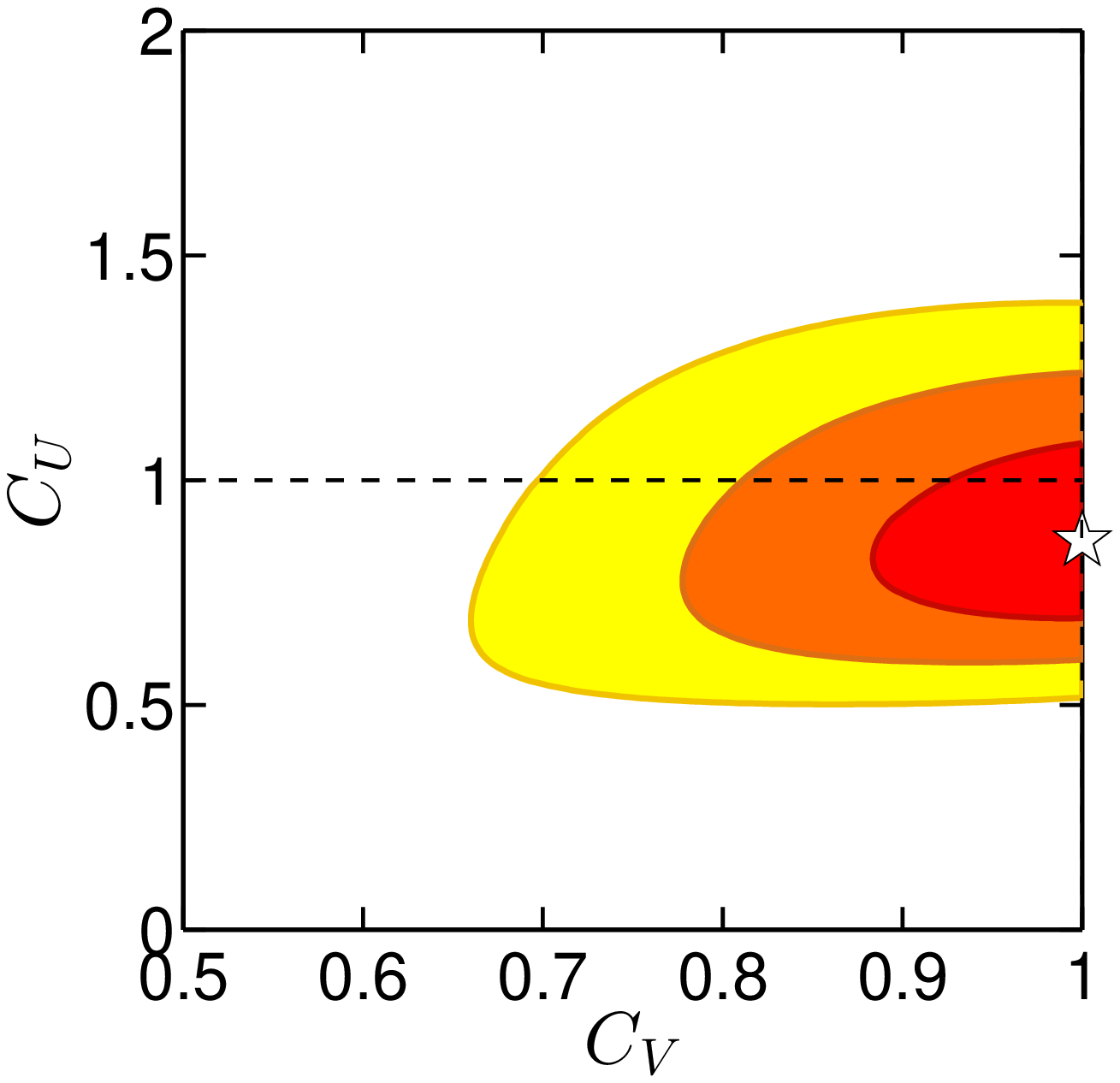}\quad
\includegraphics[width=5cm]{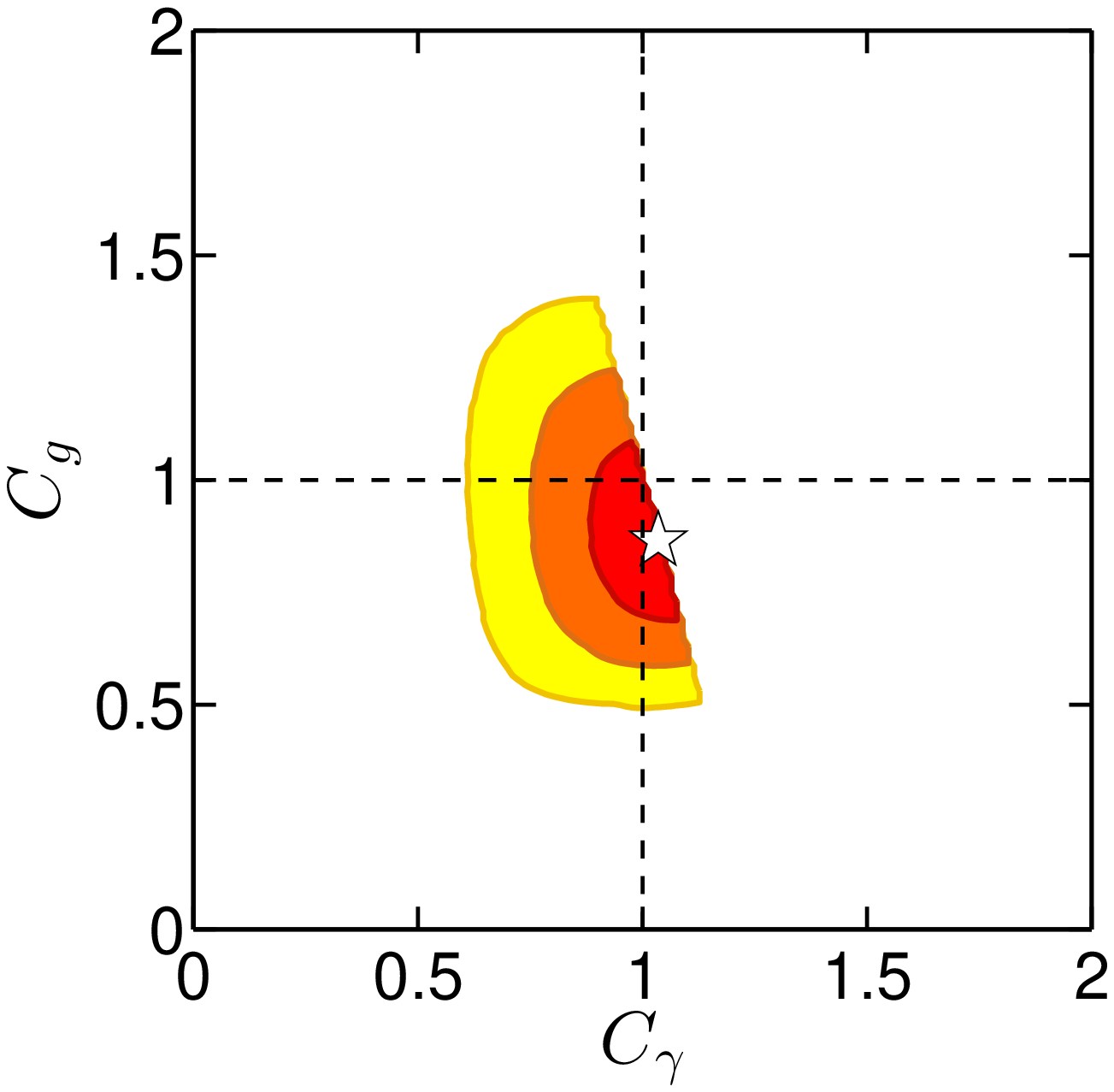}\quad
\caption{As in Fig.~\ref{fig:CUpos-CDpos-CV} but for $\CV\leq 1$.
\label{fig:CUpos-CDpos-CVle1} }
\end{figure}

In models where the Higgs sector consists of doublets+singlets only one always
obtains  $\cv\le1$.   Results for this
case are shown in Fig.~\ref{fig:CUpos-CDpos-CVle1}. Given the slight
preference for $\cv>1$ in the previous free-$\cv$ plots, it is no
surprise the $\cv=1$ provides the best fit along with $\CU=\cg=0.87$,
$\CD=0.88$ and $\cp=1.03$. Of course, the SM is again well within the
$68\%$ CL zone.

\begin{figure}[t!]\centering
\includegraphics[width=5cm]{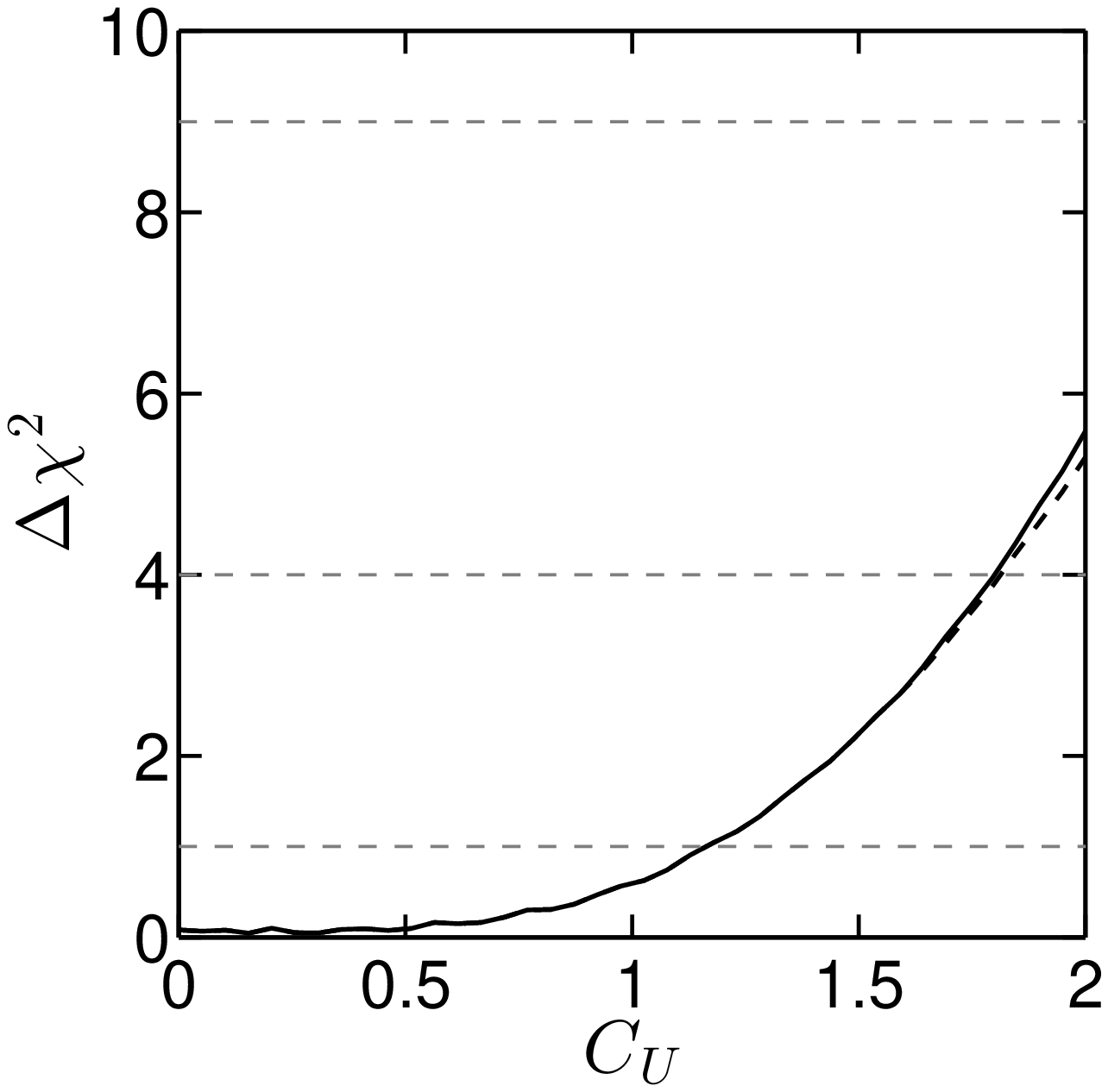}\quad
\includegraphics[width=5cm]{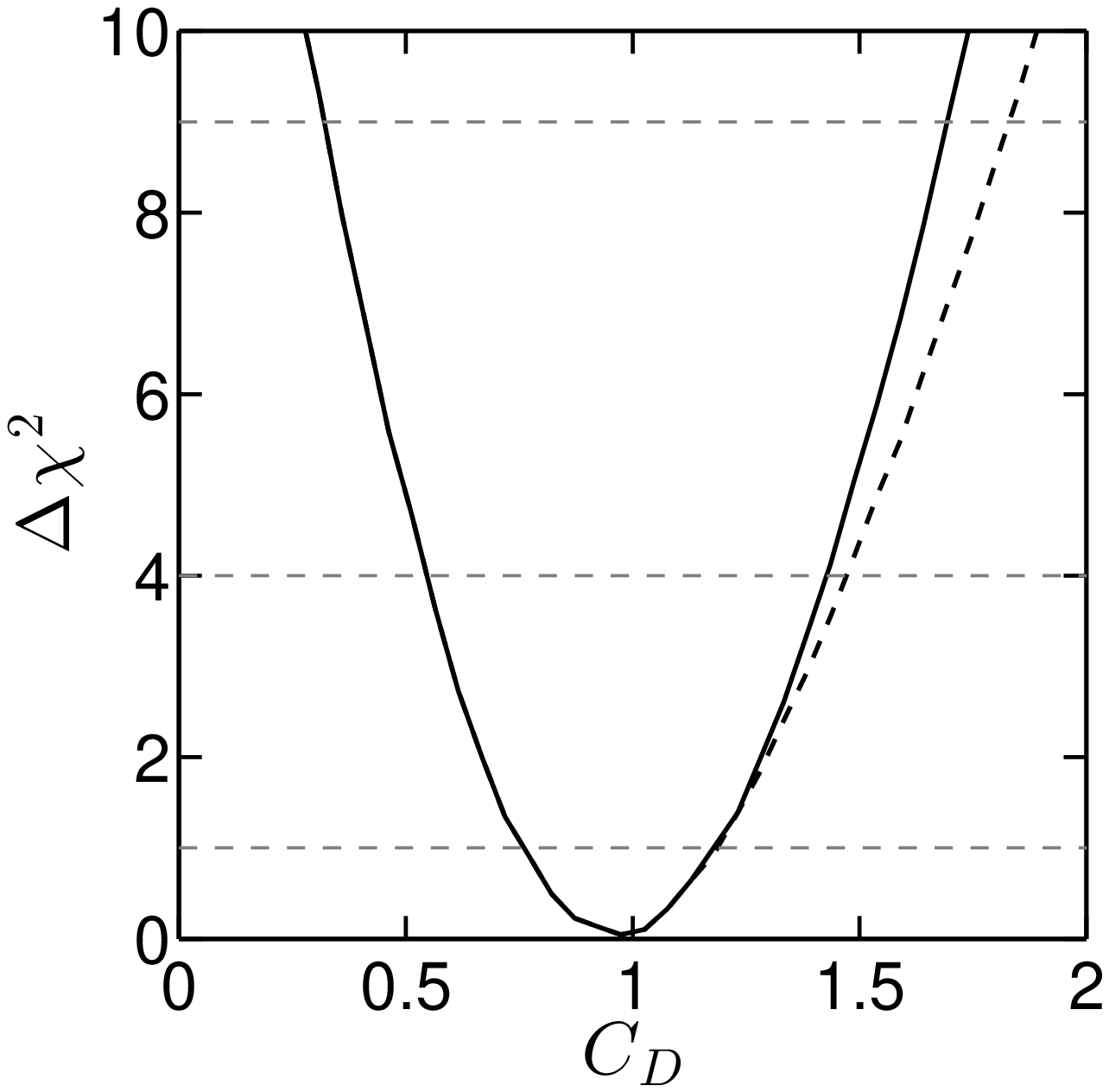}\quad
\includegraphics[width=5cm]{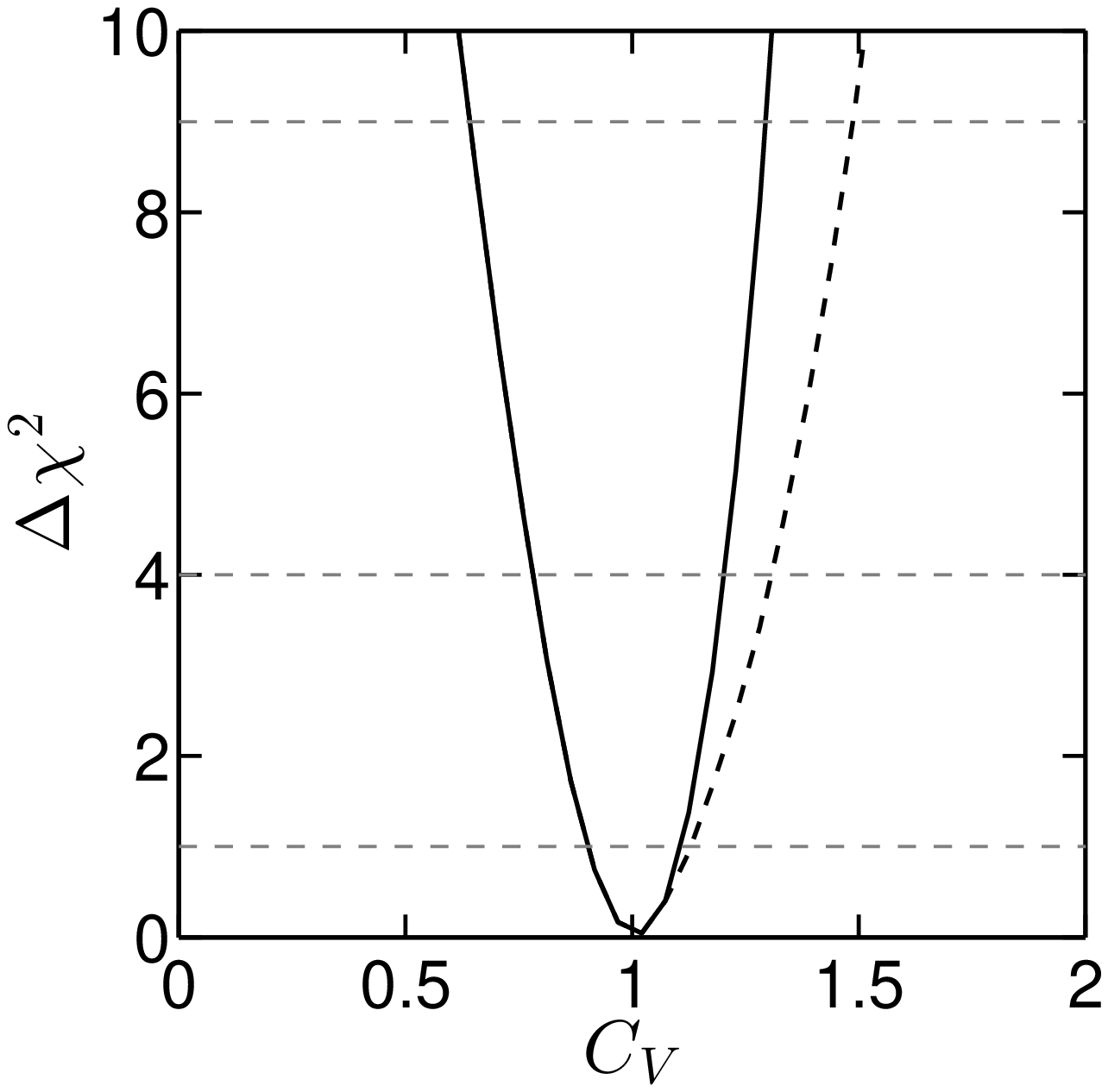}\quad
\includegraphics[width=5cm]{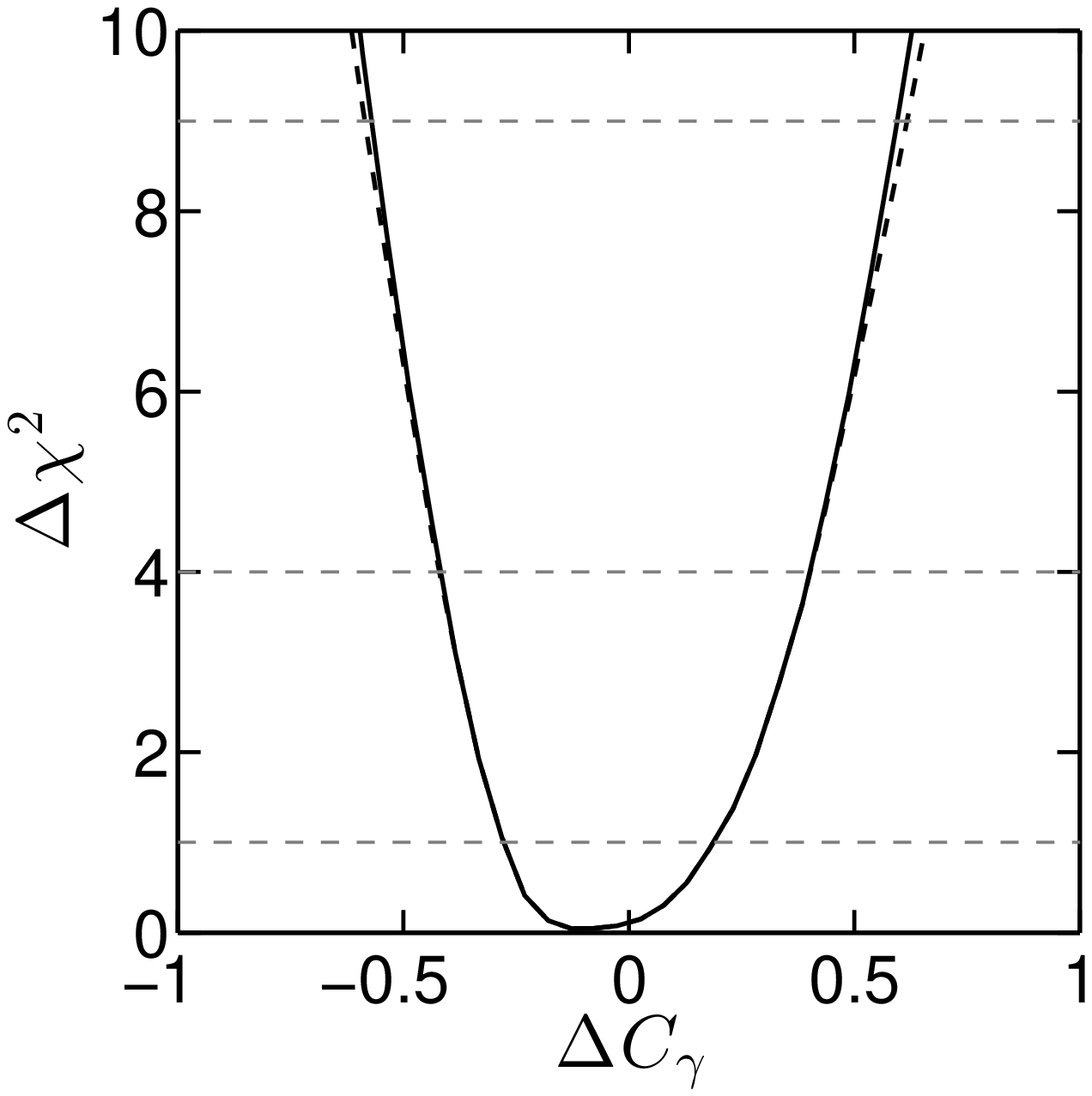}\quad
\includegraphics[width=5cm]{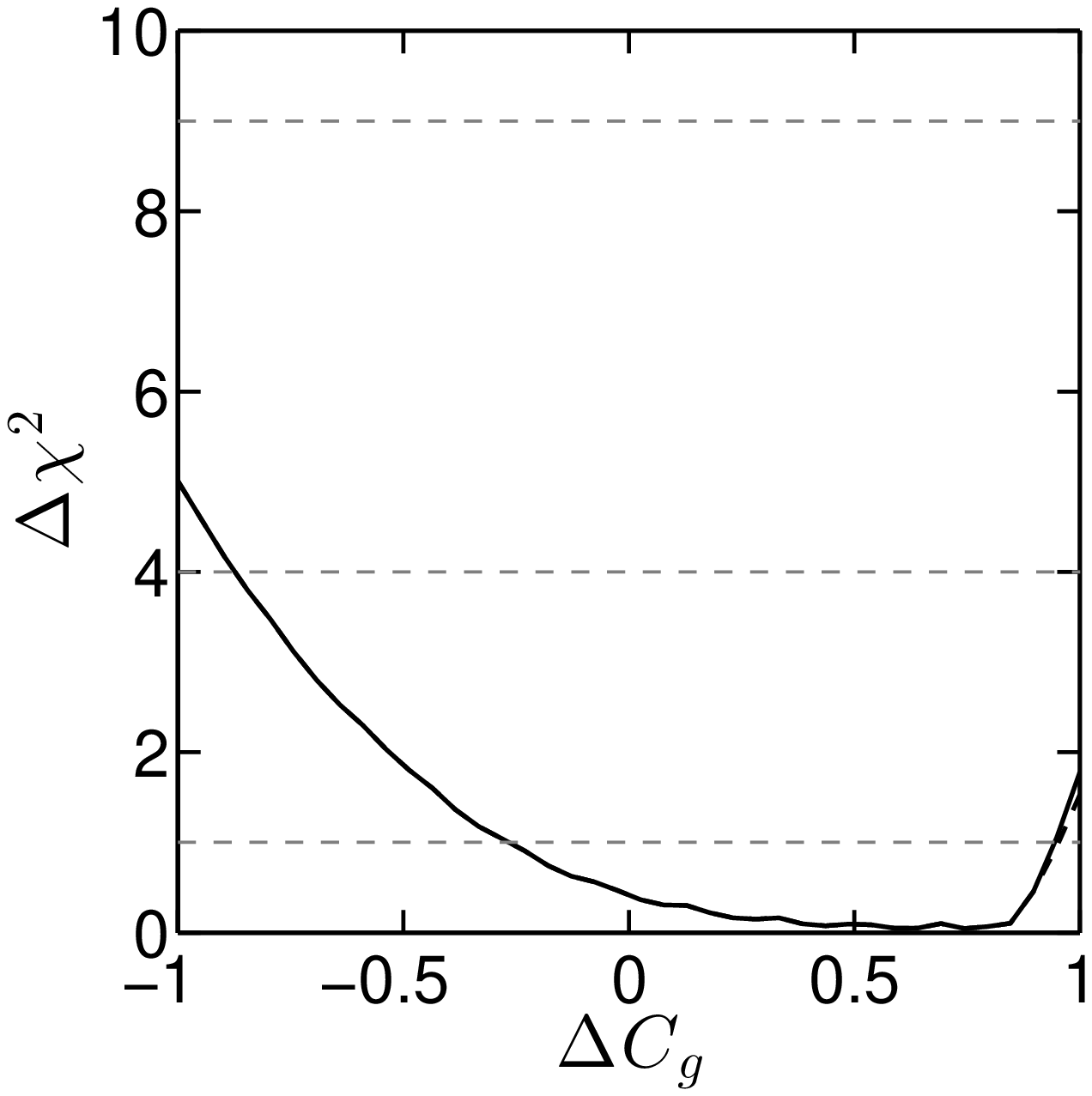}\\
\caption{Five (six) parameter fit of $\CU$, $\CD$, $\CV$,  $\dcg$ and $\dcp$; the solid (dashed) curves 
are those obtained when invisible/unseen decay modes are not allowed (allowed) for.
\label{fig:5param} }
\end{figure}

The general case of free parameters $\cu$, $\cd$, $\cv$, $\dcg$ and $\dcp$
is illustrated in Fig.~\ref{fig:5param}, where we show the 1D $\dchisq$
distributions for these five parameters (each time profiling over the
other four parameters).  As before, the solid (dashed) lines indicate
results not allowing for (allowing for) invisible/unseen decay modes of
the Higgs.  Allowing for invisible/unseen decay modes again relaxes the
$\dchisq$ behavior only modestly.  The best fit point always corresponds
to $\brinv=0$.  

\begin{figure}[t!]\centering
\includegraphics[width=6cm]{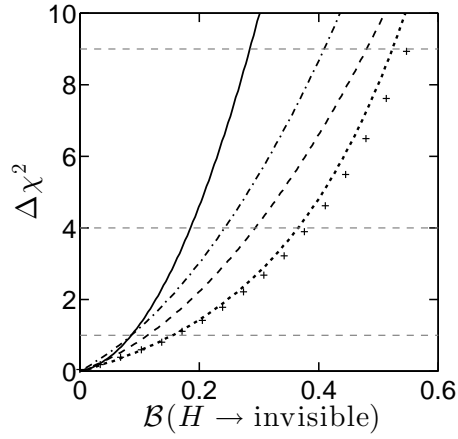}\\
\caption{$\Delta\chi^2$ distributions for the branching ratio of
invisible Higgs decays for various cases.  
Solid: SM+invisible. 
Dashed: varying $\dcg$ and $\dcp$ for $\CU=\CD=\CV=1$.
Dotted: varying $\CU$, $\CD$ and $\CV$ for $\dcg=\dcp=0$. 
Dot-dashed: varying $\CU$, $\CD$ and $\CV\leq 1$ for $\dcg=\dcp=0$.
Crosses: varying $\CU$, $\CD$, $\CV$,  $\dcg$ and $\dcp$.
\label{fig:BRinv} }
\end{figure}

An overview of the current status of invisible decays is given in Fig.~\ref{fig:BRinv}, which shows 
the behavior of $\dchisq$ as a function of $\brinv$ for various different cases of interest: \\
\indent a) SM Higgs with allowance for invisible decays --- one finds $\brinv<0.09$ (0.19); \\
\indent b) $\cu=\cd=\cv=1$ but $\dcp,\dcg$ allowed for --- $\brinv< 0.11$ (0.29); \\
\indent c) $\cu,\cd,\cv$ free, $\dcp=\dcg=0$, --- $\brinv<0.15$ (0.36); \\
\indent d) $\cu,\cd$ free, $\cv\leq 1$, $\dcp=\dcg=0$ --- $\brinv<0.09$ (0.24); \\
\indent e) $\cu,\cd,\cv,\dcg,\dcp$ free --- $\brinv<0.16$ (0.38).  \\
(All $\brinv$ limits are given at 68\% (95\%) CL.) 
Thus, while $\brinv$ is certainly significantly limited by the current data set, there remains ample room for invisible/unseen decays.  At 95\% CL, $\brinv$ as large as $\sim 0.38$ is possible.
Here, we remind the reader that the above results are obtained after fitting the $125.5\gev$ data {\em and} inputting the experimental results for the $(Z \to \ell^+\ell^-) \;+$ invisible direct searches. When $C_V \leq 1$, $H\to\,$invisible is much more constrained by the global fits to the $H$ properties than by the direct searches for invisible decays, {\it cf.}\ the solid, dashed and dash-dotted lines in Fig.~\ref{fig:BRinv}. 
For unconstrained  $C_U$, $C_D$ and $C_V$, on the other hand, {\it cf.}\ dotted line and crosses in Fig.~\ref{fig:BRinv}, the limit comes from the direct search for invisible decays in the $ZH$ channel. 
 
A comment is in order here. In principle there is a flat direction in the unconstrained LHC Higgs coupling fit when unobserved decay modes are present: setting $\cu = \cd = \cv \equiv C$, so that ratios of rates remain fixed, all the Higgs production$\times$decay rates can be kept fixed to the SM ones by scaling up $C$ while adding a new, unseen decay mode with branching ratio ${\cal B}_{\rm new}$ according to $C^2 = 1/(1 - {\cal B}_{\rm new}$)~\cite{Zeppenfeld:2000td,Djouadi:2000gu}, see also \cite{Duhrssen:2004cv}.\footnote{We thank Heather Logan for pointing this out.} 
In \cite{Belanger:2013kya} we found that it is mainly $\cv$ which is critical here, because of the rather well measured ${\rm VBF}\to H\to VV$ channel.  Therefore limiting $\cv\le 1$ gives a strong constraint on ${\cal B}_{\rm new}$, 
similar to the case of truly invisible decays. Concretely we find at 95\%~CL:  
{\it i)}  ${\cal B}_{\rm new}<0.21$ for a SM Higgs with allowance for unseen decays; 
{\it ii)}  ${\cal B}_{\rm new}<0.39$ for $\cu=\cd=\cv=1$ but $\dcp,\dcg$ allowed for; and 
{\it iii)}  ${\cal B}_{\rm new}<0.31$ for $\cu,\cd$ free, $\cv\leq 1$ and $\dcp=\dcg=0$. 
For unconstrained  $C_U$, $C_D$ and $C_V$, however, there is no limit on ${\cal B}_{\rm new}$.

\begin{figure}[t!]\centering
\includegraphics[width=6cm]{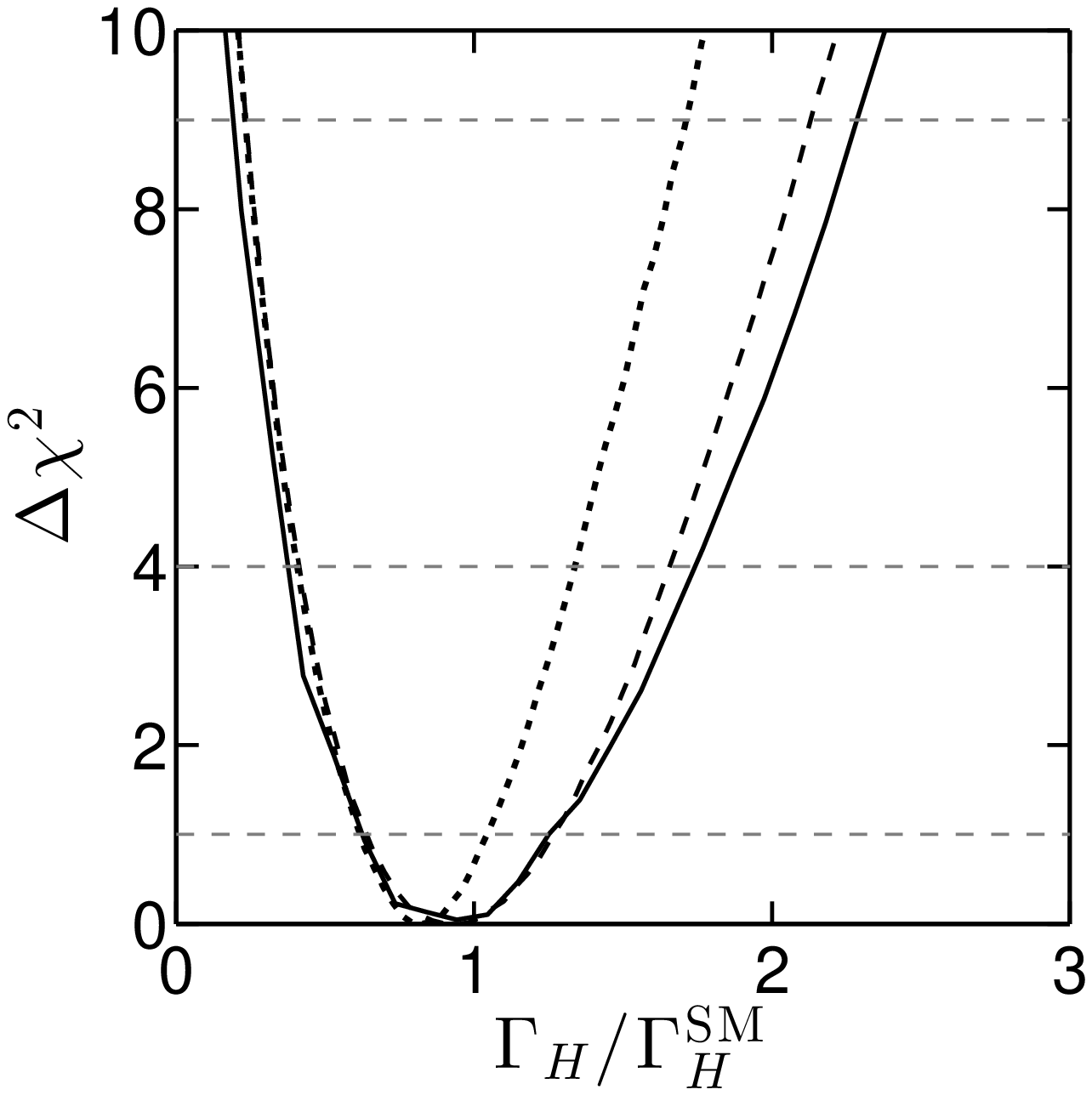}\quad
\includegraphics[width=6cm]{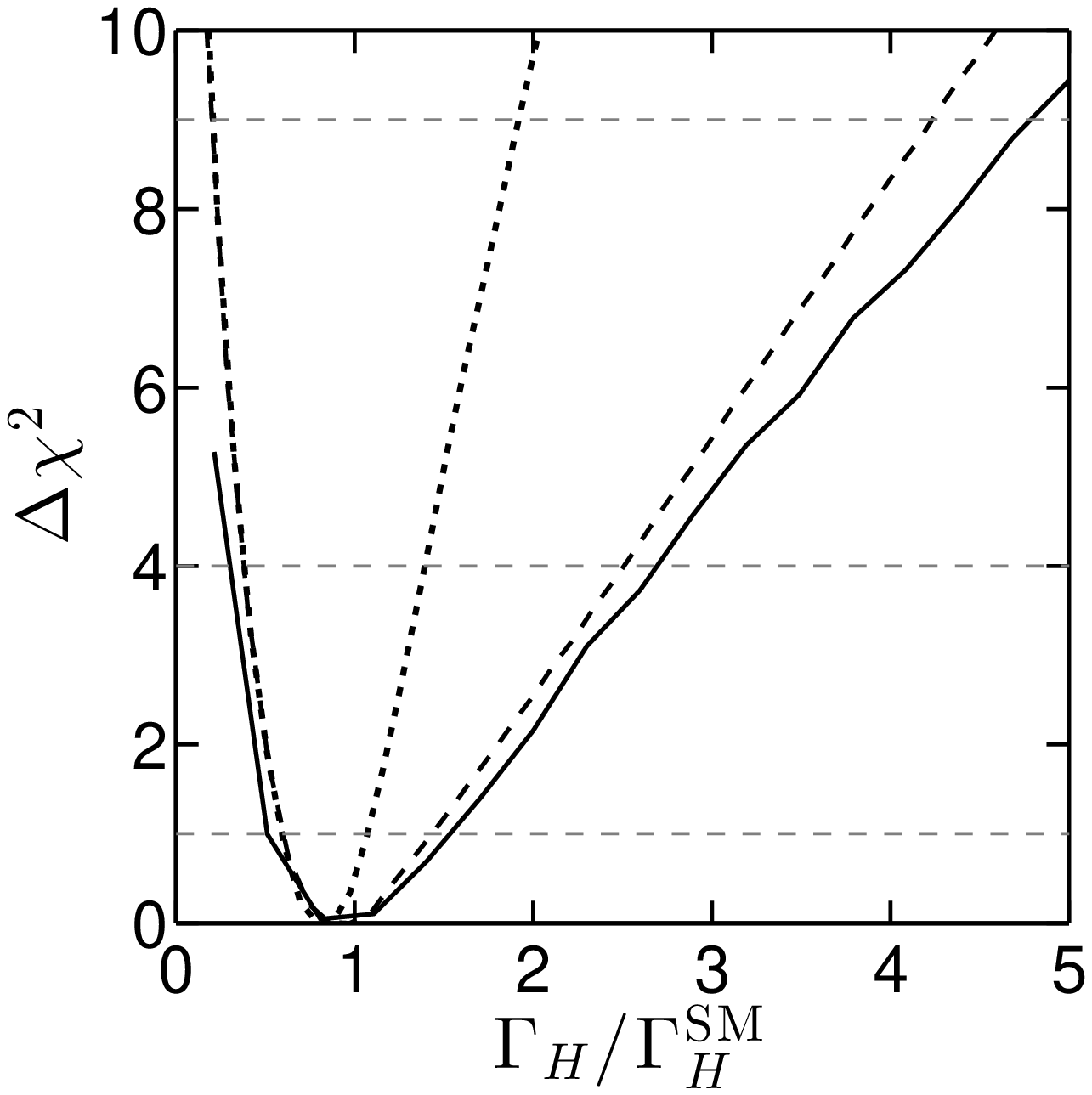}
\caption{$\Delta\chi^2$ distributions for the total Higgs decay width relative to SM, $\Gamma_{\rm tot}/\Gamma_{\rm tot}^{\rm SM}$, on the left without invisible decays, on the right including $\brinv$ as a free parameter in the fit. The lines are for:   
$\CU$, $\CD$ and $\CV\le1$ (dotted);
$\CU$, $\CD$ and free $\CV$ (dashed); and 
$\CU$, $\CD$, free $\CV$, $\dcg$, $\dcp$ (solid).
\label{fig:Rwidth} }
\end{figure}

With this in mind, 
the global fit we perform here also makes it possible to constrain the Higgs boson's total decay width, $\Gamma_{\rm tot}$, 
a quantity which is not directly measurable at the LHC. 
For SM + invisible decays, we find 
$\Gamma_{\rm tot}/\Gamma_{\rm tot}^{\rm SM}<1.11$ (1.25) at 68\% (95\%) CL. 
Figure~\ref{fig:Rwidth} shows the $\Delta\chi^2$ as function of 
$\Gamma_{\rm tot}/\Gamma_{\rm tot}^{\rm SM}$ for the fits of:  $\CU$, $\CD$, and $\CV\le1$; 
$\CU$, $\CD$, and $\CV$ free; and $\CU$, $\CD$, $\CV$, $\dcg$, $\dcp$.  
The case of $\dcg$, $\dcp$ with $\cu=\cd=\cv=1$ is not shown;  without invisible decays we find 
$\Gamma_{\rm tot}/\Gamma_{\rm tot}^{\rm SM}=[0.98,1.0]$ ($[0.97,1.02]$) at 68\% (95\%) CL 
in this case. Allowing for invisible decays this changes to 
$\Gamma_{\rm tot}/\Gamma_{\rm tot}^{\rm SM}=[0.97,1.14]$, ($[0.96,1.46]$), \ie\ it is very close 
to the line for $\CU$, $\CD$, $\CV\le1$ in the right plot of Fig.~\ref{fig:Rwidth}.

\clearpage
\section{Application to specific models}

So far our fits have been largely model-independent, relying only on assuming the 
Lagrangian structure of the SM. Let us now apply our fits to some concrete examples 
of specific models in which there are relations between some of the coupling factors $C_I$.

\subsection{Two-Higgs-Doublet Models}

\begin{table}[b]
\begin{center}
\begin{tabular}{|c|c|c|c|c|c|}
\hline
\ & Type I and II  & \multicolumn{2}{c|}  {Type I} & \multicolumn{2}{c|}{Type II} \cr
\hline
Higgs & VV & up quarks & down quarks \& & up quarks & down quarks \&  \cr
&  &  &  leptons &  &  leptons \cr
\hline
 $h$ & $\sin(\beta-\alpha)$ & $\cosa/ \sinb$ & $\cosa/ \sinb$  &  $\cosa/\sinb$ & $-{\sina/\cosb}$   \cr
\hline
 $H$ & $\cos(\beta-\alpha)$ & $\sina/ \sinb$ &  $\sina/ \sinb$ &  $\sina/ \sinb$ & $\cosa/\cosb$ \cr
\hline
 $A$ & 0 & $\cotb$ & $-\cotb$ & $\cotb$  & $\tanb$ \cr
\hline 
\end{tabular}
\end{center}
\vspace{-.15in}
\caption{Tree-level vector boson couplings $C_V$ ($V=W,Z$) and fermionic couplings $C_{F}$
normalized to their SM values for the two scalars $h,H$ and the pseudoscalar $A$ 
in Type I and Type II Two-Higgs-doublet models.}
\label{tab:2hdm-couplings}
\end{table}

As a first example, we consider Two-Higgs-Doublet Models (2HDMs) of Type~I and Type~II 
(see also \cite{Altmannshofer:2012ar,Chang:2012ve,Chen:2013kt,Celis:2013rcs,Grinstein:2013npa,Coleppa:2013dya,Chen:2013rba,Eberhardt:2013uba,Craig:2013hca,Maiani:2013nga} 
for other 2HDM analyses in the light of recent LHC data). 
In both cases, the basic parameters describing the coupling of either the light $h$ or heavy $H$ CP-even 
Higgs boson are only two: $\alpha$ (the CP-even Higgs mixing angle) and $\tanb=v_u/v_d$, where $v_u$ 
and $v_d$ are the vacuum expectation values of the Higgs field that couples to up-type quarks and down-type 
quarks, respectively.  The Type~I and Type~II models are distinguished by the pattern of their fermionic couplings 
as given in Table~\ref{tab:2hdm-couplings}.  The SM limit for the $h$ ($H$) in the case of both Type~I and Type~II models corresponds to $\alpha=\beta-\pi/2$ ($\alpha=\beta$).  
We implicitly assume that there are no contributions from non-SM particles to the loop 
diagrams for $\cp$ and $\cg$.  In particular, this means our results correspond to the case where the charged 
Higgs boson, whose loop might contribute to $\cp$, is heavy.  

\begin{figure}[t!]\centering
\includegraphics[width=6.1cm]{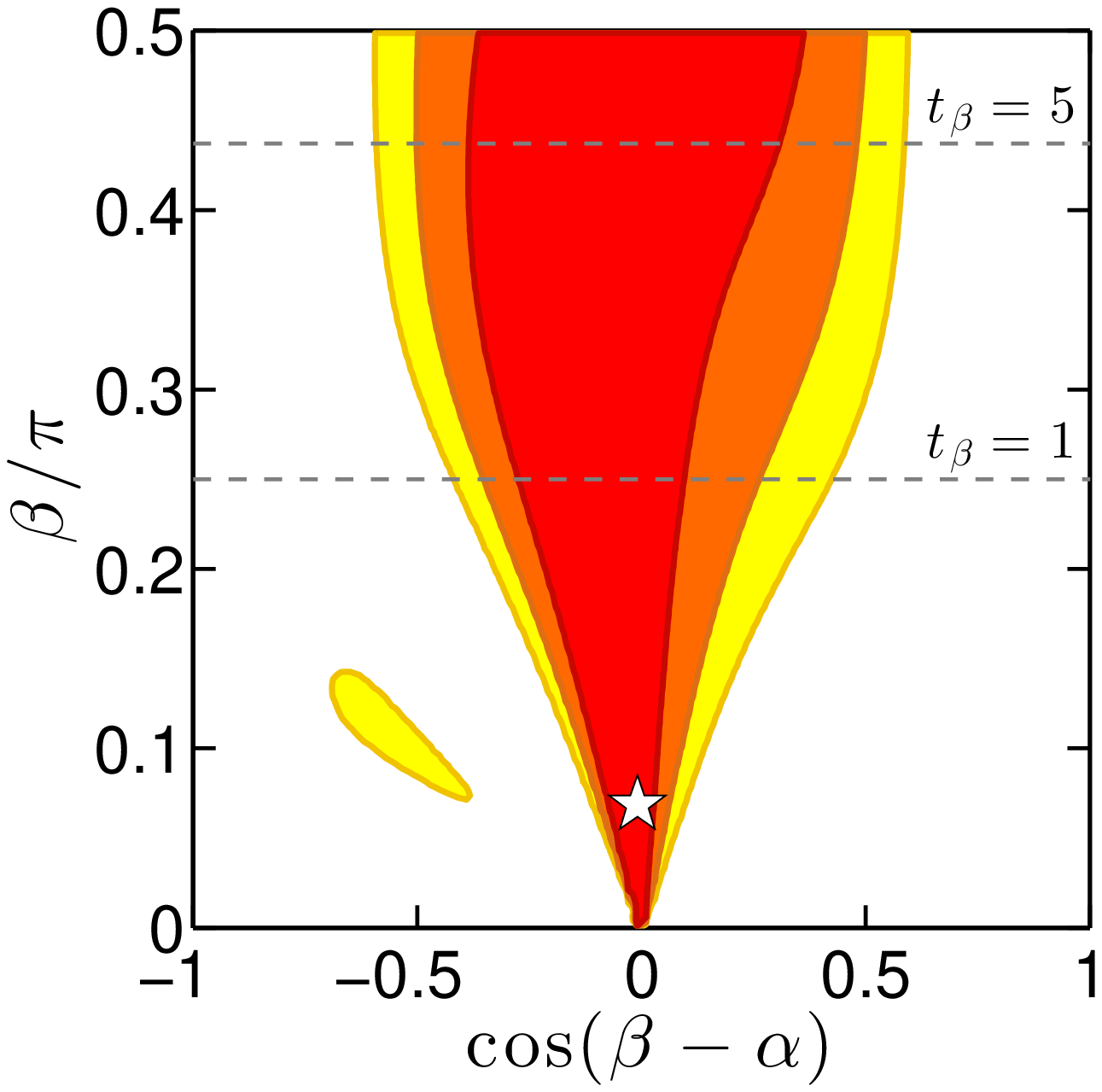}\quad
\includegraphics[width=6.1cm]{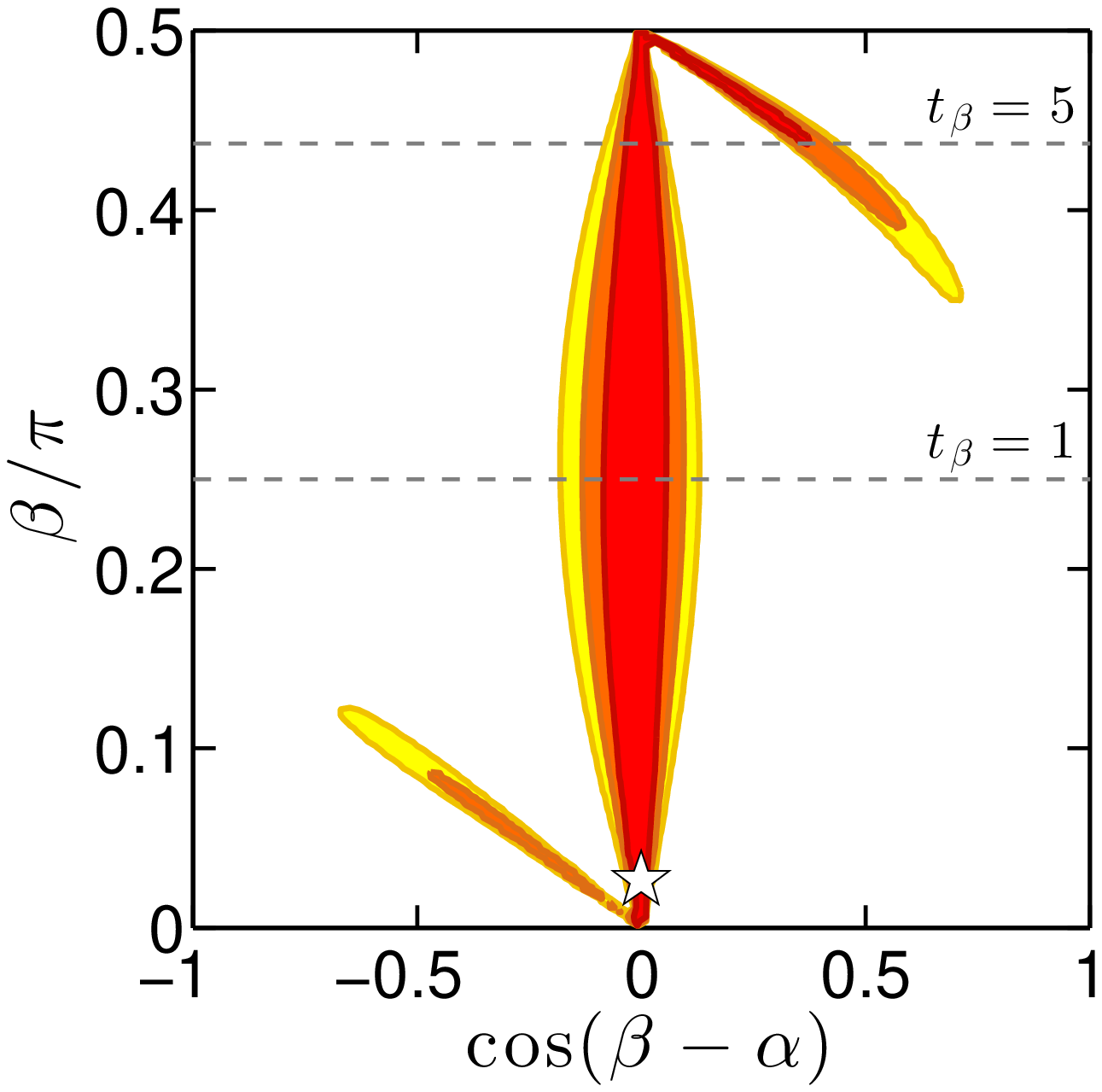}\\
\includegraphics[width=6cm]{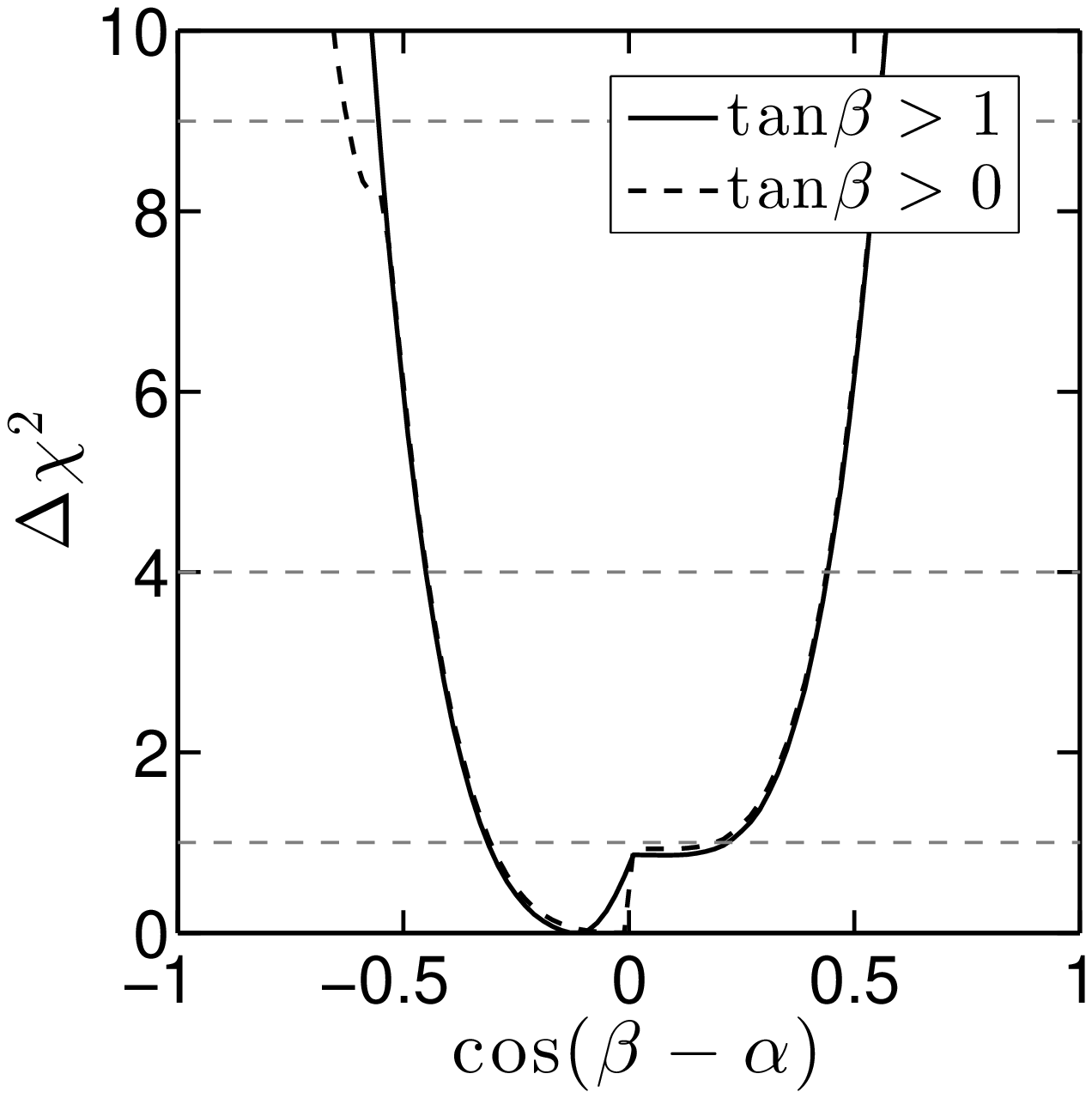}\quad
\includegraphics[width=6cm]{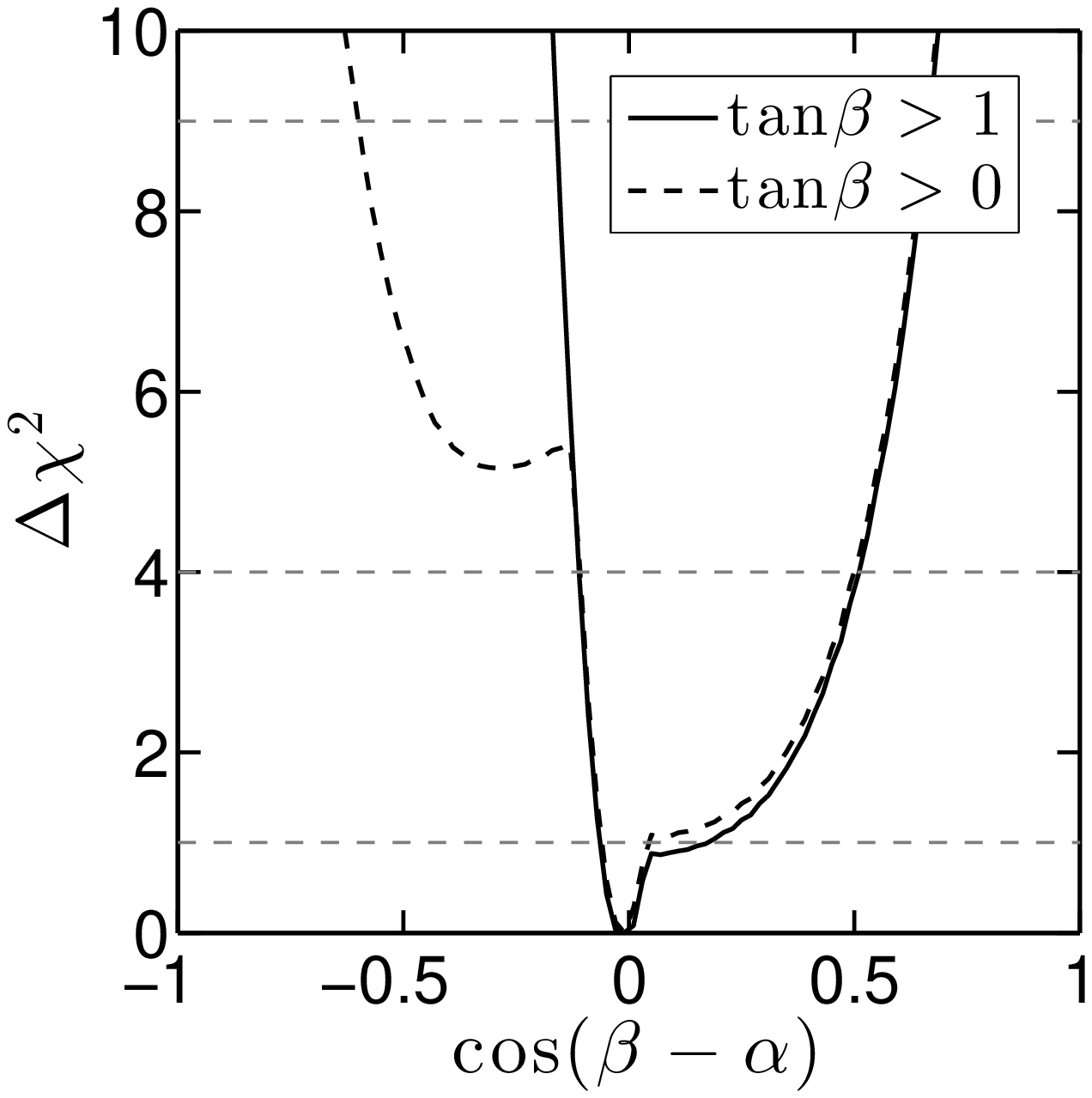}
\caption{Fits for the  2HDM Type~I (left) and type~II (right) models for
$m_h=125.5$~GeV.   See text for details. \label{fig:2hdm} 
}
\end{figure}

The results of the 2HDM fits are shown in Fig.~\ref{fig:2hdm} for the case that the state near 125~GeV
is the lighter CP-even $h$. To be precise, the top row shows $\Delta\chi^2$ contours in the 
$\beta$ versus $\cos(\beta-\alpha)$ plane while the bottom row shows the 1D projection of   
$\Delta\chi^2$ onto  $\cos(\beta-\alpha)$ with $\beta$ profiled over. 
For identifying the heavier $H$ with the state near 125~GeV, replace $\cos(\beta-\alpha)$ by $\sin(\beta-\alpha)$ in the 1D plots. (Since the $\sim 125\gev$ state clearly couples to $WW,ZZ$ we do not consider the case where the $A$ is the only state at $\sim 125\gev$.)

In the case of the Type~I model, we note a rather broad valley along the SM limit of $\cos(\beta-\alpha)=0$, 
which is rather flat in $\tan\beta$; the 68\% (95\%) CL region extends to 
$\cos(\beta-\alpha) = [-0.31,\, 0.19]$ ($[-0.45,\,0.44]$). 
The best fit point lies at $\beta\simeq0.02\pi$ and $\alpha\simeq 1.52\pi$ 
with $\chimin=18.01$ for 21 d.o.f.\ (to be compared to the SM $\chimin=18.95$). 
Requiring $\tan\beta>1$, this moves to $\beta\simeq0.25\pi$, 
\ie\ $\tan\beta$ just above 1, with  $\alpha\simeq 1.71\pi$ and $\chimin=18.08$.
At 99.7\% CL, there is also a small island at $\cos(\beta-\alpha)\approx -0.5$ and $\tan\beta<1$, 
which corresponds to the $\CU<0$ solution. (This is responsible for the splitting of the two lines at $\cos(\beta-\alpha)\lesssim-0.5$ in the 1D plot.)

In contrast, for the Type~II  model, we observe two narrow 68\%~CL valleys in the $\beta$ versus 
$\cos(\beta-\alpha)$ plane, one along the SM solution with the minimum again very close 
to $\beta\approx 0$ and a second banana-shaped one with $\tan\beta\gtrsim 5$ (3) and 
$\cos(\beta-\alpha)\lesssim 0.4$ (0.6) at 68\% (95\%) CL. 
This second valley is the degenerate solution with $\CD\approx-1$; it does not appear in Fig.~3 
of \cite{Craig:2013hca} because there $\CU,\CD>0$ was implicitly assumed.
The best fit point is very similar to that for Type~I: $\beta\simeq0.01\pi$ ($0.25\pi$) and $\alpha\simeq 1.5\pi$ ($1.75\pi$) with $\chimin=18.68$ ($18.86$) for 21 d.o.f.\ for arbitrary $\tan\beta$ ($\tan\beta>1$).  
Again, there is an additional valley very close to $\beta\sim 0$, extending into the negative $\cos(\beta-\alpha)$ direction, which however does not have a 68\%~CL region. 
In 1D, we find $\cos(\beta-\alpha) = [-0.11,\,0.50]$ at 95\% CL.

Let us end the 2HDM discussion with some comments regarding the ``other''
scalar and/or the pseudoscalar $A$. To simplify the discussion, we will focus on the $m_h=125.5\gev$ case.
First, we note that if the $H$ and $A$ are heavy enough (having masses greater than roughly $600\gev$) then their properties are unconstrained by LHC data and the global fits for the $h$ will be unaffected. If they are lighter then it becomes interesting to consider constraints that might arise from not having observed them. Such constraints will, of course, depend upon their postulated masses, both of which are independent parameters in the general 2HDM. For purposes of discussion, let us neglect the possibly very important $H,A\to hh$ decays.  The most relevant final states are then $H\to VV$ and $H,A\to \tau\tau$.

With regard to observing the heavy Higgs in the $H \to VV$ channels, we note that for the $H$ our fits predict the $VV$ coupling to be very much suppressed in a large part (but not all) of the 95\% CL allowed region. While this implies suppression of the VBF production mode for the $H$ it does not affect the ggF production mode and except for very small $VV$ coupling the branching ratio of the $H$ to $VV$ final states declines only modestly. As a result, the limits in the $ZZ\to4\ell$ channel \cite{ATLAS-CONF-2013-013}, 
which already extend down to about $0.1\times$SM 
in the mass range $m_H\approx 180-400$~GeV, and to about  $0.8\times$SM 
at $m_H\approx 600$~GeV, can be quite relevant. 
For instance, for a heavy scalar $H$ of mass $m_H=300$~GeV, in the 95\%~CL region of our fits 
the signal strength in the $gg\to H\to ZZ$
channel ranges from 0 to 5.4 in Type~I and from 0 to 33 in Type~II. For $m_H=600$~GeV, 
we find $\mu(gg\to H\to ZZ)\lesssim 1.1$ (0.6) in Type~I (II).  Further, at the best-fit point for $\tanb>1$, $\mu(gg\to H\to ZZ)=1.10~(0.08)$ at $m_H=300~(600)\gev$ in Type~I and  $\mu(gg\to H\to ZZ)=0.12~(0.001)$ at $m_H=300~(600)\gev$ in Type~II, which violate the nominal limits at $m_H=300\gev$ in both models.  
Note, however, that it is possible to completely evade the $4\ell$ bounds
if $H\rightarrow hh$ decays are dominant.

Moreover, both the $H$ and the $A$, which has no tree-level couplings to $VV$, may 
show up in the $\tau\tau$ final state through ggF. Limits from ATLAS \cite{Aad:2012yfa} 
range (roughly) from $\mu(gg\to H,A \to \tau\tau)$   $< 2500$ at $m_{H,A}=300\gev$ to $<21000$ at $m_{H,A}=500\gev$. 
These may seem rather weak limits, but in fact the signal strengths for $H\to\tau\tau$ and $A\to\tau\tau$ 
(relative to $H_{\rm SM}$) can be extremely large. In the case of the $A$, this is because the $A\to\tau\tau$ branching ratio is generically much larger than the $H_{\rm SM}\to\tau\tau$ branching ratio, the latter being dominated by $VV$ final states at high mass.  In the case of the $H$, the same statement applies whenever its $VV$ coupling  is greatly suppressed.
We find that only the Type~I model with $\tan\beta>1$ completely evades the $\tau\tau$ 
bounds throughout the 95\% CL region of the $h$ fit since both the fermionic couplings of $H$ and $A$ 
are suppressed by large $\tan\beta$.
In the Type~II model, $gg\to A\to\tau\tau$ satisfies the $\tau\tau$ bounds at 95\%~CL, but 
$gg\to H\to\tau\tau$  can give a very large signal. However, the best fit $h$ point for $\tanb>1$ in Type~II 
predicts  $\mu(gg\to H\to\tau\tau)$ values of $674$ and $6.4$ at $300$ and $500\gev$, both of which  
satisfy the earlier-stated bounds. We also stress that no bounds are available in the $\tau\tau$ 
channel above 500~GeV. 

Clearly, a full study is needed to ascertain the extent to which limits in the $H\to ZZ$ and $H,A\to \tau\tau$ 
channels will impact the portion of the $\alpha$ --- $\beta$ plane allowed at 95\%~CL after taking into account  
Higgs-to-Higgs decays, which are typically substantial.
This is beyond the scope of this paper and will be presented elsewhere~\cite{2hdmyun}.

\subsection{Inert Doublet Model}

In the Inert Doublet Model (IDM)~\cite{Deshpande:1977rw}, 
a Higgs doublet $\tilde H_2$ which is odd 
under a $Z_2$ symmetry is added to the SM leading to four new particles: a scalar $\tilde{H}$, 
a pseudoscalar $\tilde{A}$,  and two charged states $\tilde{H}^\pm$ in addition to the SM-like Higgs $h$.\footnote{For distinction with the 2HDM, 
we denote all IDM particles odd under $Z_2$ with a tilde.}
All other fields being even, this discrete symmetry not only guarantees that the lightest inert Higgs particle is stable, and thus a suitable dark matter candidate~\cite{Ma:2006km,Barbieri:2006dq,LopezHonorez:2006gr,Krawczyk:2013jta}, but also  prevents the coupling of any of the inert doublet particles to pairs of SM particles.
Therefore, the only modification to the SM-like Higgs couplings is through the charged Higgs contribution 
to $\Delta C_\gamma$. The scalar potential of the IDM  is given by
\begin{eqnarray}
V&=&\mu_1^2 |H_1|^2 + \mu_2^2 |\tilde{H}_2|^2 +\lambda_1 |H_1|^4+\lambda_2|\tilde{H}_2|^4+\lambda_3|H_1|^2 |\tilde{H}_2|^2\nonumber\\
& &  +\, \lambda_4 |H_1^\dagger \tilde{H}_2|^2 
         +\frac{\lambda_5}{2} \left[  \left( H_1^\dagger \tilde{H}_2 \right)^2 + {\rm h.c.}\right] \,,
\end{eqnarray}
where $\mu_2^2>-v^2$ is required in order that $\tilde H_2^0$ not acquire a non-zero vev (which would violate the symmetry needed for $\tilde H$ to be a dark matter particle).
The crucial interactions implied by this potential are those coupling the light Higgs $h$ associated with the $H_1$ field to pairs of Higgs bosons coming from the $\tilde H_2$ field.
These are given by:  $-(2 m_W/g) \lambda_3 h \tilde{H}^+ \tilde{H}^-$,
 $-(2 m_W/g) \lambda_L h \tilde{H} \tilde{H}$ and  $-(2 m_W/g) \lambda_S h \tilde{A} \tilde{A}$ for the charged, scalar and pseudo scalar, respectively, where
\begin{equation}
\lambda_{L,S}= \frac{1}{2}(\lambda_3+\lambda_4\pm\lambda_5)\,.
\end{equation}
With these abbreviations, the Higgs masses at tree-level can be written as 
\beq
   m^2_h = \mu_1^2 + 3\lambda_1v^2,\quad
   m^2_{\tilde H,(\tilde A)} = \mu_2^2 + \lambda_{L(S)}\,v^2,\quad
   m^2_{\tilde H^\pm} = \mu_2^2 + \frac{1}{2}\lambda_3v^2\,.
   \label{eq:idm-mh}
\eeq
Moreover, the couplings to the inert charged and neutral Higgses are related by
\begin{equation}
\frac{\lambda_3}{2}=\frac{1}{v^2}\left( m_{\tilde{H}^+}^2-m_{\tilde{H}}^2 \right) + \lambda_L \,.
\label{eq:lambda3}
\end{equation}
It is important to note that a priori  $m^2_{\tilde H,\tilde A,\tilde H^+}$ are each free parameters and could be small enough that $h$ decays to a pair of the dark sector states would be present and possibly very important.  The $h\to \tilde H\tilde H$ and $h\to \tilde A\tilde A$ decays would be invisible and contribute to $\brinv$ for the $h$; $h\to \tilde H^+\tilde H^-$ decays would generally be visible so long as the $\tilde H^+$ was not closely degenerate with the $\tilde H$.

Theoretical constraints  impose some conditions on the couplings. Concretely, we assume a generic 
perturbativity upper bound $|\lambda_i|<4\pi$, which, when coupled with the vacuum stability 
and perturbative unitarity conditions on the potential, leads to $\lambda_3>-1.5$ and 
$\mu_2^2\gtrsim-4.5\times 10^4$~GeV$^2$~\cite{Krawczyk:2013jta,Swiezewska:2012ej}. We also adopt
a lower bound of $m_{\tilde{H}^\pm}>70$~GeV, as  derived from chargino limits at LEP~\cite{Pierce:2007ut,Lundstrom:2008ai}.
Note however that LHC exclusions for the SM Higgs do not apply to members of the inert doublet because 
{\em i)} they do not couple to fermions and {\em ii)} trilinear and quartic couplings to gauge bosons involve two inert Higgses. 

\begin{figure}[t]\centering
\includegraphics[width=6cm]{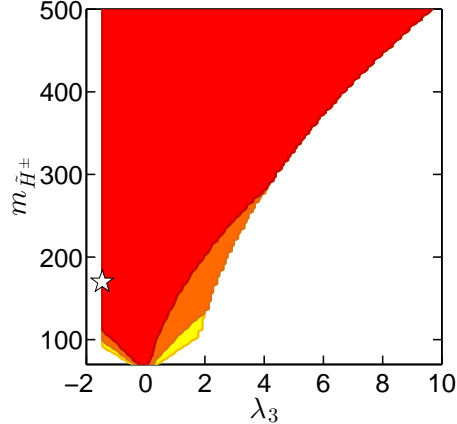}
\caption{Contours of 68\%, 95\%, 99.7\% CL in the $m_{\tilde{H}^\pm}$ versus $\lambda_3$ plane for the 
IDM assuming that there are no invisible decays of the SM-like Higgs $h$. 
\label{fig:idm1} }
\end{figure}

Let us now turn to the fit results.\footnote{In our IDM fits, the $h\gamma\gamma$ coupling is computed with {\tt micrOMEGAs\,3}~\cite{Belanger:2013oya}.} 
First, we consider the case where $m_{\tilde{H}},m_{\tilde{A}}>m_h/2$---the only deviation from the SM then 
arises from the charged Higgs contribution to  $\Delta C_\gamma$ parametrized by $\lambda_3$ and $m_{\tilde{H}^\pm}$. 
The general one-parameter  fit to the Higgs couplings leads to the bounds  
$-0.02~(-0.13) <\Delta C_\gamma < 0.17~(0.26)$ at $1\sigma$ $(2\sigma)$. 
The corresponding contours in the $m_{\tilde{H}^\pm}$ versus $\lambda_3$ plane are shown in Fig.~\ref{fig:idm1}.  
Note that the 3rd equality of Eq.~(\ref{eq:idm-mh}) and the lower bound of $\mu_2^2\gtrsim-4.5\times 10^4$~GeV$^2 $ imply an upper bound on $\lam_3$ for any  given $m_{\tilde{H}^\pm}$.  This excludes the large-$\lam_3$ region when $m_{\tilde H^+}\gsim 130\gev$.  
The impact of the global fit is confined to the region $m_{\tilde H^+}\lsim 130\gev$ and $|\lam_3|\lsim 2$ (at 95\% CL). The best fit point lies at $m_{\tilde H^+}=170$~GeV and $\lam_3=-1.47$.

\begin{figure}[t]\centering
\includegraphics[width=5.2cm]{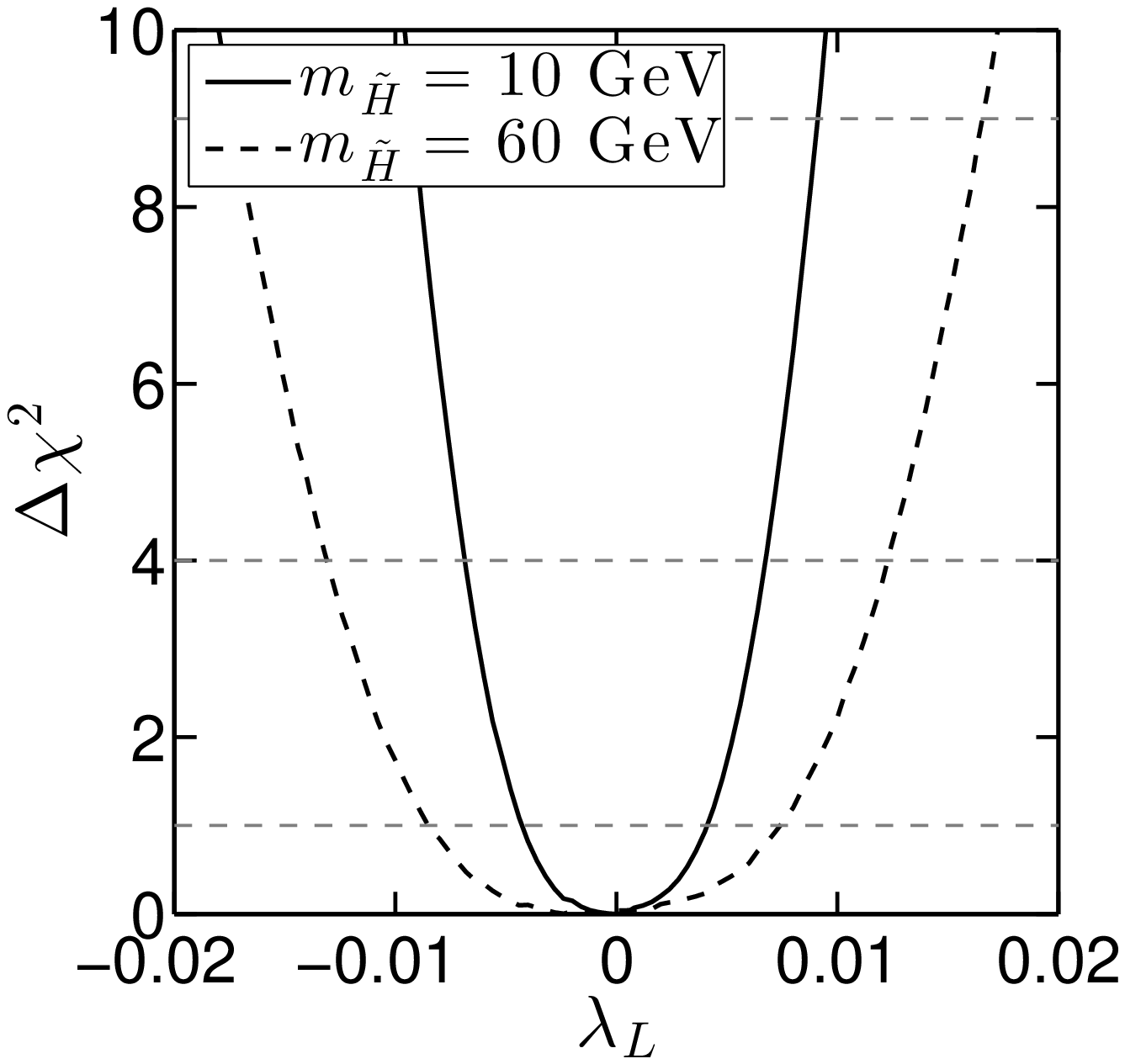}
\includegraphics[width=5cm]{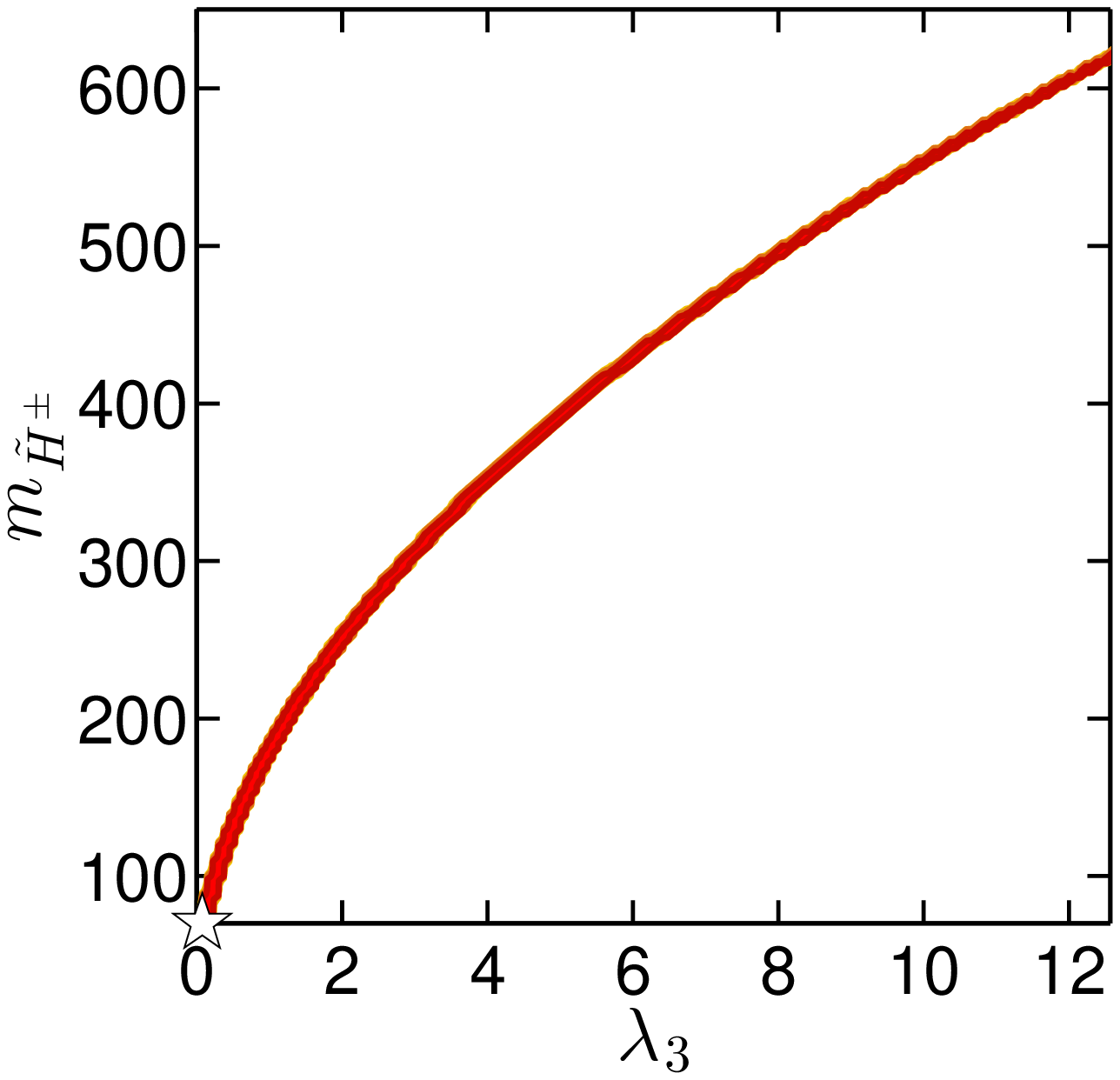}
\includegraphics[width=5cm]{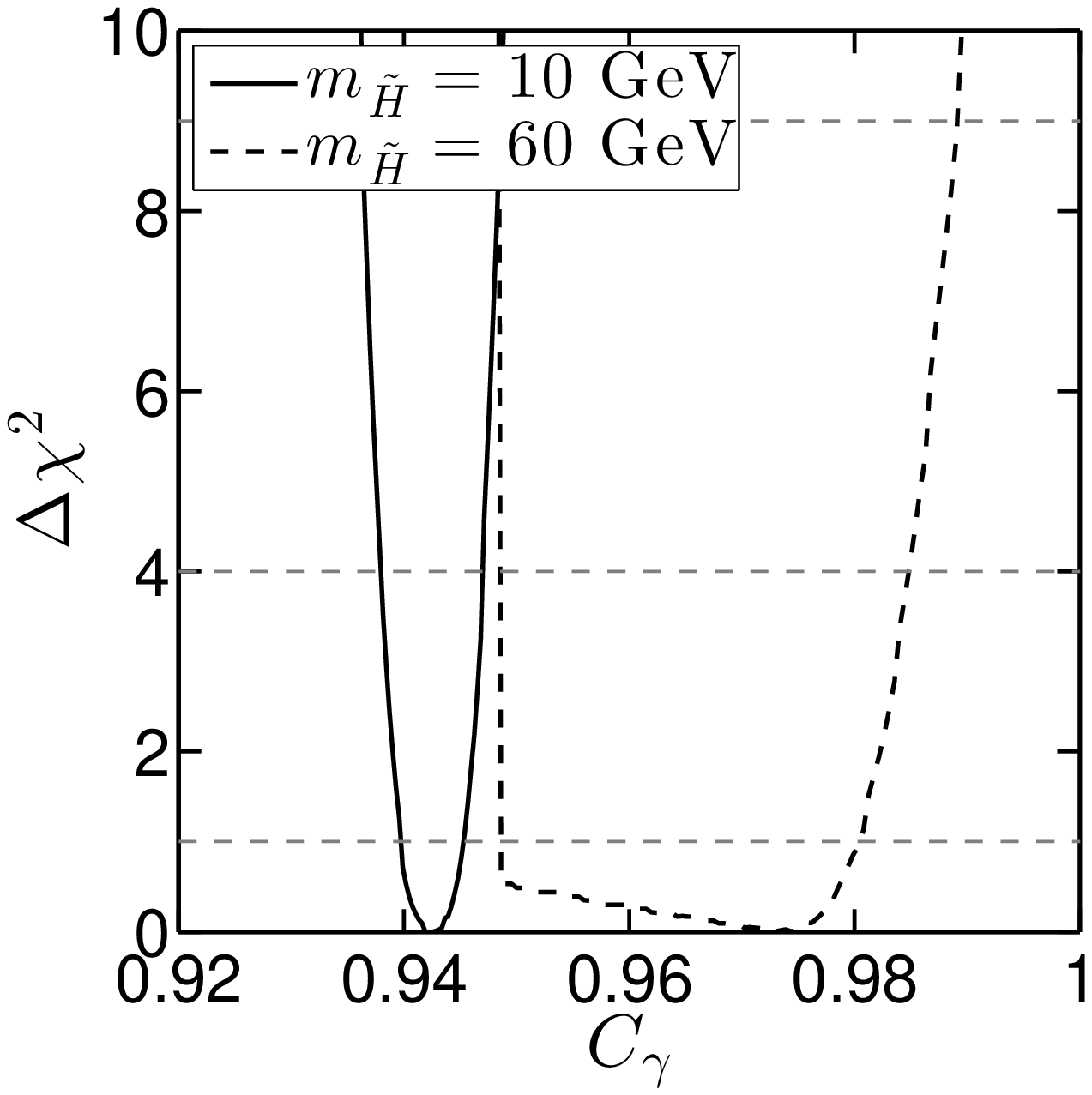}
\caption{Left panel: $\Delta\chi^2$ distribution of $\lambda_L$ for $m_{\tilde H}=10$~GeV  (full line)  and 60~GeV (dashed line) with $m_{\tilde{H^+}}$ profiled over its whole allowed range. 
Middle panel: relation between $m_{\tilde{H}^\pm}$ and $\lambda_3$ with $m_{\tilde H}$ profiled over from 1 to 60~GeV. 
Right panel: $\Delta\chi^2$ as function of $C_\gamma$ for $m_{\tilde H}=10$~GeV  (full line)  and 60~GeV (dashed line) with $m_{\tilde{H}^\pm}$ profiled over.  
\label{fig:idm2} }
\end{figure}

Second, we consider the case where the inert scalar is light and examine how invisible 
$h\to \tilde H\tilde H$ decays further constrain the parameters.
The bounds on the invisible width actually lead to a strong constraint on the coupling $\lambda_L$.  
The $1\sigma$ ($2\sigma$) allowed range is roughly $\lambda_L \times 10^3=\pm 4$  ($\pm 7$) 
for $m_{\tilde{H}}=10~{\rm GeV}$. This bound weakens only when the invisible decay is suppressed by 
kinematics; for  $m_{\tilde{H}}=60~{\rm GeV}$, we find $\lambda_L\times 10^{3}=[-9,7]$ ($[-13,12]$) at $1\sigma$ ($2\sigma$).  
The $\Delta\chi^2$ distributions of $\lambda_L$ for $m_{\tilde{H}}=10$ and 60~GeV are shown in the left panel 
in Fig.~\ref{fig:idm2}, with $m_{\tilde H^\pm}$ profiled over from 70~GeV to about 650~GeV (the concrete upper 
limit being determined by the perturbativity constraint). 
This strong constraint on $\lambda_L$ implies that it can be neglected in Eq.~(\ref{eq:lambda3}) and 
that the charged Higgs coupling $\lambda_3$ is directly related to $m_{\tilde{H}^\pm}$ for a given 
$m_{\tilde H}$,  as illustrated in the middle panel of Fig.~\ref{fig:idm2} 
(here, the mass of the inert scalar is profiled over in the range $m_{\tilde H} \in [1,\,60]$~GeV). 
As a result the value of $C_\gamma$ is also strongly constrained from the upper bound on the invisible width. 
For example for $m_{\tilde{H}}=10$~GeV, we find that $C_\gamma =[0.940,0.945]$ at 68\% CL. 
Note that because $m_{\tilde{H}^\pm}> m_{\tilde{H}}$ is needed in order to have a neutral dark matter candidate, $\lambda_3$ is 
always positive  and therefore $C_\gamma<1$. 
To approach $C_\gamma\simeq 1$, the inert Higgs mass has to be close to the kinematic threshold,  $m_{\tilde H}\to m_h/2$ so that the constraint on $\lambda_L$ is relaxed. For illustration, see the right panel in Fig.~\ref{fig:idm2}. 
These results imply that with an improved accuracy on the measurements of the Higgs coupling, for example showing that $C_\gamma> 0.95$, it would be possible to exclude light dark matter  ($m_{\tilde H}<10$~GeV) in the IDM.
Another consequence is that for a given $m_{\tilde H}$ the perturbativity limit $\lambda_3<4\pi$ implies an upper bound on the charged Higgs mass. For $m_{\tilde H} \in [1,\,60]$~GeV we obtain $m_{\tilde{H}^\pm}<620$~GeV.

Finally note that the  case where $\tilde{A}$ is the lightest neutral state and $m_{\tilde{A}}< m_h/2$ 
is analogous to the $\tilde H$ case just discussed, with $m_{\tilde{H}}\rightarrow
m_{\tilde{A}}$ and $\lambda_L \rightarrow \lambda_S$ and leads to analogous conclusions. 
Analyses of the Higgs sector of the Inert Doublet Model were also performed recently  in~\cite{Krawczyk:2013jta,Goudelis:2013uca,Arhrib:2012ia,Gustafsson:2012aj,Swiezewska:2012eh}.

\subsection{Triplet Higgs model}

In this section we consider the model of \cite{Georgi:1985nv} which combines a single Higgs doublet field with  $Y=0$ and $Y=\pm 1$ triplet fields in such a way that custodial symmetry is preserved at tree level.
The phenomenology of this model was developed in detail
in~\cite{Gunion:1989ci,Englert:2013zpa}.  In this model, the neutral doublet and triplet
fields acquire vacuum expectation values given by
$\vev{\phi^0}=a/\sqrt 2$ and 
$\vev{\chi^0}=\vev{\xi^0}=b$, respectively.  It is the presence of the two triplet fields and their neutral members having the same vev, $b$, that guarantees $\rho=1$ at tree level.
The value of 
$v^2\equiv a^2+8b^2=(246\gev)^2$ is determined by the $W,Z$ masses.  However, the relative magnitude of $a$ and $b$ is a parameter of the model.  The relative mixture is defined by the doublet-triplet mixing angle $\theta_H$ with cosine and sine given by $ c_H ={a\over\sqrt{a^2+8b^2}}$ and 
$s_H= \sqrt{{8b^2\over a^2+8b^2}}$.
The angle $\theta_H$ is reminiscent of the $\beta$ angle of a 2HDM.  Just like $\beta$, $\theta_H$ can be taken to lie in the 1st quadrant so that both $c_H$ and $s_H$ are positive.

In this model, it is most natural to choose a Higgs sector potential that preserves the custodial symmetry. ln this case, the Higgs eigenstates comprise a five-plet, a triplet and two singlets, $\hone$ and $\honepr$. The Higgs bosons of the five-plet couple only to vector boson pairs and those of the triplet couple only to fermion pairs.  Further, the neutral members of the five-plet and the triplet cannot mix (without violating the custodial symmetry).  As a result, they cannot describe the Higgs-boson like state seen at the LHC.  In contrast, the $\hone$  and $\honepr$ can mix.  Further, their reduced couplings are given by
\beq
C_F(\hone)={1\over c_H},\quad C_V(\hone)=c_H,\quad C_F(\honepr)=0,\quad C_V(\honepr)={2\sqrt 2\over \sqrt 3}s_H\,,
\eeq	
where all fermionic couplings scale with the common factor $C_F$. 
We see that in the limit $c_H\to 1$ the $\hone$ looks exactly like the SM Higgs boson and the $\honepr$ has no tree-level couplings.
More generally, from these expressions, it is clear that only a Higgs state that is primarily $\hone$ can provide the SM-like signal rates that typify the $\sim 125.5\gev$ state observed at the LHC. 	

The mixing of the $\hone$ and $\honepr$ is determined by 
the mass-squared matrix:
\beq
{\cal M}^2_{\hone,\honepr}=
\left( \begin{array}{cc} c_H^2\bar\lam_{13} & s_Hc_H\bar \lam_3 \cr
           s_Hc_H\bar \lam_3 & s_H^2\bar \lam_{23} \cr\end{array}\right)v^2 \,,
           \label{monematrix}
\eeq
where we have defined 
\beq
\bar\lam_{13}\equiv 8(\lam_1+\lam_3)\,,\quad \bar\lam_{23}\equiv 3(\lam_2+\lam_3)\,,\quad \bar\lam_{3}\equiv 2\sqrt 6 \lam_3\,,
\label{lambardefs}
\eeq
where  $\lam_{1,2,3}$ are couplings appearing in the full Higgs sector potential (see \cite{Gunion:1989ci}), with $\lam_1+\lam_3>0$ and $\lam_2+\lam_3>0$ required for stability in the asymptotic $\phi$ and $\chi$ directions, respectively, and $\lam_1\lam_2+\lam_1\lam_3+\lam_2\lam_3>0$ required  for positive mass-squared for the mass eigenstates coming from the $\hone$--$\honepr$ sector.
Clearly, the mixing between $\hone$ and $\honepr$ vanishes in
the limit of $\lam_3\rta 0$.   More generally, the above mass-squared matrix will be diagonalized by a rotation matrix specified by an angle for which we use the 2HDM-like notation,  $\alpha$.
We define $\alpha$ using the convention in which the Higgs boson mass eigenstates are given by
\beq
H=\cosa \hone+\sina\honepr\,, \quad H'=-\sina \hone+\cosa\honepr\,.
\eeq
We can solve for the $\bar\lam$'s in terms of $m_H^2$ and $m_{H'}^2$ and the mixing angle $\alpha$:
\beq
\bar\lam_{13}={m_H^2\ca^2+m_{H'}^2\sa^2\over  \ch^2v^2}\,,\quad
\bar\lam_{23}={m_H^2\sa^2+m_{H'}^2\ca^2\over  \sh^2v^2}\,,\quad
 \bar\lam_3 = {(m_H^2-m_{H'}^2)\sa\ca
\over \ch\sh v^2}\,,
\eeq
valid regardless of the relative size of $m_H^2$ and $\mhpr^2$.

As regards the masses of the triplet members and of the five-plet members, we have degeneracy at tree-level within the two representations with
\beq
m_{H_5}^2=3(\lam_5 \sh^2+\lam_4 \ch^2)v^2\,,\quad m_{H_3}^2=\lam_4 v^2\,,
\label{mhfivethree}
\eeq
implying that these masses can be chosen independently of the $\hone$--$\honepr$ sector.

The couplings of the $H$ relative to the SM are:
\beq
C_F={\cosa\over \ch}\,,\qquad C_V=\ch\cosa+{2\sqrt 2\over \sqrt 3}\sh\sina\,.
\label{tripcoup}
\eeq
Note that if $\sh$ is sizable, then $C_V$ will be enhanced relative the SM value of $1$ and the fermonic couplings will also be enhanced.
As noted earlier, the angle $\theta_H$ can be chosen to be in the first quadrant: $0\leq \theta_H\leq \pi/2$.  For a full range of possible phenomenology, we must  explore $0\leq \alpha\leq 2\pi$. 
In passing, we note that if we require $C_F=1$ then $\cosa=\ch$, and plugging into the expression for $\cv$ we find 
that $\ch^2=1$ is required if we demand also that $C_V=1$.

\begin{figure}[t]\centering
\includegraphics[width=6cm]{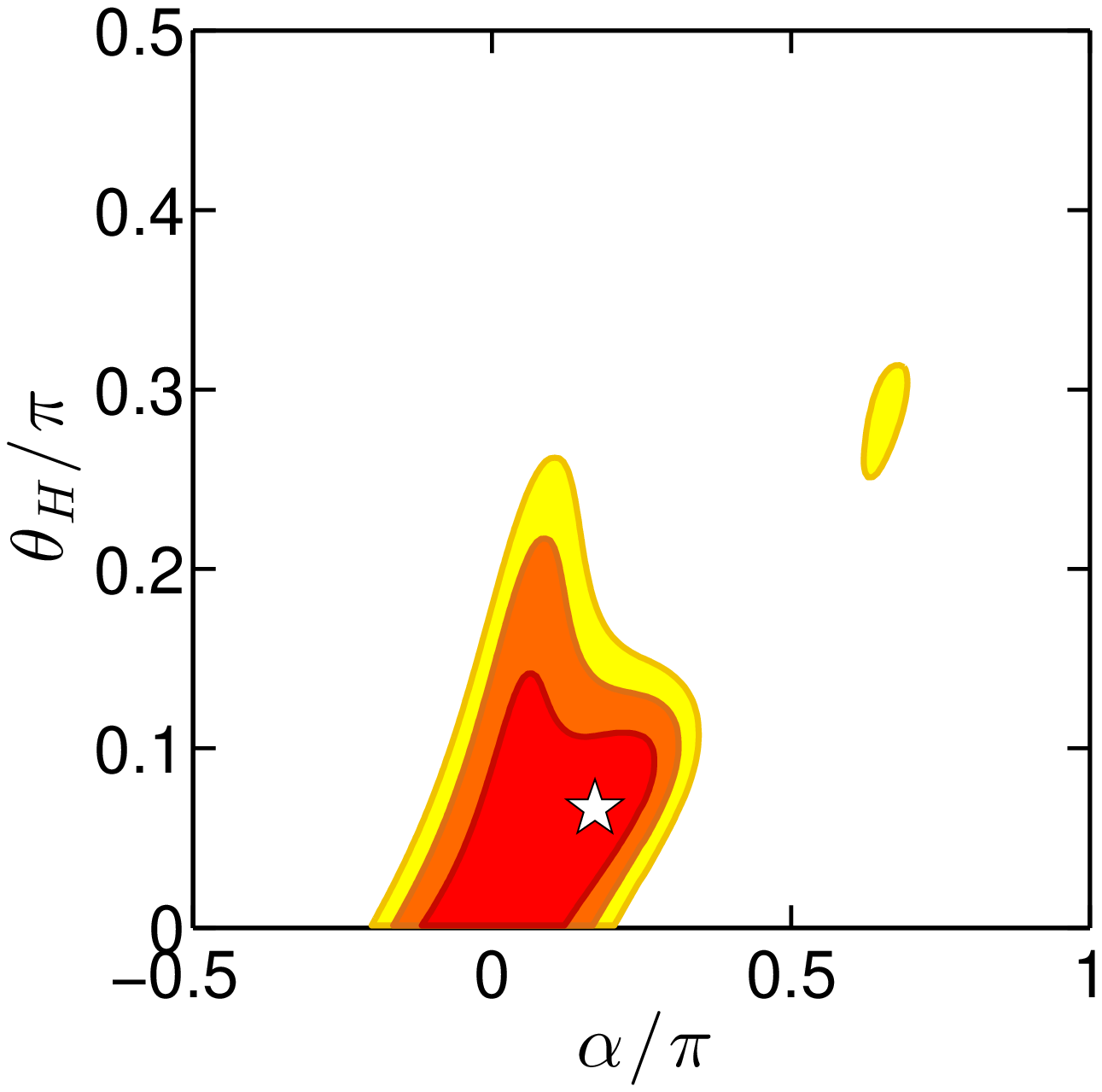}\quad
\includegraphics[width=6cm]{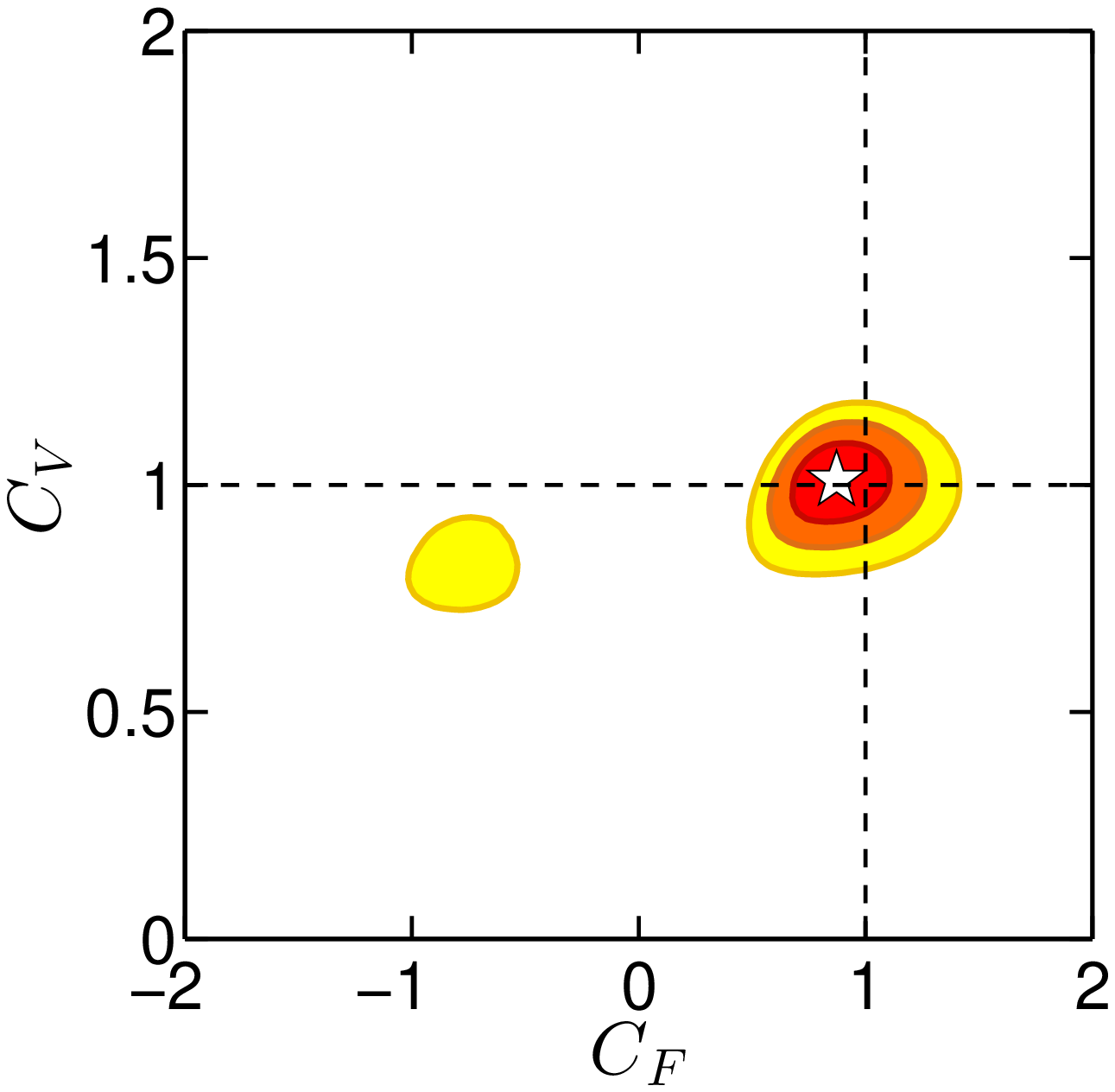} 
\caption{Fit for the Georgi--Machacek triplet model \cite{Georgi:1985nv}
assuming that $H=\cosa \hone+\sina\honepr$ is the observed state at
$125.5$~GeV. (The right plot is also valid for 2HDMs of Type~I.)  See text for details. \label{fig:tripletH} }
\end{figure}

The interesting question we want to answer is what does the LHC data allow for $\theta_H$ and $\alpha$. 
The result is shown in Fig.~\ref{fig:tripletH}, on the left in the $\theta_H$ versus $\alpha$ plane and 
on the right in the $\CV$ versus $\cf$ plane.
As expected, the preferred region lies at small $\alpha$ and small $\theta_H$,  roughly  $\alpha \in [0,\,\pi/4]$ and $\theta_H \in [0,\,0.1\pi]$, 
leading to a very SM-like picture in the  $\CV$ versus $\cf$ plane. 

At the best-fit point, $\alpha\sim 0.2\pi$, $\theta_H\sim 0.07\pi$,  and taking $m_{H'}=m_H/2$ (so as to avoid $H\to \hpr\hpr$ decays, see below), we find:
\beq
\bar\lam_{13}\sim 0.21\,,\quad\bar\lam_{23}\sim 2.93\,,\quad \bar\lam_3 \sim 0.42\,,
\eeq
perfectly consistent with the vacuum stability conditions given earlier and with perturbativity for the couplings themselves.  As $m_{H'}$ increases, $\bar\lam_{23}$ increases (when holding $\alpha$ and $\theta_H$ at their best-fit values). For example, at $\mhpr=400\gev$, we have $\bar\lam_{13}=1.14$, $\bar\lam_{23}=38.25$ and $\bar\lam_3=-5.32$. From \Eq{lambardefs} we see that this is still within the perturbative limits defined as $|\lam_i|<4\pi$.

We have seen that the SM-like nature of the observed $125.5\gev$ state requires small $\theta_H$ and $\alpha$, implying  that the  $H$ state will be mostly $\hone$ and that the $H'$ state will be mostly $\honepr$.  Further, from \Eq{monematrix} and the above results for the $\bar\lam_i$ we see that it is most natural  for  the mass of the $H'$ to be smaller than the mass of the $H$ for moderate values of the $\lam_i$.  This brings up the possibility that $H\to H'H'$ decays could be possible.  If present, they would significantly deplete the SM decay modes of the $H$ and the fit to the data would be bad. The $HH'H' $ coupling is given by (using Eq. (2.21) of \cite{Gunion:1989ci} and the notation $\ca\equiv\cos\alpha$, $\sa\equiv \sin\alpha$)
\beq
-2 H H'^2 \left[4 \lam_3 \ca^3 \ch+ 
   4 [9 (\lam_1 + \lam_3 )- 2 \lam_3] \ca\ch\sa^2 + 
   \sqrt 6 \lam_3 \ch \sa^3 - \sqrt 6 \lam_3
   \ca^2\sa(2\ch-9\sh)\right] v  \,.
\eeq
At the best fit point, the coefficient of $HH'^2$ is $\sim -0.57v$ for $\mhpr=m_H/2$, falling slowly as $\mhpr$ decreases. Since this is a large coupling, $\br(H\to H'H')$ would be large at the best-fit point  if this decay is allowed. Thus, our fitting results must be taken to apply only to the situation where $\mhpr>m_H/2$.  As discussed above, this presents no particular problem in the context of the model. 
 
 There are also couplings of the $H$ to pairs of five-plet or triplet members (Eq.~(2.22) of \cite{Gunion:1989ci}).  Thus, to avoid the associated decays of the $H$ we need to require $\mhfive>m_H/2$ and $\mhthree>m_H/2$, as \Eq{mhfivethree} shows is easily arranged for appropriate choices of (the independent parameters) $\lam_4$ and $\lam_5$. In fact, experimental limits on the charged Higgs members of the five-plet and triplet from LEP \cite{Abbiendi:2013hk} are of order $80\gev$ and from LHC  of order $120\gev$ \cite{Chatrchyan:2012vca} assuming decay to $\tau^+\nu$.  Limits 
on the doubly-charged Higgs of the 5-plet from the LHC \cite{CMS:double} are of order $300\gev$ (for decays to two charged leptons).  Thus it seems certain that  the (degenerate) masses of all the five-plet and all the triplet Higgses are necessarily  $>m_H/2$.  Note that this automatically means that the $H\to \hthree Z$ and $H\to \hthreepm\wmp$ decays which could be significant (see Eq.~(2.15) of \cite{Gunion:1989ci}) will also be forbidden.

\begin{figure}[t]\centering
\includegraphics[width=6cm]{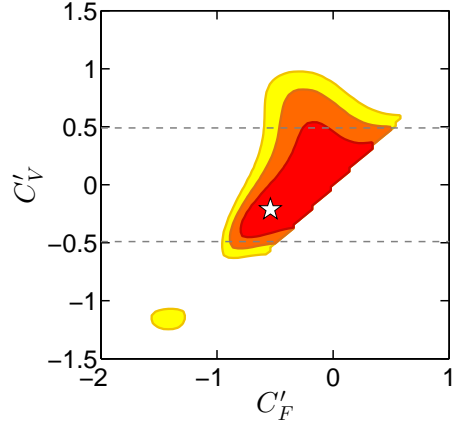}
\caption{Fit of $C_V^{\,\prime}$ versus $C_F^{\,\prime}$ of the $H'$ in the Georgi--Machacek triplet Higgs model with $m_H=125.5$~GeV.  
The regions above and below the dashed lines are excluded by LEP constraints for $m_{H'}=100$~GeV.
\label{fig:triplet-alt} }
\end{figure}

Of course, it is certainly interesting to consider the $H'$ itself.  Its couplings relative to the SM are
\beq
C_F^{\,\prime}=-{\sina\over \ch}\,,\qquad C_V^{\,\prime}={2\sqrt 2\over \sqrt 3}\sh\cosa-\ch\sina\,.
\label{tripcoupHprime}
\eeq
For small $\alpha$ and $\theta_H$, both will be small --- the $H'$ will be weakly coupled to both fermions and vector bosons.  This is illustrated by plotting the preferred regions in the
 $C_V^{\,\prime}$ versus $C_F^{\,\prime}$ plane,  displayed in Fig.~\ref{fig:triplet-alt} (where we are assuming, as above, that $H\to H'H'$ decays are forbidden).

In the mass region $\mhpr \in[m_H/2,m_H]$ only LEP2 data could potentially yield direct constraints on the $\hpr$.  For the lower portion of this mass range, $\epem\to Z^*\to ZH'$ limits are significant and would eliminate some portion of the larger $|\cv'|$ region of Fig.~\ref{fig:triplet-alt}. Using Table 14 of \cite{Schael:2006cr} and noting that $\br(H' \to b\anti b)$ will be approximately the same as for a SM Higgs boson, we see that $\cv'^2$ is limited to $\lsim 0.028$ at $\mhpr \sim 63\gev$ rising to $\lsim 0.044$ at $\mhpr=80\gev$ and $\lsim 0.24$ at $\mhpr=100\gev$. Thus, for the plot of Fig.~\ref{fig:triplet-alt} to not conflict with LEP2 95\% CL limits over the full $68\%$ ($95\%$) CL regions of the plot would require $\mhpr>100\gev$ ($>112\gev$).  For the best fit point, the LEP constraints are obeyed so long as $\mhpr>85\gev$.  

The fact that the $\hpr \hthree Z$  coupling is large (in fact, enhanced) for small $\theta_H$ will imply constraints coming  from $\epem\to Z^* \to \hpr \hthree$ on the lower $\hpr$ masses  should $\mhthree$ be small enough. For the $\mhpr=85\gev$ lower bound associated with the best-fit point (see above), Table~18 of \cite{Schael:2006cr} shows that $\mhthree>110\gev$ is required, whereas for $\mhpr$ close to $m_H$, there are no constraints for any $\mhthree>m_H/2$.  In any case, all such constraints are avoided if $\mhthree$ is above $m_H$, as easily arranged given \Eq{mhfivethree}, and almost guaranteed given the strong limits on its degenerate charged Higgs partner discussed above.  

Of course, to avoid LEP2 limits on the $\hpr$ the easiest choice  is to take $\mhpr>m_H$.  In this region, LHC constraints derived from the $ZZ\to 4\ell$ channel must be examined. To do so, we need to first recall that $\mu(X\to H \to 4\ell)$ scales as $C_X^2\br(H\to ZZ)/ \br (H_{\rm SM} \to ZZ)$, where $X=$~ggF or VBF. Even though $\cv'^2$ is typically suppressed, \eg\ $\cv'^2\sim 0.06$ at the best fit point for the $H$, since both the partial width and total width are typically dominated by the $VV$ final state $\br(H\to ZZ)/ \br (H_{\rm SM} \to ZZ)$ is typically of order 1. In this approximation $\mu({\rm VBF}\to H'\to ZZ)$ will be suppressed because of the suppressed $\cv'^2$ and  $\mu({\rm ggF}\to H'\to ZZ)$  will be suppressed by the small $C_F'^2$ values. Thus, except for the large $\cv'$ region of Fig.~\ref{fig:triplet-alt},  we expect that the LHC bounds are satisfied for any $\mhpr>m_H$.  Further, $\hpr\to 4\ell$  estimates should also take into account $\hpr\to HH$  decays, present when $\mhpr>2m_H$.  These decays would deplete the $4\ell$ channel, making it easier to satisfy the $4\ell$ constraints.  LHC constraints on the $HH$ type final state are not currently available from ATLAS and CMS.  Finally, the $H'\to\tau\tau$ partial width will not be enhanced in this model since $|C_F'|<1$ (see Fig.~\ref{fig:triplet-alt}). Coupled with the reduced ggF rate, this will mean (unlike the 2HDM models) that the constraints from this channel will not impact the 95\% CL region of the $H$ fit even before allowing for $\hpr\to HH$ decays.

\section{Conclusions}

Using all publicly available results from the LHC and Tevatron experiments, 
we determined combined likelihood ellipses for the Higgs signal around $125.5$~GeV 
in the (ggF+ttH) versus (VBF+VH) production plane for various independent final states: 
$\gam\gam$, $ZZ$, $WW$, $\tau\tau$ and $b\anti b$. We presented parameterizations 
of these ellipses that should be of general utility for exploring different types of models.

Any model in which the Lagrangian structure has a SM-like form can be parameterized  via scaling factors, $\cu$, $\cd$ and $\cv$, for the up-quark, down-quark, lepton and vector boson couplings (relative to SM values), respectively.  Additional New Physics contributions to the one-loop gluon and photon couplings can be allowed for by writing scaling factors for the $gg$ and $\gam\gam$ couplings in the form  $C_g=\overline \cg+\dcg$ and $C_\gam=\overline C_\gam+\dcp$ where the $\overline C_{g,\gam}$ values are those predicted for given $\cu$, $\cd$ and $\cv$ using SM particle loops only.  We can also allow for invisible/unseen decay modes of the Higgs by adding an invisible component to Higgs decays parameterized by $\brinv$. In terms of these input parameters, the $\chisq$ associated with each ellipse can be calculated.  In this way, we were able to explore the behavior of the total
$\chisq$ as a function of any one parameter (profiling over the other parameters that were allowed to vary freely in a given case) and also to determine the 68\%, 95\% and 99.7\% contours in various 2-D planes of any two of the freely varying parameters.  

The most general fits considered were those in which $\cu,\cd,\cv,\dcg,\dcp$ were all allowed to vary freely. 
If there are no unseen (as opposed to truly invisible) decay modes of the Higgs,
one finds that the observed $125.5\gev$ state prefers to have quite SM-like couplings 
whether or not  $\brinv=0$ is imposed --- 
more constrained fits, for example taking $\dcg=\dcp=0$ while allowing $\cu,\cd,\cv$ to vary, inevitably imply that the other parameters must lie even closer to their SM values.   

Allowing for invisible decays of the $125.5\gev$ state through $\brinv>0$ does not change the best-fit parameter values but does widen the $\dchisq$ distributions somewhat leading to important implications, \eg, 
for decays into dark matter particles.  In particular, we found that at 95\% CL there is  still considerable room for such Higgs decays, up to $\brinv\sim 0.38$ when $\cu,\cd,\cv,\dcg,\dcp$ are all allowed to vary independently of one another. In comparison, a fit for which $\cu,\cd$ are allowed to vary freely, but $\cv\leq 1$ is required (as appropriate for any doublets+singlets model) and $\dcg=\dcp=0$ is imposed, yields $\brinv\lsim 0.24$ at 95\% CL. Even requiring completely SM couplings for the Higgs ($\cu=\cd=\cv=1$, $\dcg=\dcp=0$) still allows $\brinv\leq 0.19$ at 95\%~CL.
It is worthwhile noting that for $\cv \leq 1$, the limits on $\brinv$ from global coupling fits are currently more constraining than those from direct searches for invisible decays, \eg, in the $ZH\to\ell^+\ell^- + E_T^{\rm miss}$ mode;  
thus for $\cv \leq 1$ the limits on  merely unseen (\ie\ not strictly invisible) decays are similar to the ones on $\brinv$.

As part of the fitting procedure, the total width of the Higgs relative to the SM prediction is computed as a function of the parameters and a $\dchisq$ distribution for $\Gamma_{\rm tot}/ \Gamma_{\rm tot}^{\rm SM}$ is obtained.  Assuming no unseen, but potentially visible, decays, we found $\Gamma_{\rm tot}/ \Gamma_{\rm tot}^{\rm SM}\in[0.5,2]$ at 95\% CL for the case where $\cu,\cd,\cv,\dcg,\dcp$ and $\brinv$ are all allowed to vary freely, while $\Gamma_{\rm tot}/ \Gamma_{\rm tot}^{\rm SM}\in[1,1.25]$ at 95\% CL if $\cu=\cd=\cv=1$, $\dcg=\dcp=0$ are imposed and only $\brinv\geq 0$ is allowed for.  These are useful limits given the inability to directly measure $\Gamma_{\rm tot}$ at the LHC.
Of course, if there are unseen (but not invisible) decays, there is a flat direction that would prevent setting limits on the total width.

In the second part of the paper, we then examined implications of these results in the context of some simple concrete models with an extended Higgs sector: the Type I and Type II Two-Higgs-doublet models; the Inert Doublet Model; and the custodially symmetric triplet Higgs model.  
Concretely, we used the combined likelihood ellipses to constrain the parameter spaces with 
corresponding implications for the properties of the other Higgs boson(s) of the model. 
In particular, the ability to discover a 2nd neutral Higgs boson with mass above $125.5\gev$ in, 
\eg, the $4\ell$ mode can be quantified.

In the 2HDM, enhancement of the signal strength for a 2nd neutral (scalar or pseudoscalar) Higgs boson with mass above $125.5\gev$ can occur in both the $4\ell$ and $\tau\tau$ channels. Therefore additional constraints on $\alpha$ and $\beta$ can be set unless the decay of the heavier Higgs to a pair of the $125.5\gev$ states dominates.  Generally the signals in both channels can be  at a level of interest for future LHC runs.
In the triplet model, when the second Higgs, $H'$, is heavy  the LHC bounds in both the $H'\to 4\ell$ and $H'\to \tau\tau$ channel are generally satisfied even without
taking into account the heavy Higgs decays into pairs of $125.5$~GeV Higgses. 
Only the region of parameter space with large $C'_V$ requires a large branching fraction into Higgs pairs to deplete the $4\ell$ signal. 
We stress that in both these  models the heavy Higgs to Higgs pair decays are generically important when allowed, implying that ways must be found to be sensitive to the $4b$, $b\bar{b}\tau\tau$ and $4\tau$ final states resulting therefrom.

In the Inert Doublet Model, the inert Higgs states can only be pair-produced and therefore are not currently constrained. However, we showed that the bound on the invisible decay of the $125.5$~GeV SM-like Higgs, relevant when one inert Higgs is lighter than $\approx 60$~GeV,  constrains the allowed range for the two-photon width.  Thus, a precise determination of $C_\gamma$  could rule out light inert Higgs dark matter.

\section*{Acknowledgements} 

This work originated from the workshops ``Implications of the 125~GeV Higgs boson'', 
which was held  18--22 March at LPSC Grenoble, and 
``The LHC Higgs Signal: Characterization, Interpretation and BSM Model Implications'', 
held 22--26 April 2013 at UC Davis. 
Partial support by US DOE grant DE-FG03-91ER40674 and by IN2P3 under contract PICS FR--USA No.~5872 is gratefully acknowledged. 
GB, BD, SK acknowledge partial support from the French ANR~DMAstroLHC.
UE acknowledges partial support from the French ANR~LFV-CPV-LHC, ANR~STR-COSMO and the European Union FP7 ITN INVISIBLES (Marie Curie Actions,~PITN-GA-2011-289442).

\appendix

\section{Combining likelihoods of different experiments in the Gaussian
approximation}

As function of the model-dependent signal rates $\mu_i$ (where $i$
stands for $\gam\gam$, $VV^{(*)}$, $b\bar{b}$ and $\tau\tau$ (or
$b\bar{b}=\tau\tau$)), the likelihoods in the Gaussian approximation in
the $\mu({\rm ggF + ttH})$ versus $\mu({\rm VBF + VH})$ plane obtained
by the experiment $j$ (where $j$ stands for ATLAS, CMS or the Tevatron)
can be expressed as $\chi^2_{i,j}$ with
\bea\label{eq:a1}
\chi_{i,j}^2 &=& a_{i,j}(\mu_i^{\rm{ggF}}-\hat{\mu}_{i,j}^{\rm{ggF}})^2
+2b_{i,j}(\mu_i^{\rm{ggF}}-\hat{\mu}_{i,j}^{\rm{ggF}})
(\mu_i^{\rm{VBF}}-\hat{\mu}_{i,j}^{\rm{VBF}})
+c_{i,j}(\mu_i^{\rm{VBF}}-\hat{\mu}_{i,j}^{\rm{VBF}})^2\nonumber\\
&\equiv& a_{i,j}(\mu_i^{\rm{ggF}})^2+c_{i,j}(\mu_i^{\rm{VBF}})^2
+2b_{i,j}\mu_i^{\rm{ggF}}\mu_i^{\rm{VBF}}
+d_{i,j}\mu_i^{\rm{ggF}}+e_{i,j}\mu_i^{\rm{VBF}}+\dots
 \,,
\eea
where $\hat{\mu}_{i,j}^{\rm{ggF}}$ and $\hat{\mu}_{i,j}^{\rm{VBF}}$
denote the best-fit points of the experiment $j$.\footnote{Of course $\chi_{i,j}^2$ defined in this way 
is not an absolute $\chi^2$, but rather a $\Delta\chi^2$ relative to the best fit value of the experiment in a given channel.} 
The dots denote terms
independent of $\mu_i$, which are irrelevant for $\chi_{i}^2$ relative
to the best-fit points as defined in Eq.~\eqref{eq:1}. $d_{i,j}$ and
$e_{i,j}$ are given by
\beq\label{eq:a2}
d_{i,j} = -2a_{i,j}\hat{\mu}_{i,j}^{\rm{ggF}}
          -2b_{i,j}\hat{\mu}_{i,j}^{\rm{VBF}},
\qquad
e_{i,j} = -2c_{i,j}\hat{\mu}_{i,j}^{\rm{VBF}}
          -2b_{i,j}\hat{\mu}_{i,j}^{\rm{ggF}}\; .
\eeq
Combining experiments leads to
\beq\label{eq:a3}
\chi_{i}^2=a_{i}(\mu_i^{\rm{ggF}})^2+c_{i}(\mu_i^{\rm{VBF}})^2
+2b_{i}\mu_i^{\rm{ggF}}\mu_i^{\rm{VBF}}
+d_{i}\mu_i^{\rm{ggF}}+e_{i}\mu_i^{\rm{VBF}} \,,
\eeq
with
\beq\label{eq:a4}
a_i = \sum_{j}a_{i,j},\ b_i = \sum_{j}b_{i,j},\
c_i = \sum_{j}c_{i,j},\ d_i = \sum_{j}d_{i,j},\
e_i = \sum_{j}e_{i,j}\; .
\eeq
From \eqref{eq:a3} one obtains Eq.~\eqref{eq:1} with
\beq\label{eq:a5}
\hat{\mu}_{i}^{\rm{ggF}}= \frac{b_i e_i - c_i d_i}{2(a_i c_i - b_i^2)},
\quad
\hat{\mu}_{i}^{\rm{VBF}}= \frac{b_i d_i - a_i e_i}{2(a_i c_i - b_i^2)}
\; .
\eeq

\section{Comparison with ATLAS and CMS couplings fits}

Coupling fits using all available results up to the Moriond 2013 conference have been performed individually by ATLAS and CMS~\cite{ATLAS-CONF-2013-034,CMS-PAS-HIG-13-005}. While the present paper aims at presenting combined results from ATLAS, CMS and Tevatron using parameterizations motivated by various models of New Physics, the coupling fits made by ATLAS and CMS that combine the information from different channels can be used to check the robustness of the implementation of the experimental searches as presented in Section~\ref{ssellipse}. In particular, deviations of our results from those obtained by the ATLAS and/or CMS 
give a measure for the importance of the missing correlations mentioned at the end of Section~\ref{sintro}. 

\begin{figure}[!h]\centering
\includegraphics[width=6cm]{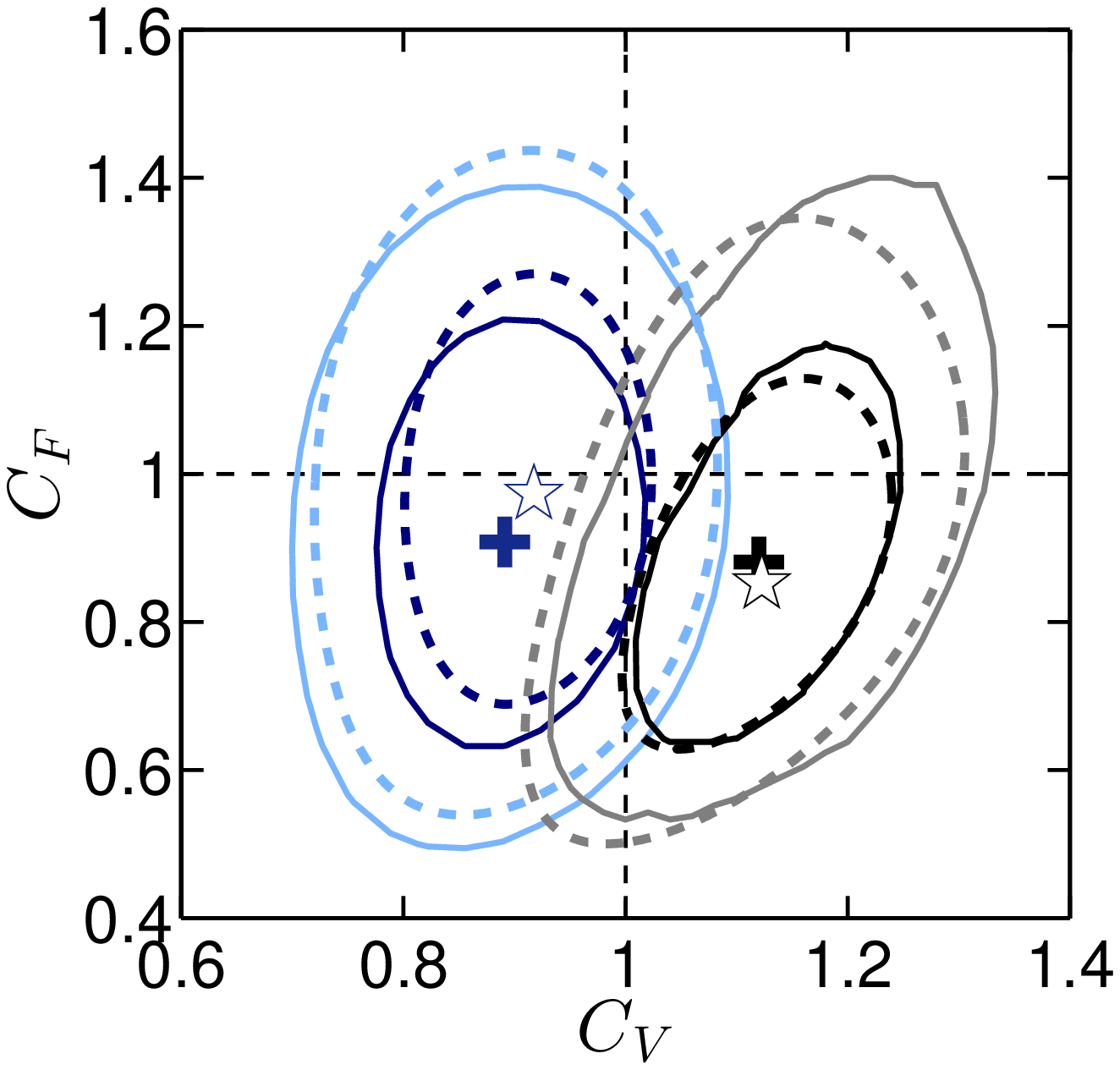}\quad
\includegraphics[width=5.85cm]{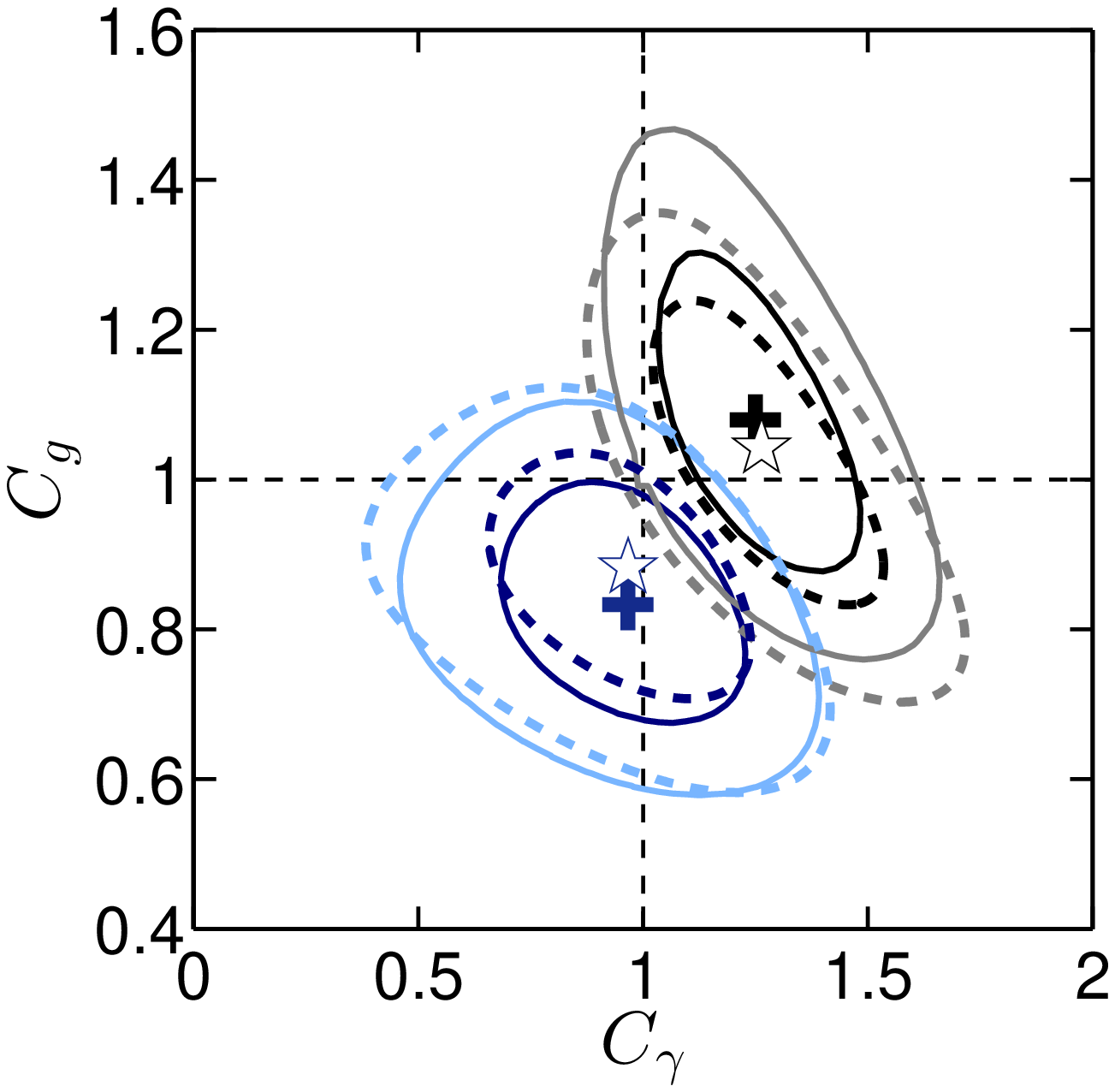}
\caption{Fit to the couplings $(C_F,C_V)$ (left) and $(C_g,C_\gamma)$
(right) using separately results from ATLAS and CMS up to the Moriond
2013 conference. The black and grey (dark blue and light blue) contours
show the 68\% and 95\% CL regions for ATLAS (CMS), respectively. The
solid contours correspond to the results published by the experimental
collaborations, while dashed contours have been obtained using the
fitted signal strength ellipses as determined using the separate data
for ATLAS (CMS) in the manner described in
Section~\ref{ssellipse}.\label{fig:lhc-check} }
\end{figure}

For the aim of comparison, 
we have performed fits to the $(C_F,\,C_V)$ and $(C_g,\,C_\gamma)$ couplings, using separately only ATLAS or CMS data up to the Moriond 2013 conference. Figure~\ref{fig:lhc-check} compares our results to those published by ATLAS~\cite{ATLAS-CONF-2013-034} and CMS~\cite{CMS-PAS-HIG-13-005}. 
We obtain good agreement  in all four cases. The ATLAS (CMS) best fit points are at distances of $\sqrt{(\Delta C_V)^2 + (\Delta C_F)^2} = 0.03$ (0.07) and $\sqrt{(\Delta C_\gamma)^2 + (\Delta C_g)^2} = 0.04$ (0.05) from the reconstructed best fit points, and good coverage of the 68\% and 95\% CL regions is observed.

For completeness, we note that our fit for $(C_F,\,C_V)$ combining ATLAS
and CMS results up to the LHCP 2013 conference can be seen in the  right
plot of Fig.~\ref{fig:tripletH}, and the one for  $(C_g,\,C_\gamma)$ in
the middle plot in Fig.~\ref{fig:CPadd-CGadd}, taking
$C_{g,\gamma}=1+\Delta C_{g,\gamma}$. 

\end{document}